\pdfoutput=1
\documentclass[12pt,reqno]{amsart}
\usepackage{amsmath,amsfonts,amssymb,amscd,amsthm,amsbsy,color}
\usepackage[pdftex]{graphicx}
\usepackage{comment}
\usepackage[format=plain,
justification=raggedright,singlelinecheck=false]{caption}

\usepackage{wrapfig}
\usepackage{fancybox}
\usepackage[maxfloats=100]{morefloats}

\usepackage{multimedia}
\usepackage{graphicx,epsfig}

\numberwithin{equation}{section}

\newtheorem{theorem}{Theorem}%[section]
\newtheorem{meta-thm}[theorem]{Meta-Theorem}

\newtheorem{remark}[theorem]{Remark}
\newtheorem{definition}[theorem]{Definition}

\newcommand\beq[1]{ \begin{equation}\label{#1} }
\newcommand{\eeq}{ \end{equation} }

\newcommand\beqa[1]{ \begin{eqnarray} \label{#1}}
\newcommand{\red}{\textcolor{black}}

\newcommand{\eeqa}{ \end{eqnarray} }

\newcommand{\beqano}{ \begin{eqnarray*} }

\newcommand{\eeqano}{ \end{eqnarray*} }

\newcommand\equ[1]{{\rm (\ref{#1})}}

\def\L{{\mathcal L}}

\def\G{{\mathcal G}}

\def\H{{\mathcal H}}

\def\R{{\mathcal R}}

\def\integer{{\mathbb Z}}

\def\real{{\mathbb R}}

\def\zed{{\mathbb Z}}

\begin{document}

\title[Bifurcations of lunisolar secular resonances for space debris orbits]
{Bifurcation of lunisolar secular resonances for space debris orbits}

\author[A. Celletti]{Alessandra Celletti}
\address{
Department of Mathematics, University of Roma Tor Vergata, Via della Ricerca Scientifica 1,
00133 Roma (Italy)}
\email{celletti@mat.uniroma2.it}

\author[C. Gale\c s]{C\u at\u alin Gale\c s}
\address{
Department of Mathematics, Al. I. Cuza University, Bd. Carol I 11,
700506 Iasi (Romania)}
\email{cgales@uaic.ro}

\author[G. Pucacco]{Giuseppe Pucacco}
\address{
Department of Physics, University of Roma Tor Vergata, Via della Ricerca Scientifica 1,
00133 Roma (Italy)}
\email{pucacco@roma2.infn.it}

\thanks{A.C. was partially supported by the European Grant MC-ITN Stardust, PRIN-MIUR 2010JJ4KPA$\_$009 and GNFM/INdAM.
C.G. was supported by a grant of the Romanian National Authority for
Scientific Research and Innovation, CNCS - UEFISCDI, project number
PN-II-RU-TE-2014-4-0320 and by GNFM/INdAM.
G.P. was partially supported by the European Grant MC-ITN Stardust and GNFM/INdAM}

%%%%%%%%%%%%%%%%%%%%%%%%%

\baselineskip=18pt              %% DRAFT MODE -- double spaced.

%%%%%%%%%%%%%%%%%%%%%%%%%

%\address{}

%\email{}

\begin{abstract}
Using bifurcation theory, we study the secular resonances induced
by the Sun and Moon on space debris orbits around the Earth. In
particular, we concentrate on a special class of secular
resonances, which depend only on the debris' orbital inclination.
This class is typically subdivided into three distinct types of
secular resonances: those occurring at the critical inclination,
those corresponding to polar orbits, and a third type resulting
from a linear combination of the rates of variation of the
argument of perigee and the longitude of the ascending node.

The model describing the dynamics of space debris includes the
effects of the geopotential, as well as the Sun's and Moon's
attractions, and it is defined in terms of suitable action-angle
variables. We consider the system averaged over both the mean
anomaly of the debris and those of the Sun and Moon. Such
multiply-averaged Hamiltonian is used to study the lunisolar
resonances which depend just on the inclination.

Borrowing the technique from the theory of bifurcations of
Hamiltonian normal forms, we study the birth of periodic orbits
and we determine the energy thresholds at which the bifurcations
of lunisolar secular resonances take place. This approach gives us
physically relevant information on the existence and location of
the equilibria, which help us to identify stable and unstable
regions in the phase space. Besides their physical interest, the
study of inclination dependent resonances offers interesting
insights from the dynamical point of view, since it sheds light on
different phenomena related to bifurcation theory.
\end{abstract}

\subjclass[2010]{37N05,37L10,70F15}
\keywords{Bifurcations, Secular resonances, Space debris}

\maketitle

%\tableofcontents

\section{Introduction}\label{intro}
The gravitational effects of the Sun and Moon in the study of the
dynamics of satellites and space debris have recently gained a
renewed interest. Indeed, the awareness that a careful
investigation of the dynamics of space debris is nowadays
mandatory leads to a thorough study of the influence of the Sun
and Moon, which are known to provoke important effects in specific
regions around the Earth (\cite{RARV15, DRADVR15}).

At the present state, one estimates that about $3 \times 10^8$
objects with size larger than 1 mm and about $3.5 \times 10^5$
objects larger than 1 cm orbit around the Earth. Leaving all these
objects without control could provoke extremely dangerous events.
In particular, dramatic scenarios can directly involve operative
satellites, or even manned space missions, like the International
Space Station. This threat certainly suffices to motivate the
efforts which are currently done toward understanding the dynamics
of such objects. A mathematical approach to study the dynamics of
space debris will strongly depend upon the location of the object
one wants to study, since all possible forces (geopotential,
atmospheric drag, lunar attraction, solar influences, etc.)
contribute to the model in a different way, according to the
altitude of the object under investigation.

Adopting the nowadays widespread classification, \red{the
circumterrestrial space}  is divided into following
regions:

$\bullet$ LEO (low--Earth orbit), running from 90 to 2\,000 km in altitude;

$\bullet$ MEO (medium--Earth orbit), spanning the region between 2\,000 and 30\,000 km in altitude;

$\bullet$ GEO (geostationary--Earth orbit), used for
trajectories around the geosynchronous orbit at altitude of 35\,786 km;

$\bullet$ HEO (high--Earth orbit), corresponding to the region above GEO.

In each of the above regions some forces prevail with respect to
others. For example, in LEO, beside the gravitational attraction
of the Earth, one definitely must consider the effects of the
atmospheric drag (thereby dealing with a dissipative dynamical
system). In MEO, the gravitational \red{effects of the point--mass Earth} and that of
the terms corresponding to \red{the second degree and order gravity--field}
coefficients $J_2$ and $J_{22}$ of the spherical harmonics are
very important, but also the influence of the Moon and the Sun,
including the effect of the solar radiation pressure, become increasingly
relevant. Finally, \red{at GEO, as well as in HEO, the lunisolar perturbations
become of the same order as the Earth's oblateness $J_2$, and both
are more important than the $J_{22}$ (tesseral terms) for long-term orbital
dynamics}  (see \cite{chaogick,deleflie2011,LDV,rossi2008,VL,VLD}).
We claim that the model we will introduce in
Section~\ref{sec:model} is conceived to describe objects in MEO, GEO, HEO, but
not in LEO, and therefore
it does not include the atmospheric drag. \\

In this work we concentrate on the effect of lunisolar
perturbations (\cite{BG,B2,ElyHowell}) and precisely on some
secular resonances, which depend only on the inclination (see
\cite{HughesI}, compare also with \cite{HughesII}). With the terms
of \sl lunar \rm and \sl solar gravitational resonance \rm we mean
that there exists a commensurability relation of the type
\red{\beq{secres} k_1\dot\omega+k_2\dot\Omega+k_3\dot\omega_b+k_4\dot\Omega_b=0 \eeq for  some integers
$(k_1,k_2,k_3,k_4)$,} not all identically zero; in \equ{secres}
the quantity $\omega$ denotes the argument of perigee of the
debris and $\Omega$ is the longitude of the ascending node,  while $\omega_b$
and $\Omega_b$ are, respectively, the argument of perigee and the
longitude of the ascending node of the perturber. When the
third--body perturber is the Sun, the suffix will be $b=S$, while
it will be $b=M$ for the Moon. As pointed out in
\cite{HughesI}, taking into account only the secular effects due to
$J_2$, one can give simple expressions for the rates of variation
of $\omega$ and $\Omega$. Inserting such expressions in
\equ{secres} one obtains resonant relations involving the orbital
elements $(a,e,i)$ of the debris, where $a$ is the semimajor axis,
$e$ the eccentricity and $i$ the inclination. As we will see in
Section~\ref{sec:secres}, some of these resonances depend only on
the inclination and not on $a$ and $e$ (\cite{HughesI}). The
dynamics of such inclination--dependent--only resonances is the
object of study of the present work.

\vskip.1in

To enter the details, one can introduce three main classes of
secular resonances depending only on inclination (see
Section~\ref{sec:secres}): $(i)$ those corresponding to the
so-called critical inclination, $(ii)$ polar orbits, and $(iii)$
the resonances arising from linear combinations of the type
\red{$k_1 \dot\omega+ k_2 \dot\Omega=0$ with $k_1$,
$k_2 \in\integer \backslash \{0\}$}. The critical inclination
is the well-known value $i=63.4^o$; polar orbits correspond to
$i=90^o$; the resonances of type $(iii)$ occur at two specific
values of the inclination as given later in \equ{inc}. \red{These inclinations are referred  to the equatorial plane.}

A simple, but exhaustive, mathematical model describing the cases
$(i)$-$(ii)$-$(iii)$ can be obtained as follows. We introduce the
Hamiltonian function including the Keplerian part, the secular
part of the geopotential, limited to the most significant term
(the so-called \red{$J_2$ approximation), and the} contributions due to the
Moon and the Sun. Following \cite{HughesI,Lane,Giacaglia1974}, we
will conveniently express the elements of the debris with respect
to the \sl equatorial \rm plane and those of the Moon with respect
to the \sl ecliptic \rm plane; the advantage is that the
inclination of the Moon becomes nearly constant and the argument
of the perigee of the Moon as well as the longitude of the lunar
ascending node vary linearly with time. We refer the reader to
\cite{CGPRnote} for a thorough mathematical derivation of the
expansion of the lunisolar Hamiltonian.

We analyze each case (critical inclination, polar case, and linear
combination) by properly introducing an adapted system of resonant
coordinates; we are thus led to a resonant 2 degrees-of-freedom
Hamiltonian that we proceed to average with respect to the fast
angle. The action conjugated to the fast angle, say $T$, becomes
an integral of motion. Using a technique common in bifurcation
theory (see \cite{MP14,PM14,Pucacco}), and which has been
successfully applied to the study of the bifurcations of halo
orbits around the collinear Lagrangian points (\cite{CPS}), we
investigate the occurrence of bifurcations associated to the
lunisolar secular resonances. This leads to an exhaustive study of
the generation of families of periodic orbits as the level of the
integral $T$, obtained after the averaging process, is varied. We
stress that the study of lunisolar resonances is markedly
different with respect to the analysis of halo orbits used in
\cite{CPS}. In fact, besides the intrinsic constraints on some
dynamical quantities (e.g., positivity of the action variables),
the conditions for the existence of bifurcating families
of lunisolar resonances include
also physical limitations on the orbital elements (most notably
the eccentricity) in such a way that their values are compatible
with the fact that the perigee of the debris cannot be lower than
the Earth's radius (\cite{B1}).
%(see also \cite{PM14}, \cite{MP14}, \cite{Pucacco}) is that, beside the limitations
%intrinsic in the dynamics, we also need to respect the physical constraint due to the fact
%that our planet has a given radius. This will mean to limit the variation of some quantities
%(most notably the eccentricity)
%in such a way that their minimum value is compatible with the fact that the perigee of the debris
%cannot be lower than the Earth's radius. \\

Beside obtaining a physically relevant description of the
occurrence of periodic orbits, the results presented in the
current work also provide interesting insights into the phenomenon
of bifurcations (see also \cite{Breiter1999,B2}). In particular,
we will see that the various types $(i)$-$(ii)$-$(iii)$ of secular
resonances must be studied using different variants of the method
and that each case leads to a different dynamical behavior. For
example, in some cases we need to refine the method and perform
very accurate computations to reconstruct the true dynamics, since
several bifurcating families coexist (this will be achieved by
implementing some expansions to suitable higher orders). Moreover,
a connection with the geometric approach to integrable Hamiltonian
systems based on their invariants is performed in analogy with the
methods used to describe the phenomenon of critical inclination
in the most effective way (\cite{CDMcrit,CDDHcrit}).
This leads to remark that the methods and results presented in this
work have a twofold interest: we obtain information on the
behavior of lunisolar secular resonances which might be used
to study the dynamics of space debris, and we implement bifurcation
theory on some case studies, which show different behaviors from the
dynamical perspective.\\

As described before, our study of the bifurcations is based on a multiply averaged system.
To understand what happens in the non-averaged model, we provide qualitative
arguments: the basic (averaged) model admits whiskered tori and
normally hyperbolic invariant manifolds (hereafter, NHIM); the
non-averaged system in which the rates of variation of the longitudes of the ascending
node of Sun and Moon are assumed to be constant is described by a 2--dimensional
Hamiltonian system, for which KAM invariant tori exist, provided that suitable
assumptions and non-degeneracy conditions are satisfied; if we consider also
the rates of variation of the longitudes of the ascending
node of Sun and Moon, then we obtain a non--autonomous, 2--dimensional
Hamiltonian model, which might show the phenomenon of Arnold's diffusion
through transition chains between invariant tori.
The investigation of such non--averaged models might be the object of study of a future work.

\vskip.1in

This paper is organized as follows. In Section~\ref{sec:model} we
introduce a model describing lunisolar secular resonances; in
particular, we determine suitable series expansions obtained after
a multiple averaging process. Following \cite{HughesI}, we compute
in Section~\ref{sec:secres} the secular resonances that depend
just on the inclination of the satellite and which lead to the
cases $(i)$-$(iii)$ mentioned before. In
Section~\ref{sec:casestudy1} we perform a detailed investigation
of the generation of equilibria for the case study
\red{$\dot\omega+\dot\Omega=0$, a sample} of class $(iii)$. The
analytical study of the resonance $\dot\omega+\dot\Omega=0$ will
be complemented by a numerical investigation through the so-called
Fast Lyapunov Indicators (FLIs).   The other secular resonances of
type $(iii)$ are investigated in Section~\ref{sec:other}. Secular
resonances corresponding to the critical inclination are studied
in Section~\ref{sec:critical} and those corresponding to polar
orbits in Section~\ref{sec:polar}.
A qualitative description of the dynamics in the non--averaged problem
is provided in Section~\ref{sec:nonaveraged}.
Some conclusions are drawn in Section~\ref{sec:conclusions}.

\section{A model including the lunisolar effect}\label{sec:model}
\red{We consider the motion of a small body, say $D$, that we identify with a space debris. We assume that $D$ is subject to the gravitational influence of the Earth, Sun and Moon. The Hamiltonian describing de dynamics of $D$ has the form (see, e.g., \cite{Kaula, CGPRnote, DRADVR15}):}
%\red{
%$\begin{equation}\label{full_hamiltonian}
%\H=\H_{Kep}+\H_{geo}+\H_{Moon}+\H_{Sun}\ ,
%\end{equation}}
$$\H=\H_{Kep}+\H_{geo}+\H_{Moon}+\H_{Sun}\ ,$$
where $\H_{Kep}$
represents the Keplerian part, $\H_{geo}$ describes the perturbation due to the Earth, $\H_{Moon}$ and $\H_{Sun}$ denote the
contributions due to the Moon and Sun, respectively.

%\red{The Hamiltonian parts $\H_{Moon}$ and $\H_{Sun}$ are related to the well known disturbing functions $\R_{Moon}$, $\R_{Sun}$ (see %\cite{Kaula1962, Lane, CGPRnote}) through the following relations:}
%\red{\begin{equation}
%  \H_{Moon}=- \R_{Moon}\,, \qquad  \H_{Sun}=-\R_{Sun}.
%\end{equation}}

We will express the Hamiltonian $\H$ in terms of Delaunay action-angle variables, usually denoted as
$(L,G,H,M,\omega,\Omega)$, where the actions are defined by
\begin{equation}\label{Delaunayvar}
L=\sqrt{\mu_E a}\ ,\qquad G=L\sqrt{1-e^2}\ ,\qquad H=G\cos i
\end{equation}
with $\mu_E=\G m_E$ the product of the gravitational constant $\G$ and the mass $m_E$ of the Earth,
$a$ the semimajor axis, $e$ the orbital eccentricity, $i$ the inclination,
while the angle variables are the mean anomaly $M$, the argument of perigee $\omega$, and the longitude of the
ascending node $\Omega$.

\red{The quantities $a$, $e$, $i$, $\omega$, $\Omega$ and $M$, called orbital elements,
describe the dynamics of $D$. Under the effect of the Keplerian part, all orbital
elements, but the mean anomaly $M$ which varies linearly in time, are constant.
When a perturbation is considered, the orbital elements are no longer constant,
but rather change in time. The long--term variation of the orbital elements can be
studied by means of perturbation theory. The first step consists in expanding the
perturbation in Fourier series in terms of the orbital elements.
The coefficients of the series expansion depend on the semi--major axes, inclinations and eccentricities,
while the trigonometric arguments involve linear combinations of the following angles: the mean anomalies,
the longitudes of the nodes, the arguments of periapsides. Moreover, the disturbing function depends also
on the rotation, hence on the hour angle, of the disturbing body.}

\red{Within the infinite number of terms of the series expansion, only some
terms are really relevant for the long--term
evolution of the orbital elements.
In general, the terms of the expansion of the disturbing force are classified as follows:
\sl short periodic \rm terms (involving fast angles),
\sl secular \rm terms (independent of the fast angles), \sl resonant \rm terms (implying commensurabilities between
the fast angles). According to the averaging principle (see, e.g., \cite{MD1999}),
the effects of the short periodic terms average out over a long-time.
Hence, such terms can be dropped from the expansion and we can focus only on the most
relevant terms.}

\red{Two types of resonance affect the motion of the space debris (and artificial satellites):
$(a)$ \sl tesseral \rm resonances, occurring when there is a commensurability between the Earth's rotation period and
the orbital period of the space debris (\cite{CGmajor, CGext,VLD}), and $(b)$ \sl lunisolar \rm resonances,
which involve commensurabilities among the slow frequencies
of orbital precession of a satellite and the perturbing body
(\cite{HughesI, HughesII, ElyHowell, CGPRnote, DRADVR15}). Tesseral resonances provoke
variations of the semi--major axis on a time scale of the order of hundreds of days,
while lunisolar resonances influence the evolution of the eccentricity
and inclination on a much longer time scale, of the order of tens (or hundreds) of years.
In this work, we focus on the effects induced by the lunisolar resonances. Thus,
in the following sections we average the
disturbing functions over the mean anomalies of both the space debris and the perturbing bodies.}

As it is well known (see, e.g., \cite{Alebook}), the Keplerian
part of the Hamiltonian is given by
$$
\H_{Kep}(L)=-{{\mu_E^2}\over {2L^2}}\ .
$$

For later convenience, it is important to underline that the units
of length and time are normalized so that the geostationary
distance is unity (it amounts to $42\,164.1696$ km) and that the period
of Earth's rotation is equal to $2 \pi$. As a consequence, from
Kepler's third law it follows that $\mu_E=1$. Therefore, unless
the units are explicitly specified, the numbers appearing in the
following sections will be expressed in the above units.

\subsection{The perturbing function $\H_{geo}$ }\label{sec:geo}
As for $\H_{geo}$, we limit ourselves to the most
important contribution, corresponding to the $J_2$ gravity
coefficient of the secular part (see, e.g., \cite{CGmajor},
compare also with \cite{CGext}), precisely:
$$
\H_{geo}(L,G,H)={{R_E^2 J_2 \mu_E^4}\over {4}}\ {{1}\over
{L^3G^3}}\ (1-3{H^2\over G^2})\ ,
$$
where $R_E$ is the mean equatorial radius of the Earth and
$J_2=1.08263\times 10^{-3}$. This expression of the geopotential
corresponds to taking an average of the Hamiltonian over the mean
anomaly of the space debris as well as over the sidereal time of
the Earth, and to consider only the most important term of the
expansion in the spherical harmonics of the geopotential.

\subsection{The lunar potential}\label{sec:lunar}
As far as the lunar contribution is concerned, following
\cite{Lane,Giacaglia1974,HughesI}, it is
convenient to express the orbital elements of the satellite with
reference to the equatorial plane and the orbital elements of the
Moon with respect to the ecliptic plane. In fact, since the main
perturbing effect is due to the Sun, the motion of the lunar's
elements with respect to the celestial equator, in particular the
argument of perigee and the longitude of the ascending node, are
such that their changes are nonlinear. For instance, the longitude
of the ascending node varies between $-13^{\circ}$ and
$+13^{\circ}$ with a period of 18.6 years.
 On the contrary, if we consider the
elements of the Moon with respect to the ecliptic plane, then the
inclination $i_M$ is close to a constant (in analogy to $a_M$ and
$e_M$), while the variations of the argument of perihelion
$\omega_M$ and the longitude of the ascending node $\Omega_M$ are
approximately linear (see for example \cite{Simon1994}) with rates
$\dot\omega_M\simeq 0.164^{\circ}/day$, $\dot\Omega_M\simeq
-0.053^{\circ}/day$. \red{The rate of variation of the lunar mean anomaly is  $\dot M_M\simeq 13.06^{\circ}/day$.} This
implies that the change of $\omega_M+\Omega_M$ has a period of
8.85 years, while the variation of $\Omega_M$ has a period of 18.6
years.

%In the light of the above remark, it is convenient to express $\R_{Moon}$ in terms of the space debris elements referred to the
%celestial equator and to express the elements of the Moon with respect to the ecliptic plane. This approach
%has been adopted, e.g., in \cite{Lane}, \cite{Giacaglia1974}, \cite{HughesI}. As stated in \cite{HughesI},
%the advantage of this approach is that the elements
%of the Moon, $a_M$, $e_M$, $i_M$, are nearly constant, while $\omega_M$, $\Omega_M$, $M_M$ vary almost linearly with time (see for example \cite{Simon1994})
%with rates $\dot\omega_M\simeq 0.164^{\circ}/day$,
%$\dot\Omega_M\simeq -0.053^{\circ}/day$, $\dot M_M\simeq 13.06^{\circ}/day$, which imply that
%the change of $\omega_M+\Omega_M$ has a period of 8.85 years,
%while the variation of $\Omega_M$ has a period of 18.6 years.

In a first approximation, we assume that the Moon moves on an
elliptic orbit with semimajor axis $a_M=384\,748/42\,164.1696$
(expressed in units of the geostationary radius), eccentricity
$e_M=0.0549006$ and inclination $i_M=5^{\circ}15'$; the mass $m_M$
of the Moon, expressed in Earth's masses, is about equal to
0.0123. \red{The potential due to the lunar attraction will be computed as a truncation of the series expansion to the second
order in the ratio of the semi-major axes, so that the lunar potential is approximated
by quadrupole fields.} Moreover, we will consider the averages of the potential over the
mean anomalies of the debris and the Moon (see
Section~\ref{sec:casestudy1}). In fact, the mean anomaly of the
debris is a fast angle; on the other hand, the mean anomaly of the Moon can
be neglected, since the secular resonances we shall consider will not depend upon the mean anomaly of
the Moon as well as on that of the Sun (see Remark~\ref{ave} below).

We proceed now to give an explicit expansion of the lunar potential; to this end, we follow the approach of
\cite{Lane} and \cite{Giacaglia1974}, recently revisited in \cite{CGPRnote}.

We recall that we denote by $a_M$, $\omega_M$, $\Omega_M$ the semimajor axis, the argument of perigee,
the longitude of the ascending node of the Moon referred to the ecliptic plane, respectively.
After some computations for which we address the reader to \cite{CGPRnote}, we obtain that $\H_{Moon}$ is given by the following expression:
\begin{equation}\label{HR_Moon}
\H_{Moon}=-\R_{Moon}\,,
\end{equation}
where
\red{\beqa{Rmoon}
\R_{Moon}&=&{1\over 2}\ \G m_M\sum_{m=0}^2 \sum_{s=0}^2 \sum_{p=0}^2  {a^2\over a_M^3}
(-1)^{[{m\over 2}]}\ K_m\, K_s\ {{(2-s)!}\over {(2+m)!}}\nonumber\\
&&\ \times F_{2mp}(i)\ F_{2s1}(i_M)\ H_{2\  p\ 2p-2}(e)\, G_{2 1 0}(e_M)\nonumber\\
&& \times \Big\{ U_2^{m,-s} \cos\Bigl((2-2p)\omega+ m\Omega+s\Omega_M-s{\pi\over 2}-y_s\pi\Bigr)\nonumber\\
&&+ U_2^{m,s} \cos\Bigl((2-2p)\omega+m\Omega-s\Omega_M+s{\pi\over 2}-y_s\pi \Bigr)\Big\}\ .
\eeqa}
We consider the expansion up to degree 2 in $a/a_M$
and we retain only the terms of the expansion \red{which provide the average over the fast orbital phases (i.e., the mean anomalies of both the
debris and the Moon).}
%As a consequence, we define the functions $G_{npj}$ and $H_{nqr}$ as follows:
%$$
%G_{npj}(e)=\left\{%
%\begin{array}{ll}
%  X_0^{n, n-2p}(e) & \textrm{if\ } n-2p+j=0  \\
%  0 & \textrm{if\ } n-2p+j\neq 0 \\
% \end{array} \right.
%$$
%and
%$$
%H_{nqr}(e_M)=\left\{%
%\begin{array}{ll}
%  X_0^{-(n+1), n-2q}(e_M) & \textrm{if\ } n-2q+r=0  \\
%  0 & \textrm{if\ } n-2q+r\neq 0 \\
% \end{array} \right.\ ,
%$$
%respectively, where $X_0^{n, n-2p}(e)$ and $X_0^{-(n+1), n-2q}(e_M)$ are the Hansen coefficients (see \cite{Giacaglia1976}).
In \eqref{Rmoon} the following notation has been introduced: \red{$H_{2\  p\ 2p-2}(e)$ and $G_{2 1 0}(e_M)$ are,
respectively, the Hansen coefficients  $X_0^{2, 2-2p}(e)$ and
$X_0^{-3, 0}(e_M)=(1-e_M)^{-3/2}$  (see \cite{Giacaglia1976})}; $[\cdot]$ denotes the integer part; $K_0=1$, otherwise $K_m=2$ if $m\not=0$;
$F_{2mp}(i)$ is the \sl Kaula inclination function \rm (\cite{Kaula}, see also \cite{chao}) that we define for the generic case $F_{nmp}(i)$ as
\beqano
F_{nmp}(i)&=&\sum_{t=0}^{\min{\{p,[{{n-m}\over 2}]}\}} {{(2n-2t)!}\over {t!(n-t)!(n-m-2t)!\ 2^{2n-2t}}} \sin^{n-m-2t}i\ \sum_{s=0}^m\left(\begin{array}{c}
  m \\
  s \\
 \end{array}\right)
 \cos^si\nonumber\\
 &&\times \sum_c \left(\begin{array}{c}
  n-m-2t+s \\
  c \\
 \end{array}\right)
\left(\begin{array}{c}
  m-s \\
  p-t-c \\
 \end{array}\right)
 (-1)^{c-[{{n-m}\over 2}]}\ ,
\eeqano
where $c$ is summed over all values for which the binomial coefficients are not zero;
%under the restriction assumed before that
%$2-2p+j=0$, the function $G_{2pj}(e)$ is given by the Hansen coefficient $X(2,2-2p;e)$ (see \cite{Giacaglia1976}) and it is defined
%in the generic case $G_{npj}(e)$ as
%\beqa{G}
%G_{npj}(e)&=&(1+({e\over {1+\sqrt{1-e^2}}})^2)^{-n-1}\ \sum_{t=0}^{|2n-2p+1|} \ \sum_{s=0}^{|2p+1|} \ \Big({{2p+1}\over s}\Big)\
%\Big({{2n-2p+1}\over t}\Big)\nonumber\\
%&&(-{e\over {1+\sqrt{1-e^2}}})^{s+t}\ J_b^{-n+2p-s+t}(0)\ ,
%\eeqa
%where $J_b$ denotes the Bessel function of the first kind;
%$G_{2pj}^{(4)}(e)$ means that we consider the
%expansion of $G_{2pj}$ up to the order 4 in the eccentricity,
%under the restriction $2-2q+r=0$, the function $H_{2qr}(e_M)$ is given by the Hansen coefficient $X(2,2-2q;e_M)$ and
%the generic function $H_{nqr}(e_M)$ is defined as
%\beqa{H}
%H_{nqr}(e_M)&=&(1+({e\over {1+\sqrt{1-e^2}}})^2)^{n}\ \sum_{t=0}^{10} \ \sum_{s=0}^{10} \ \Big({{-2n+2h}\over s}\Big)\
%\Big({{-2n-2h}\over t}\Big)\nonumber\\
%&&(-{e\over {1+\sqrt{1-e^2}}})^{s+t}\ J_b^{-n+2h-s+t}(0)\ ;
%\eeqa \marginpar{The sums should go at infinity}
%while $H_{2qr}^{(2)}(e_M)$ means that we consider the
%expansion of $H_{2qr}$ up to the order 2 in the eccentricity of the Moon;
the function $U_2^{m,s}$ is defined as (compare with \cite{Giacaglia1974})
$$
U_2^{m,s}={{(-1)^{m-s}}\over {(2+s)!}}\ \cos^{s+m}({\varepsilon\over 2}) \sin^{s-m}({\varepsilon\over 2})\
{{d^{2+s}}\over {dz^{2+s}}}(z^{2-m} (z-1)^{2+m})\ ,
$$
where $z=\cos^2 ({\varepsilon\over 2})$ and  $\varepsilon$ denotes the obliquity of the ecliptic, which is equal to $\varepsilon=23^{\circ}26'21.45''$;
the quantity $y_s$ is zero for $s$ even and it is equal to $1/2$ for $s$ odd.

\subsection{The solar potential}\label{sec:solar}
The gravitational potential due to the Sun, $\R_{Sun}$, is found analogously to that of the Moon. However,
contrary to the lunar case, we use equatorial elements for both the satellite and the Sun (see \cite{Kaula1962}).
Precisely, we assume that the Sun moves on an elliptic orbit
with semimajor axis $a_S=149\,597\,871/42\,164.1696$ (expressed in units of the geostationary radius),
eccentricity $e_S=0.01671123$, inclination $i_S=23^{\circ} 26' 21.406''$, argument of perigee $\omega_S=282.94^{\circ}$,
longitude of the ascending node $\Omega_S=0^{\circ}$; the mass of the Sun $m_S$,
expressed in Earth's masses, is equal to about 333\,060.4016.

Averaging as in Section~\ref{sec:lunar} over the mean anomaly of the space debris and that of the Sun,
and truncating to \red{second order in the ratio of semi--major axes,} one obtains:
\begin{equation}\label{HR_Sun}
\H_{Sun}=-\R_{Sun}\ ,
\end{equation}
where
\red{\beqa{Rsun}
\R_{Sun}&=&\G m_S\sum_{m=0}^2 \sum_{p=0}^2 {a^2\over a_S^3}
\ K_m\, {{(2-m)!}\over {(2+m)!}}\ F_{2mp}(i)\ F_{2m1}(i_S)\nonumber\\
&&\ \times H_{2\ p\ 2p-2}(e)\, G_{210}(e_S)\ \cos\Bigl((2-2p)\omega+m(\Omega-\Omega_S)\Bigr)\ .
\eeqa}
The notation appearing in \equ{Rsun} has been already introduced in Section~\ref{sec:lunar}.

We stress that, since in $\H_{geo}$, $\H_{Moon}$ and $\H_{Sun}$ we retain only the secular terms,
then it means that we consider the Hamiltonian
averaged over $M$ and therefore its conjugated action $L$ turns out to be constant.

\section{Secular resonances depending only on inclination}\label{sec:secres}
In \cite{HughesI} it has been shown that there are 15 types of third-body (lunar and solar) resonances.
\red{This classification accounts for all possible resonances involving a third--body perturber: secular resonances, semi--secular resonances and mean motion resonances.}

\red{The semi--secular resonances and mean motion resonances will not be an object of study of the present work:
involving the mean anomalies of the Moon and Sun, whose rates of variation are $\dot M_M\simeq 13^{\circ}/day$, $\dot M_S\simeq 1^{\circ}/day$, the semi--secular resonances mostly take place in the LEO region. On the other hand, the mean motion resonances occur on a different time scale than the secular resonances.
We are indeed interested in gravity secular resonances, which are defined as follows.}

\begin{definition} \label{def:secres}
A lunar gravity secular resonance occurs whenever there exists an integer vector
\red{$(k_1,k_2,k_3,k_4)\in\integer^4\backslash\{0\}$, such that
\beq{secresmoon}
k_1\dot\omega+k_2\dot\Omega+k_3\dot\omega_M+k_4\dot\Omega_M=0\ .
\eeq}
We have a solar gravity secular resonance whenever there exist \red{$(k_1,k_2,k_3,k_4)\in\integer^4\backslash\{0\}$, such that
\beq{secressun}
k_1\dot\omega+k_2\dot\Omega+k_3\dot\omega_S+k_4\dot\Omega_S=0\ .
\eeq}
\end{definition}

\begin{remark}\label{ave}
The commensurability relations \red{\equ{secresmoon} and \eqref{secressun}}  are independent on $\dot M_M$ and $\dot M_S$;
this remark motivated the computations of Section~\ref{sec:model}, where the expansions
$\R_{Moon}$ in \equ{Rmoon} and $\R_{Sun}$ in \equ{Rsun} have been computed by averaging over
$M_M$ and $M_S$.
\end{remark}

\red{We stress that the above definition of secular resonance is as general as possible. However, given the fact that the lunar and solar expansions are truncated to the second order in the ratio of semi--major axes, in view of \eqref{Rmoon} and \eqref{Rsun}, the Hamiltonian $\H$ is independent of $\omega_M$ and $\omega_S$.
Therefore, for all resonances studied here, one has $k_3=0$. Moreover, since $\dot{\Omega}_S \simeq 0$,  the relations \eqref{secresmoon} and \eqref{secressun} may be rewritten in the particular form:
\beq{secresmoonbis}
(2-2p)\dot\omega+m\dot\Omega+\kappa \dot\Omega_M=0\ , \qquad m,p=0,1,2, \quad \kappa=-2,-1,0,1,2\,,
\eeq
and
\beq{secressunbis}
(2-2p)\dot\omega+m\dot\Omega=0\ , \qquad m,p=0,1,2\,,
\eeq
respectively.}

\red{The longitude of the lunar ascending node varies with the rate $\dot{\Omega}_M \simeq -0.053^{\circ}/day$, while
the quantities
$\dot \omega$, $\dot \Omega$ can be approximated by} the following well known formulae, which take into account only the effect of $J_2$
(\cite{HughesI}):
\beqa{omega12}
\dot\omega &\simeq& 4.98 \Bigl({R_E\over a}\Bigr)^{7\over 2}\ (1-e^2)^{-2}\ (5\cos^2 i-1)\ ^{\circ}/day\ ,\nonumber\\
\dot\Omega &\simeq& -9.97 \Bigl({R_E\over a}\Bigr)^{7\over 2}\ (1-e^2)^{-2}\ \cos i\ ^{\circ}/day\ .
\eeqa
Inserting \equ{omega12} in \equ{secresmoonbis} or \equ{secressunbis}, we obtain a relation involving the satellite's elements
$a$, $e$, $i$: this expression provides the location of the secular resonance.

As pointed out in \cite{HughesI}, some resonances turn out to be independent on $a$, $e$, and they depend
only on the inclination. The general class of resonances depending only on the inclination \red{is characterized by the relation $k_1 \dot{\omega}+k_2 \dot{\Omega}=0$, $k_1, k_2 \in \mathbb{Z}$. From this class, the most important ones are those for which $k_1, k_2 \in \{-2,-1,0,1,2\}$. In fact, under the quadrupolar approximation considered in this paper, the only possible resonances are:}
\beqa{secresi}
\dot\omega&=&0\nonumber\\
\dot\Omega&=&0\nonumber\\
\dot\omega+\dot\Omega&=&0\nonumber\\
-\dot\omega+\dot\Omega&=&0\nonumber\\
-2\dot\omega+\dot\Omega&=&0\nonumber\\
2\dot\omega+\dot\Omega&=&0\ .
\eeqa
\red{The resonances involving $k_1$ and $k_2$ with $|k_1|>2$ or/and $|k_2|>2$ occur at higher degree expansions of the lunar and solar disturbing functions, their influence being negligible in the MEO region.}

Inserting \equ{omega12} in \equ{secresi}, one readily sees that one obtains expressions involving just the inclinations.
Precisely, the specific values of the inclinations corresponding to the cases listed in \equ{secresi} are,
respectively:
\beqa{inc}
&&63.4^{\circ}\qquad {\rm or}\qquad 116.4^{\circ}\nonumber\\
&& \qquad \quad \quad 90^{\circ}\nonumber\\
&&46.4^{\circ}\qquad {\rm or}\qquad 106.9^{\circ}\nonumber\\
&&73.2^{\circ}\qquad {\rm or}\qquad 133.6^{\circ}\nonumber\\
&&69.0^{\circ}\qquad {\rm or}\qquad 123.9^{\circ}\nonumber\\
&&56.1^{\circ}\qquad {\rm or}\qquad 111.0^{\circ}\ .
\eeqa
The values in \equ{inc} are obtained as follows. Let us consider as an example the third relation in \equ{secresi}
which, together with \equ{omega12}, yields
$$
\dot\omega+\dot\Omega=\Bigl({R_E\over a} \Bigr)^{7\over 2}\ (1-e^2)^{-2}\ [4.98(5\cos^2 i-1)-9.97\cos i]\ ;
$$
the last expression is equal to zero for $i=46.4^{\circ}$ and $i=106.9^{\circ}$. In a similar way one finds
the other values listed in \equ{inc}.

\red{According to \cite{HughesI, rossi2008}, we consider the following} three classes
of lunisolar secular resonances depending only on specific values of the inclination:

\begin{enumerate}
\item[$(i)$]
$\dot\omega=0$, which occurs at the critical inclinations $i=63.4^{\circ}$, $116.4^{\circ}$;
\item[$(ii)$]
$\dot\Omega=0$, which corresponds to polar orbits;
\item[$(iii)$]
$k_1\dot\omega+k_2\dot\Omega=0$ for some $k_1$, $k_2\in \{-2,-1,1,2\}$.
\end{enumerate}

In the following we shall refer to the above cases as secular resonances of types, respectively,
$(i)$, $(ii)$, $(iii)$.

\vskip.1in

The secular resonances appearing in \equ{secresi} are the most important secular resonances,
as their amplitude is larger than the amplitude associated \red{to both the resonances involving the lunar ascending node $\Omega_M$ and the resonances of higher order.}

Recalling that we averaged over the mean anomalies of the debris and of the third body, we conclude that the
Hamiltonian we are going to study \red{(omitting the Keplerian part)} has the following form:
\beqa{H1}
\H(G,H,\omega,\Omega,\omega_M,\Omega_M,\omega_S,\Omega_S)&=&
{\alpha\over G^3}\left(1-3{H^2\over G^2}\right)-\R_{Moon}(G,H,\omega,\Omega,\omega_M,\Omega_M)\nonumber\\
&-&\R_{Sun}(G,H,\omega,\Omega,\omega_S,\Omega_S)\ ,
\eeqa
where,
for a given $L=L_0$, the quantity $\alpha$ is defined as
\beq{alfa}
\alpha\equiv {{R_E^2 J_2 \mu_E^4}\over {4L_0^3}} = 6.19325 \times 10^{-6} L_0^{-3}
\eeq
and $R_{Moon}$, $R_{Sun}$ are given, respectively, in \equ{Rmoon}, \equ{Rsun}.

%\marginpar{\red{I deleted $\sqrt{5}$ since in the notebook we use $\overline{J}_2=484.1651 $,
%while in the paper we use $J_2=1082.63$. These quantities are related by $J_2=\sqrt{5} \overline{J}_2$.}}

\section{Secular resonance: $\dot\omega+\dot\Omega=0$}\label{sec:casestudy1}
In this Section, we analyze in detail one of the cases appearing in \equ{secresi} and precisely
we concentrate on the resonance $\dot\omega+\dot\Omega=0$, which is
located at one of the inclinations $i=46.4^{\circ}$ or $i=106.9^{\circ}$.
The study of the other resonances in \equ{secresi} is deferred to Section~\ref{sec:other}.
Under some simplifying assumptions, we introduce a model (see Section~\ref{sec:model2}),
which is described by an integrable Hamiltonian function. The investigation of periodic orbits
is presented in Sections~\ref{sec:birth}, \ref{sec:alternative}, using two different methods from
bifurcation theory. The validity of the model and of the results is analyzed in Section~\ref{sec:FLI},
through a numerical technique based on the computation of suitable chaos indicators.
We remark that the model as well as the computation of periodic orbits through bifurcation
theory relies on some series expansions to given orders. The orders used in the current
Section for the resonance $\dot\omega+\dot\Omega=0$ suffice to provide reliable and stable results.
However, this might not be the case for other resonances and, in fact, while studying other
resonances of type \equ{secresi} in Section~\ref{sec:other}, we add a
discussion on the accuracy of the expansion of the Hamiltonian model
(see Section~\ref{sec:accuracy}).

\subsection{The model}\label{sec:model2}
We start by introducing resonant variables through the symplectic transformation
$(G,H,\omega,\Omega)\rightarrow (S,T,\sigma,\eta)$ defined by
\beqa{RV}
\sigma&=&\omega+\Omega\ ,\qquad\ \ S=G\ ,\nonumber\\
\eta&=&\Omega\ ,\qquad\qquad\ \ T=H-G\ .
\eeqa
From \equ{H1} we obtain that the resonant Hamiltonian in the variables \equ{RV} takes the form
\beqa{Hres}
\H_{res}(S,T,\sigma,\eta,t)&=&{\alpha\over S^3}\ \Big(1-3{{(T+S)^2}\over S^2}\Big)\nonumber\\
&-&\R_{Moon}^{(res)}(S,T,\sigma,\eta,t)-\R_{Sun}^{(res)}(S,T,\sigma,\eta,t)\ ,
\eeqa
where the dependence on time in \equ{Hres} comes from the linear variations with \red{time of
$\Omega_M$}. The functions $\R_{Moon}^{(res)}$, $\R_{Sun}^{(res)}$ denote the
functions $\R_{Moon}$, $\R_{Sun}$ in \equ{Rmoon}, \equ{Rsun} expressed in terms of the resonant
variables introduced in \equ{RV}.

Our goal will be to find energy thresholds at which the bifurcations of lunisolar resonances take place
(see Section~\ref{sec:birth}). To this end, we need to introduce an integrable Hamiltonian, which is
obtained under the following two hypotheses.

{\bf H1.} {\sl Each resonance will be viewed in isolation, not interacting with the other ones}.

This means that  on the timescale associated to
\equ{Hres}, we have that $\sigma$ is the resonant (\sl slow\rm) angle, while $\eta$ is
a \sl fast \rm angle; therefore, averaging over $\eta$ (i.e., over $\Omega$), we obtain that the contributions
due to Moon and Sun are given  (in the units specified in Section~\ref{sec:model}) by the following expressions:
\red{\begin{equation}\label{R_moon_bigOmegaMoon}
\begin{split}
&\overline \R_{Moon}^{(res)}(S,T,\sigma,t)=0.10142 \cdot 10^{-6} a^2 (2 + 3 e^2) [-1 + 3 \cos^2(i)] \Bigl[7.54884 -
    \cos(\Omega_M)  \\
    & \qquad + 0.00996 \cos(2\, \Omega_M)\Bigr] +
 0.0114 \cdot 10^{-6} a^2 e^2 [1 +  \cos(i)]^2 \Bigl[52.2396 \cos(2\, \sigma)
 \\
  & \qquad  +    2.57317 \cos(2\, \sigma - 2\, \Omega_M) +
    23.2372 \cos(2\, \sigma -  \Omega_M)
     -
    \cos(2\, \sigma + \Omega_M)\\
    & \qquad +
    0.00476 \cos(2\, \sigma + 2\, \Omega_M)\Bigr]\,,
    \end{split}
\end{equation}}
and
\begin{equation}\label{Rsunaverage}
\begin{split}
&\overline \R_{Sun}^{(res)}(S,T,\sigma, t)=0.35561 \cdot 10^{-6} a^2 (2 + 3 e^2) [-1 + 3 \cos^2(i)]  \\
    & \qquad +
 0.27666 \cdot 10^{-6} a^2 e^2 [1 +  \cos(i)]^2  \cos(2\, \sigma-2\, \Omega_S)\,,
    \end{split}
\end{equation}
where, for a given $L=L_0$, we have that $e^2=1-S^2/L_0^2$ and $\cos(i)=(T+S)/S$.

{\bf H2.} {\sl We neglect the rates of variation of $\Omega_M$ and $\Omega_S$}.

In fact, we may disregard just the rate of variation of $\Omega_M$,
since $\dot{\Omega}_S$ is indeed very small; as a consequence, we use the following constant
values, valid at epoch J2000 (\cite{Simon1994}):
\beq{const}
 \Omega_M=125.044555^{\circ}\ ,\qquad  \Omega_S=0^{\circ}\ .
\eeq
Due to the assumption {\bf H2}, we obtain that the functions $ \overline \R_{Moon}^{(res)}$, $\overline \R_{Sun}^{(res)}$ do
not depend anymore on time.

\vskip.1in

Under the hypotheses {\bf H1} and {\bf H2}, the contribution
due to the Moon is given  (in the units specified in Section~\ref{sec:model}) by the following expression:
\red{\begin{equation}\label{Rmoonaverage}
\begin{split}
&\overline R_{Moon}(S,T,\sigma)=0.82349 \cdot 10^{-6} a^2 (2 + 3 e^2) [-1 + 3 \cos^2(i)]  \\
    & \qquad +
 0.4828 \cdot 10^{-6} a^2 e^2 [1 +  \cos(i)]^2  \cos(2\, \sigma- 0.42418)\ ,
      \end{split}
\end{equation}}
while the contribution due to the Sun, say $\overline R_{Sun}(S,T,\sigma)$,
is expressed by \eqref{Rsunaverage} with $\Omega_S=0$.
We are thus led to consider the following reduced model:
$$
\overline R_{res}(S,T,\sigma)={\alpha\over S^3}\ \Big(1-3{{(T+S)^2}\over S^2}\Big)
-\overline R_{Moon}(S,T,\sigma)-\overline R_{Sun}(S,T,\sigma)\ .
$$

As remarked, e.g., in \cite{ElyHowell, RARV15}, in many regions of the phase space
there can be a delicate interaction between the resonances, leading to a complex secular dynamics.
However, the reduced model provided by $\overline R_{res}$ leads to an exhaustive description of the
bifurcation phenomena. We will discuss in detail in Section~\ref{sec:FLI} the
assumptions {\bf H1} and {\bf H2}, and we will highlight
numerically the differences between the reduced model $\overline R_{res}$ and the Hamiltonian \eqref{Hres}.

\vskip.1in

After some algebra, one can write the sum of the contributions $\overline R_{Moon}$ and $\overline R_{Sun}$
as given by the potential $\overline{\R}_{MoonSun}$, whose expression is
\red{\begin{equation}\label{Rmoonsunaverage}
\begin{split}
&\overline R_{MoonSun}(S,T,\sigma)=1.17909 \cdot 10^{-6} a^2 (2 + 3 e^2) [-1 + 3 \cos^2(i)]  \\
    & \qquad +
 0.74373 \cdot 10^{-6} a^2 e^2 [1 +  \cos(i)]^2  \cos(2\, \sigma - 0.27047)\ .
    \end{split}
\end{equation}}

\vskip.1in

In conclusion, we are led to consider three possible simplified models, which describe only the effect of the Moon, that
of the Sun, or both; they are represented by the following Hamiltonians:
\beqa{H3}
\H_{Moon}(S,T,\sigma)&=&{\alpha\over S^3}\ \Big(1-3{{(T+S)^2}\over S^2}\Big)-\overline{\R}_{Moon}(S,T,\sigma)\nonumber\\
\H_{Sun}(S,T,\sigma)&=&{\alpha\over S^3}\ \Big(1-3{{(T+S)^2}\over S^2}\Big)-\overline{\R}_{Sun}(S,T,\sigma)\nonumber\\
\H_{MoonSun}(S,T,\sigma)&=&{\alpha\over S^3}\ \Big(1-3{{(T+S)^2}\over S^2}\Big)-\overline{\R}_{MoonSun}(S,T,\sigma)\ ,
\eeqa
where $\overline{\R}_{Moon}$, $\overline{\R}_{Sun}$, $\overline{\R}_{MoonSun}$ are given, respectively, in
\equ{Rmoonaverage}, \equ{Rsunaverage} with $\Omega_S=0$, and \equ{Rmoonsunaverage}.\\

Our next task is to compute the equilibrium points associated to one of the Hamiltonians \equ{H3},
that we denote generically as $\H_r$. In particular, we need to find a solution of the equation
$$
\dot S=-{{\partial\H_r}\over {\partial\sigma}}=0\ .
$$
With reference to \equ{H3}, we can proceed to compute the equilibria for the following models:

\begin{enumerate}
\item[$(a)$] including just the contribution of the Moon, the equilibria are obtained by solving the equation:
$$
\dot S={{\partial\overline{\R}_{Moon}}\over {\partial\sigma}}=0\ ,
$$
which leads to \red{ $\sigma=0.212$ or $\sigma=1.783$} (modulo $\pi$);
\item[$(b)$] including just the contribution of the Sun, the equilibria are obtained by solving the equation:
$$
\dot S={{\partial\overline{\R}_{Sun}}\over {\partial\sigma}}=0\ ,
$$
which leads to $\sigma=0$ or $\sigma=\pi/2$ (modulo $\pi$);
\item[$(c)$] including both contributions of Moon and Sun, the equilibria are obtained by solving the equation:
$$
\dot S={{\partial\overline{\R}_{MoonSun}}\over {\partial\sigma}}=0\ ,
$$
which admits the solutions  \red{$\sigma=0.135$ or $\sigma=1.706$} (modulo $\pi$).
\end{enumerate}

\subsection{Birth of periodic orbits}\label{sec:birth}
We now want to investigate the birth of periodic orbits for the secular resonance corresponding to the case
$\dot\omega+\dot\Omega=0$. The other resonances appearing in \equ{secresi} will be
analyzed in Section~\ref{sec:other}.

According to \equ{H3}, we can in principle consider three different cases
in which only the Moon is present, only the Sun is taken into account, or
rather the joint contribution of the Moon and Sun is considered.
%\marginpar{I changed the Moon case with the case of both Moon and Sun because we present the phase portrait for the later. }
However, in the following discussion we shall
present the mathematical details related to the more complete case when both the Moon and Sun are considered.
The other two cases can be indeed treated in a similar way.

When dealing with the third model in \eqref{H3} with the potential given by \equ{Rmoonsunaverage}, it is convenient to
introduce the angle $s$ such that
\red{$$
2s=2 \sigma - 0.27047\ .
$$}
Recalling that $T=H-S=S(\cos i-1)$ and using
the value $i_{res}=46.4^{\circ}$ at which the resonance occurs, we introduce the quantity
$$
S_0= \frac{T}{\cos i_{res}-1}= -3.222\, T
$$
and we expand the third Hamiltonian in
\eqref{H3} around $S_0$ up to the second order, thus obtaining:
\beq{Hexp}
\H_{exp}(S,T,s;L_0)={{10^{-6}}\over {L_0^3 T^5}} \Big(p_1(S,T;L_0)+p_2(S,T;L_0)\cos 2s\Big)\ ,
\eeq
where, by using \eqref{alfa}, $p_1$, $p_2$ are the following polynomials of degree {\it two} in $S$:
\red{\beqano
p_1(S,T;L_0)&=&-0.34538 T^2 + 16.2362 L_0^7 T^5 + 10.6118 L_0^5 T^7\nonumber\\
&+&S^2 (-0.04081 + 0.565253 L_0^7 T^3 + 7.07456 L_0^5 T^5)\nonumber\\
&+& S (-0.26322 T + 5.99233 L_0^7 T^4 + 21.2237 L_0^5 T^6)\nonumber\\
p_2(S,T;L_0)&=&  L_0^5 T^3 \Big(-0.63474 L_0^2 T^2 + 0.743734 T^4 +
   S^2 (0.068246 L_0^2 + 2.97494 T^2)\nonumber\\
&+&S (0.68187 L_0^2 T + 2.97494 T^3)\Big)\ .
\eeqano}

Of course, we could have computed an expansion to higher order, but we will see that in the present case
the second order is sufficient to capture the relevant features of the dynamics. We refer to Section~\ref{sec:accuracy}
for the discussion of the accuracy of the expansion to finite order in $S$.

\begin{remark}\label{rem:SFM}
It is interesting to underline the analogy presented in \cite{B2} of the lunisolar secular resonances
with the ``second fundamental model for second-order resonances" (see \cite{lemaitre84}); this
model was denoted with the acronym SFM2 in \cite{B2}.

The second fundamental model for first-order resonances, say SFM1, has been introduced in \cite{HenLem};
in this case the resonant Hamiltonian can be reduced to the form
$$
\H_{SFM1}(S,s)=S^2+\beta S-2\sqrt{2S} \cos s
$$
for some constant $\beta\in\real$. On the contrary, the SFM2 describes resonances of second order (\cite{lemaitre84}) through a
Hamiltonian of the form
$$
\H_{SFM2}(S,s)=2 S^2+\beta S+S \cos 2s
$$
(compare with \cite{B2}).

As pointed out in \cite{B2}, some resonances (precisely, those of second-order in the eccentricity) admit more
critical points than those described in SFM2. An ``extended fundamental model for second order resonances",
say EFM2, can be introduced to describe more general cases through a Hamiltonian function of the form
$$
\H_{EFM2}(S,s)=\gamma S^3+2 S^2+\beta S+S \cos 2s
$$
for some constants $\beta,\gamma\in\real$.
\end{remark}

Recalling that $\sigma$ and  $\eta$ are the angle variables conjugated to $S$ and $T$, respectively, the equations of motion associated to \equ{Hexp} are given by
\beqa{Hexpeq}
\dot S&=&{{2 \cdot 10^{-6}}\over {L_0^3 T^5}} p_2(S,T,L_0)\ \sin 2s\nonumber\\
\dot s&=&{{10^{-6}}\over {L_0^3 T^5}}\ \Bigl({{\partial p_1}\over {\partial S}}(S,T,L_0)+{{\partial p_2}\over {\partial S}}(S,T,L_0)\ \cos 2s\Bigr)\nonumber\\
\dot T&=&0\nonumber\\
\dot \eta&=&-{{5 \cdot 10^{-6}}\over {L_0^3 T^6}}\ \Bigl(p_1(S,T,L_0)+p_2(S,T,L_0)\ \cos 2s\Bigr)\nonumber\\
&+&{{10^{-6}}\over {L_0^3 T^5}}\ \Bigl({{\partial p_1}\over {\partial T}}(S,T,L_0)+{{\partial p_2}\over {\partial T}}(S,T,L_0)\ \cos 2s\Bigr)\ .
\eeqa

The equations \equ{Hexpeq} show that
for the Hamiltonian \equ{Hexp} the variable
$T$ is a constant of motion. Therefore, we are led to investigate
a system with only one degree of freedom in the $(s,S)$ phase-plane parametrized by $T$. The essential information is delivered by the fixed points of the reduced flow that can be located by solving the two equations $\dot S = 0, \dot s = 0$. The first equation admits the solutions
\red{$$
s=0,\pi\ ,\qquad s=\pm{\pi\over 2}\ \ \ \mod 2\pi\ ,
$$
or equivalently $\sigma=0.135$, $\sigma=3.277$, $\sigma=1.706$, $\sigma=-1.435$ (modulo $2\pi$).}
These values will produce pairs of equilibria, which arise from pitchfork bifurcations.

Let us denote by $\dot s_A$, $\dot s_B$, the values of $\dot s$ at, respectively, $s=0$ and $s={\pi\over 2}$, and let
$S_A$, $S_B$ be the solutions of $\dot s_A=0$, $\dot s_B=0$. Using that $L=\sqrt{\mu_E a}$ and expanding $S_A$, $S_B$ in
series of $T$ up to the 6th order, one obtains the expressions
\red{\beqa{SAB}
S_A&=&-3.225\ T+31.712\ a^{7\over 2}\ T^4-497.59\ a^{5\over 2}\ T^6\ ,\nonumber\\
S_B&=&-3.225\ T+25.789\ a^{7\over 2}\ T^4-100.358\ a^{5\over 2}\ T^6\ .
\eeqa}

To determine the bifurcation thresholds according to the technique described, e.g., in \cite{Pucacco,MP14,PM14}, we need to constrain the variable $S$
within an interval, which is determined on the basis of the relation of $S$
with the integral of motion represented by $T$.
%\marginpar{The statement in blue is not clear to me.}
However, in the present situation we have also a
\sl physical \rm constraint, which leads to a bound on $S$ on the basis of
the following physical considerations (compare with \cite{B1,B2}).
We know that the maximum value of $S$ could be $S=L$, since the eccentricity must be greater than or
equal to zero. On the other hand, from the fact
that the distance at perigee cannot be smaller than the radius of the Earth, we obtain a minimum
value that $S$ can reach, say $S=S_{min}$. To be explicit, we start from the condition that the distance
at perigee should be at least equal to the Earth's radius:
$$
a(1-e)=R_E\ ;
$$
from this expression we have $e=(a-R_E)/a$. Taking into account that $e=\sqrt{1-S^2/L^2}=\sqrt{1-S^2/(\mu_E a)}$, we obtain
$$
1-{S^2\over {\mu_E a}}={{(a-R_E)^2}\over a^2}\ ,
$$
whose solution provides (compare with \cite{Breiter1999}):
\begin{equation}\label{Smin}
S_{min}=\sqrt{{(2a-R_E)\mu_E R_E}\over {a}}\ .
\end{equation}
We stress again that such limitation on $S_{min}$ is purely given by physical reasons and that a different choice
might lead to other results. From the minimum value of $S$, we obtain the corresponding value for the variable $T$,
say $T=T_0$, as
$$
T_0=S_{min} (\cos i_{res}-1)\ ,
$$
where $i_{res}$ is the value of the inclination at resonance, as given by \equ{inc}.

%Taking $i_{res}=73.2^{\circ}$, we get $T_0=S_{min}/0.775$.

\begin{figure}[h]
\centering
\vglue0.1cm
\hglue0.1cm
\includegraphics[width=5truecm,height=4truecm]{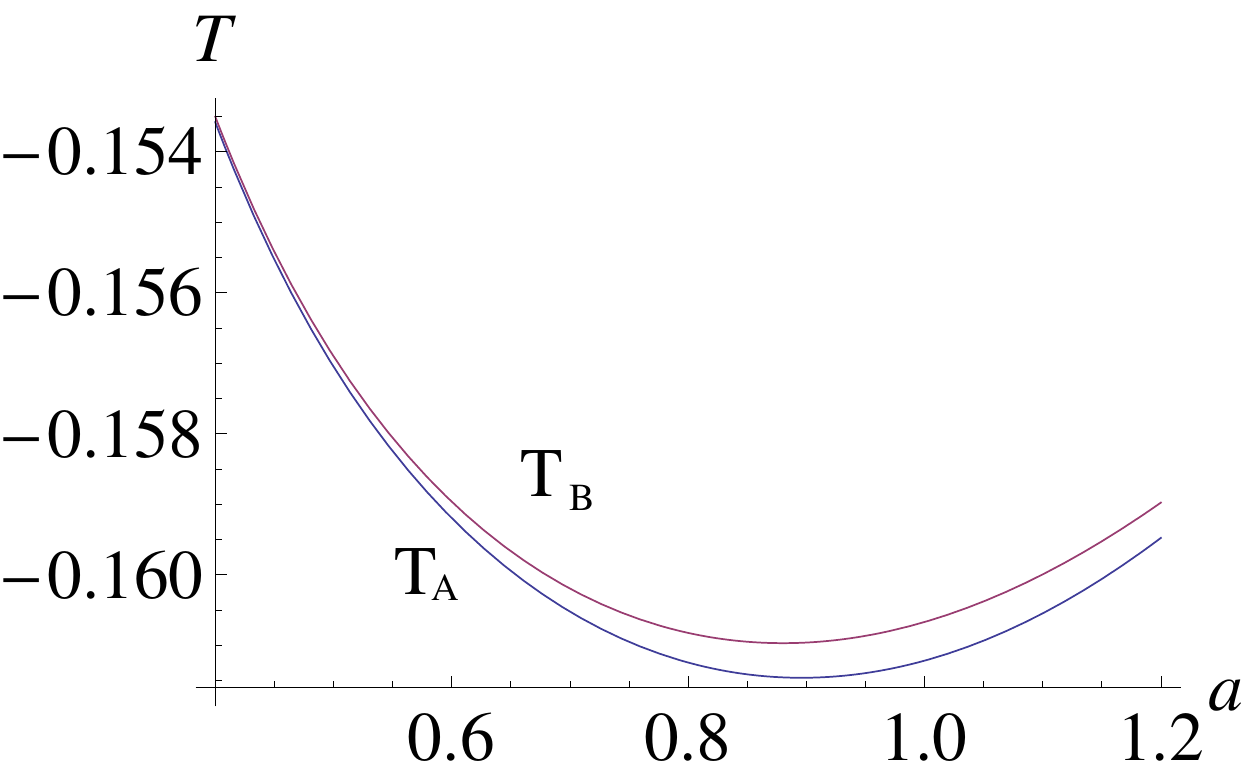}
\includegraphics[width=5truecm,height=4truecm]{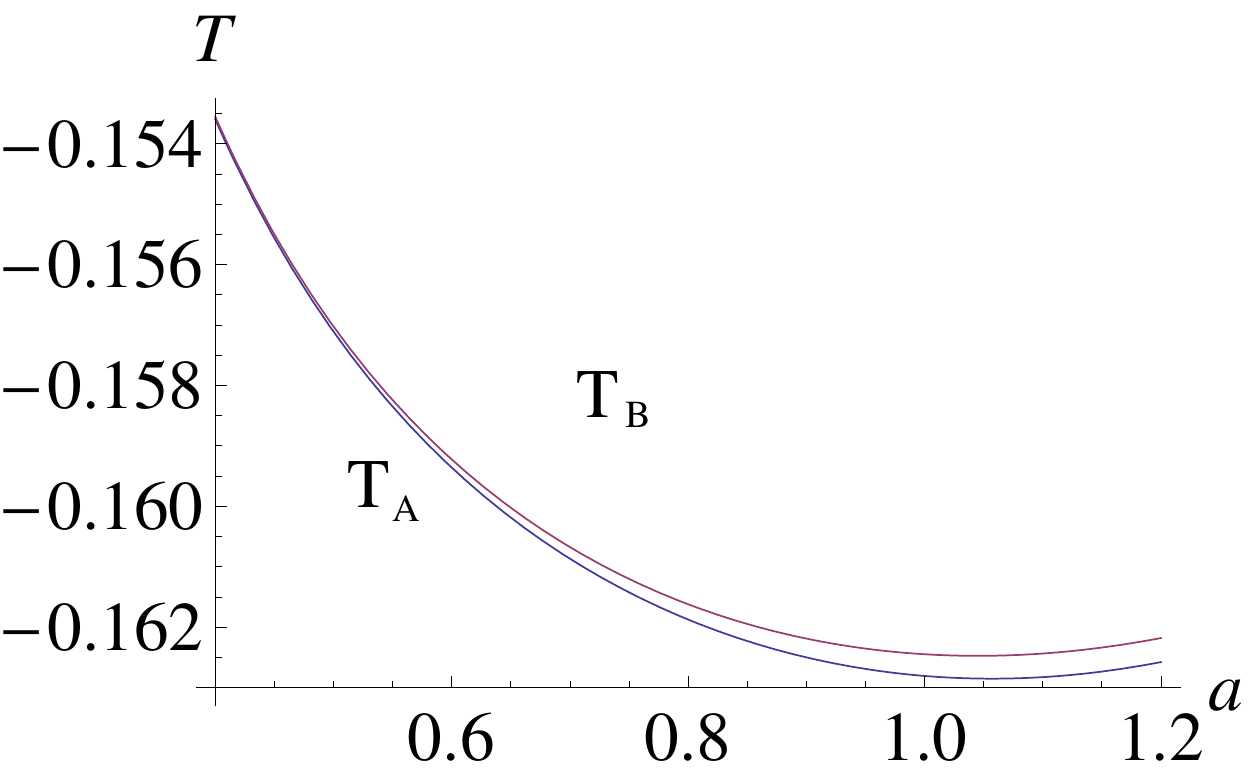}
\includegraphics[width=5truecm,height=4truecm]{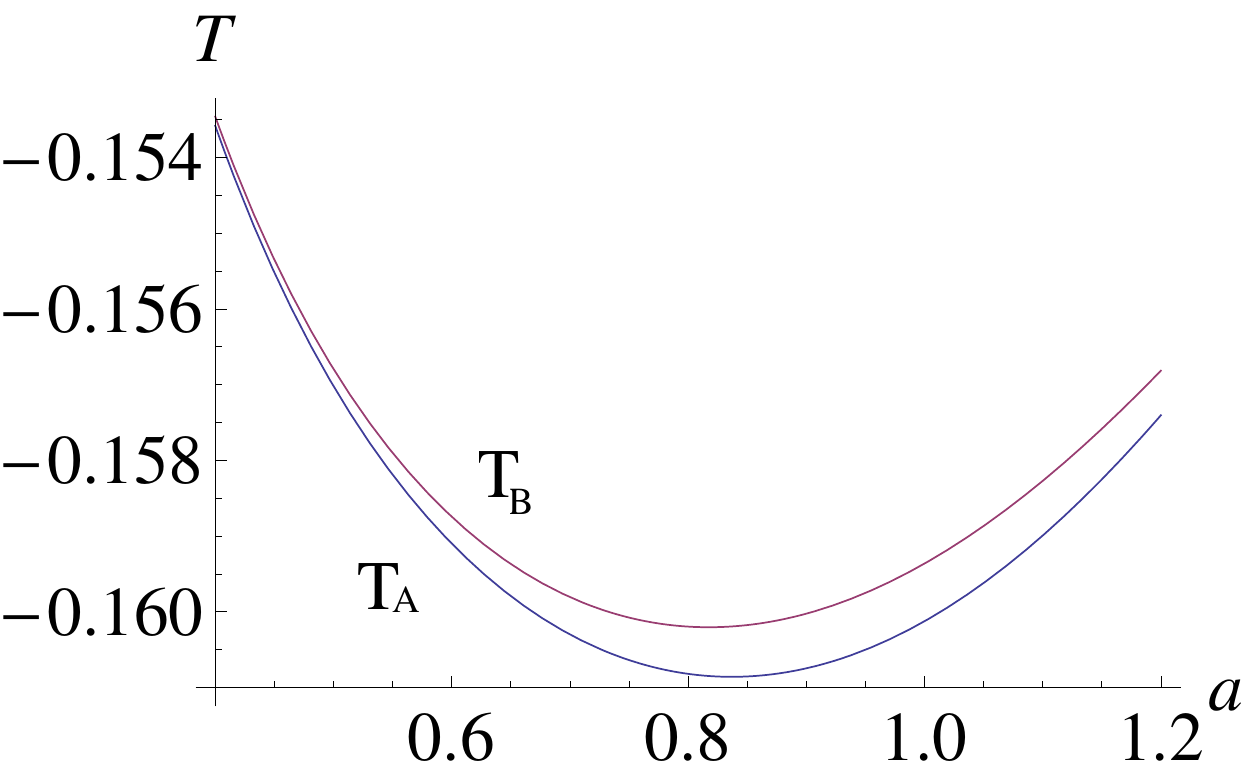}
\vglue0.1cm
\caption{Secular resonance: $\dot\omega+\dot\Omega=0$, $i=46.4^o$.
Plots of the bifurcations curves $T_A$, $T_B$ as a
function of the semi-major axis. Left: the Hamiltonian includes just the effect of the Moon;
middle: the Hamiltonian includes only the Sun; right: the Hamiltonian includes both
Moon and Sun.}
\label{fig:case1}
\end{figure}

To find explicit expressions of the bifurcation curves in the $(a,T)$-plane of the conserved quantities, we proceed by simplifying as much as possible the necessary algebra. Using a fake parameter $\lambda$, we introduce the following expansions to first order:
\beq{T11}
T_A=T_0+\lambda\ T_{A1}\ ,\qquad T_B=T_0+\lambda\ T_{B1}
\eeq
for some unknowns $T_{A1}$, $T_{B1}$. Next, we replace $T_A$, $T_B$ in the solutions $S_A$, $S_B$
given by \equ{SAB}, respectively,
and we expand to first order in $\lambda$ as
$$
S_A=\tau_0^{(A)}(a)+\lambda\ \tau_1^{(A)}(a)\ T_{A1}\ , \qquad
S_B=\tau_0^{(B)}(a)+\lambda\ \tau_1^{(B)}(a)\ T_{B1}
$$
for some known functions $\tau_0^{(A)}$, $\tau_1^{(A)}$, $\tau_0^{(B)}$, $\tau_1^{(B)}$. The quantities $T_{A1}$, $T_{B1}$
are then obtained by solving the equations
$$
S_A=S_{min}\ , \qquad S_B=S_{min}\ .
$$
Finally, we set $\lambda=1$ and we obtain explicit expressions for \eqref{T11}.
Since they have a quite cumbersome form,
we give the series expansions of $T_A$, $T_B$ around a given
semimajor axis, for example that corresponding to the GPS location, namely $a_{GPS}=26\,560/42\,164\simeq 0.6299$:
\red{\beqa{TAB}
T_A &=& -0.159538-0.0134657\ (a-a_{GPS}) + 0.039272 \ (a-a_{GPS})^2\ ,\nonumber\\
T_B &=& -0.159142-0.0120736\ (a-a_{GPS}) + 0.039572 \ (a-a_{GPS})^2\ .
\eeqa}
Figure~\ref{fig:case1}, right panel, shows the plots of the curves $T_A$, $T_B$ as a function of the semi-major axis for
the model including both effects of the Moon and Sun.
In a similar way we proceed to determine the bifurcation thresholds just due to the Sun (see Figure~\ref{fig:case1}, middle panel)
and those including just the effects of the Moon (see Figure~\ref{fig:case1}, left panel).

\vskip.2in

The bifurcation values for the GPS orbit are obtained setting $a=a_{GPS}$ in \equ{TAB}; in this way we compute
the thresholds at which the first and second bifurcations take place (see Table~\ref{table:case1}).

\vskip.1in

\red{
\begin{table}[h]
\begin{tabular}{|c|c|c|c|c|c|}
  \hline
  &inclination (in degrees)&&$T_A$ & $T_B$ \\
  \hline
$\dot\omega+\dot\Omega=0$&46.4&Moon & -0.159667 & -0.159407 \\
$\dot\omega+\dot\Omega=0$&46.4&Sun & -0.159875 & -0.159724\\
$\dot\omega+\dot\Omega=0$&46.4&Moon+Sun & -0.159538 & -0.159142\\
  \hline
$\dot\omega+\dot\Omega=0$&106.9&Moon & -0.664766 & -0.664673 \\
$\dot\omega+\dot\Omega=0$&106.9&Sun & -0.665209 & -0.665155\\
$\dot\omega+\dot\Omega=0$&106.9&Moon+Sun & -0.664441 & -0.664298\\
  \hline
 \end{tabular}
 \vskip.1in
 \caption{Bifurcation values $T_A$, $T_B$ for the GPS orbit corresponding to the cases
 in which only the Moon is considered, only the Sun is taken into account, both Sun and Moon are
 included in the model.
 The two possible inclinations, $46.4^o$ and $106.9^o$, found in \equ{inc} are considered.}\label{table:case1}
\end{table}
}

\vskip.1in

\begin{figure}[h]
\centering
\vglue0.1cm
\hglue0.1cm
\includegraphics[width=5truecm,height=4truecm]{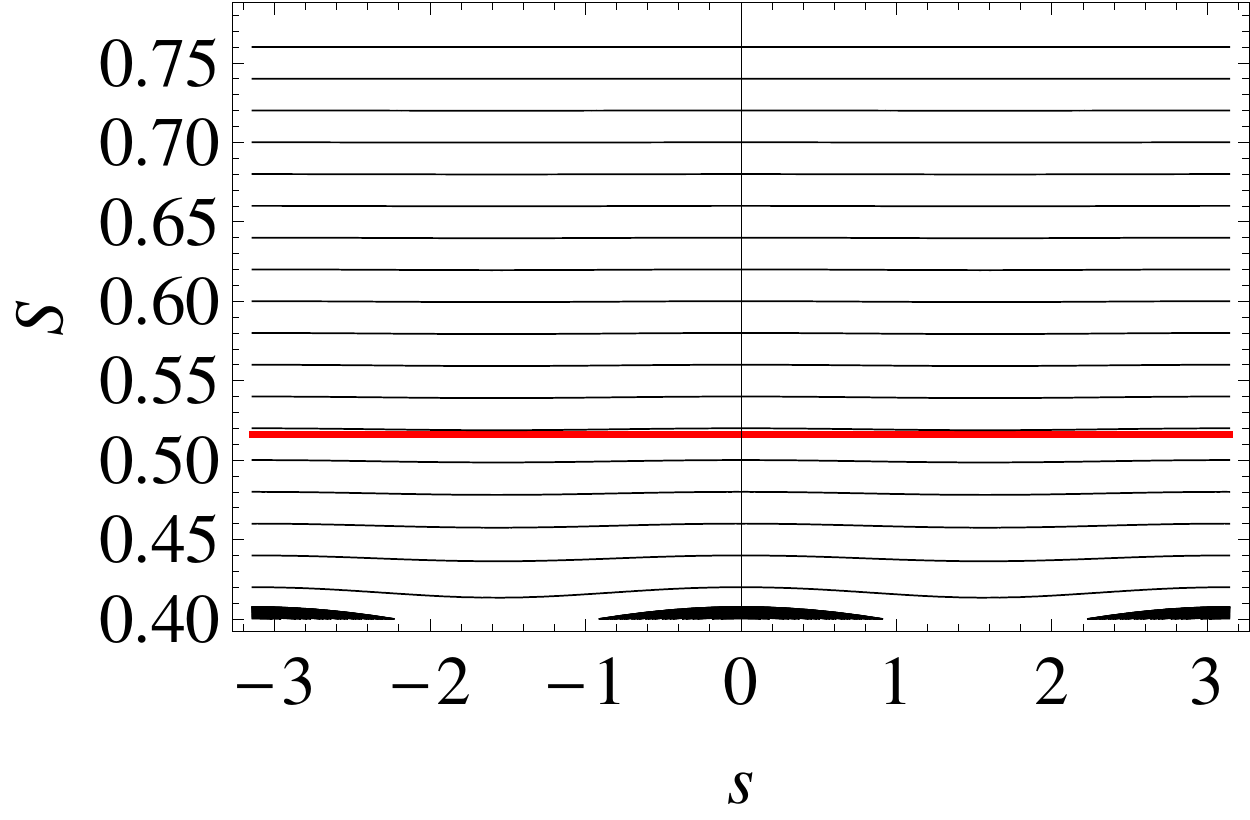}
\includegraphics[width=5truecm,height=4truecm]{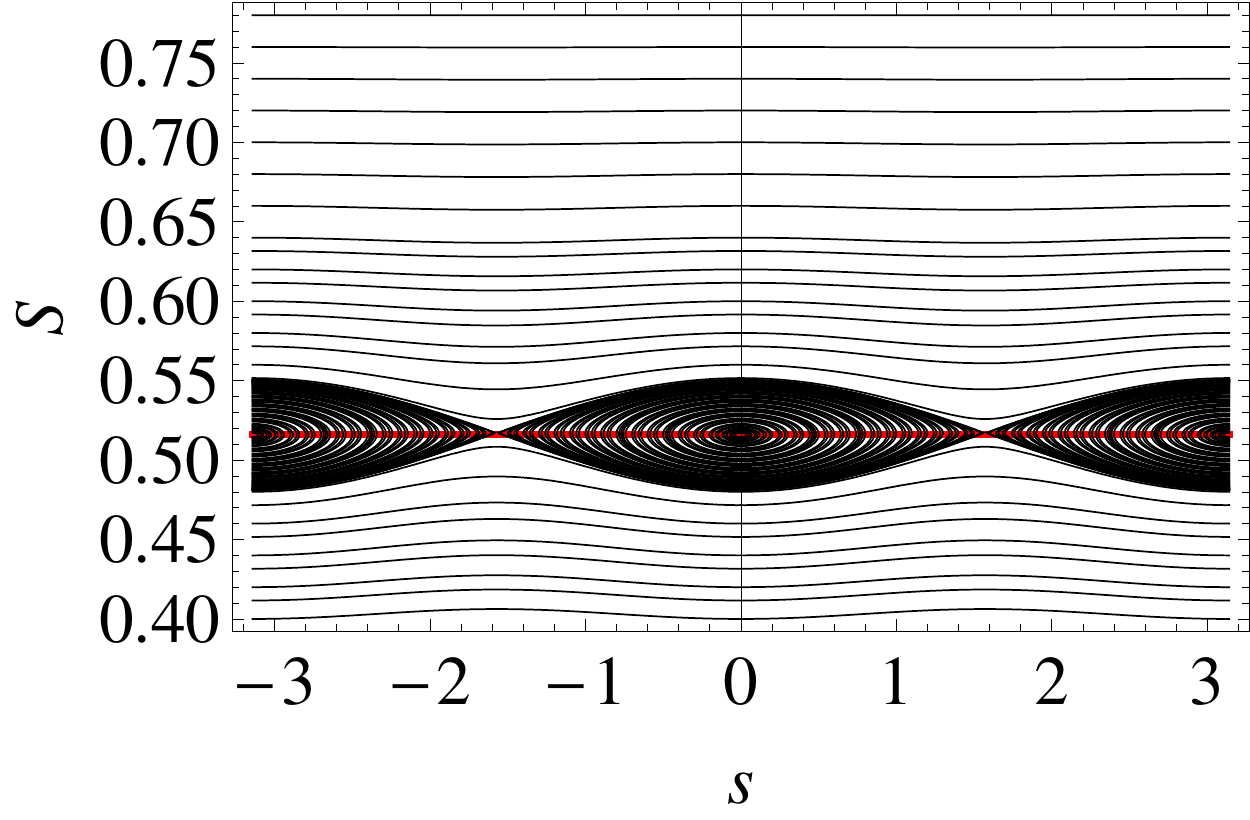}
\includegraphics[width=5truecm,height=4truecm]{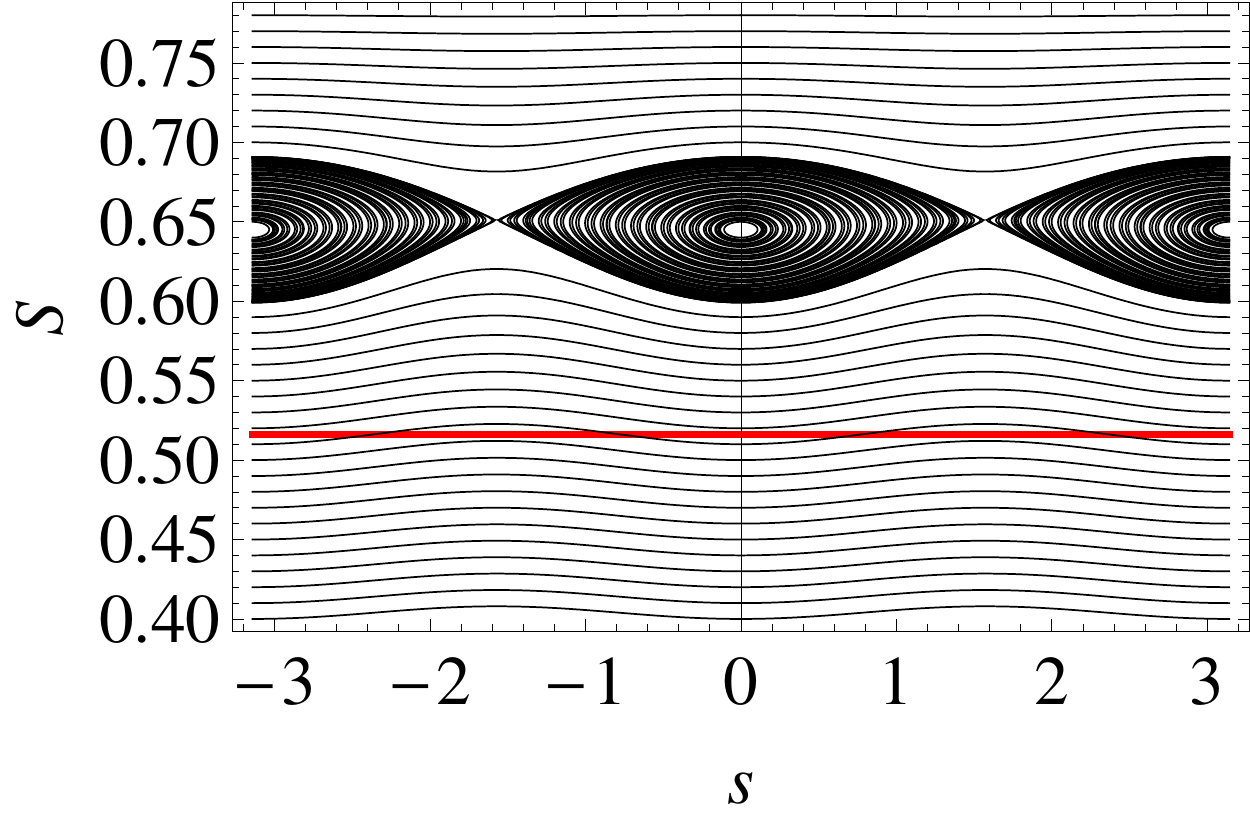}
\vglue0.5cm
\caption{Secular resonance: $\dot\omega+\dot\Omega=0$, $i=46.4^o$. The red line corresponds to the value
$S=S_{min}$.
Phase space portraits corresponding to the GPS value $a=a_{GPS}$ in the plane $(s,S)$ for the Hamiltonian including both
the Moon and Sun: $T=-0.12$ (left), \red{$T=-0.1595$ (middle),} $T=-0.2$ (right).}
\label{fig:case1plots}
\end{figure}

The interpretation of the results is the following. Take a vertical line, for example $a=1$ in Figure~\ref{fig:case1}:
below the value $T_A$ we do not find any equilibria; for values in the interval $(T_A,T_B)$ we
have one (hyperbolic) equilibrium point; above $T_B$ we find two equilibria (elliptic and hyperbolic).
This result is corroborated by Figure~\ref{fig:case1plots}, which shows the phase space portraits
in the plane $(s,S)$ with $s\in[-{\pi},{\pi})$,
$S\in[S_{min},S_{max}]$, where $S_{max}=\sqrt{\mu_E a}$, corresponding to the GPS value $a=a_{GPS}$.
The three panels show the values of $T$ below (left panel),
between (middle panel) and above (right panel) the bifurcation thresholds $T_A$, $T_B$. Given the
slight difference between all cases - Sun, Moon, Sun+Moon - as already shown in
Table~\ref{table:case1}, we just provide the results for the sample in which both the
Moon and Sun are considered.

We see that Figure~\ref{fig:case1plots}, left panel, does not exhibit any equilibrium point, while the
generation of the two elliptic equilibrium solutions at $s=\pm{\pi\over 2} $ is observed in Figure~\ref{fig:case1plots},
middle panel, for which
the value of $T$ is between those of $T_A$ and $T_B$ provided in Table~\ref{table:case1}.
For values larger than $T_B$ we observe new hyperbolic equilibria at $s=0, \pm\pi$ as shown in Figure~\ref{fig:case1plots}, right
panel.

\vskip.2in

\vskip.2in

\subsection{An alternative method for detecting bifurcations}\label{sec:alternative}
An alternative method to find bifurcations with respect to the technique presented in Section~\ref{sec:birth} relies
on the following procedure, which is based on the geometric approach used in \cite{MP14,PM14} and is related to the analysis of the critical inclination in
\cite{CDMcrit}.

Let us come back to the Hamiltonian \equ{Hexp}. We introduce Poincar\'e variables defined as
\beqa{PV}
x_1&=&L_0 e \cos s\nonumber\\
x_2&=&L_0 e \sin s\ ,
\eeqa
%where we denote by $e_0$ the value at $L_0=1$, namely $e_0=\sqrt{1-S^2}$. We notice that
and we notice that
$$
x_1^2-x_2^2=L_0^2 e^2 \cos 2s\ .
$$
Then, we can write \equ{Hexp} as
\beq{H}
\H_{exp}(x_1,x_2,T;L_0)=H_1(S,T;L_0)+H_2(S,T;L_0)\ (x_1^2-x_2^2)
\eeq
for suitable functions $H_1=H_1(S,T;L_0)$, $H_2=H_2(S,T;L_0)$. Due to the properties of the
angular momentum and of the Laplace-Runge-Lenz vector (see \cite{CDMcrit}),
%\marginpar{Giuseppe, please add details and a reference}
the variables \equ{PV} satisfy the
identity
\beq{x12}
x_1^2+x_2^2+S^2=L_0^2\ ,
\eeq
which defines a reduced phase space. Let us then introduce the variables
\beqano
X&=&x_1^2-x_2^2\nonumber\\
Y&=&2 x_1 x_2\nonumber\\
Z&=&S\ ,
\eeqano
such that \equ{x12} becomes
\beq{XY}
X^2+Y^2=(L_0^2-S^2)^2\ .
\eeq
We observe that \equ{H}, expressed in these new variables, depends on $X$, $Z$
but not on $Y$, so that its level surfaces are parabolic cylinders intersecting
the \sl lemon-shaped \rm surface  \equ{XY} on curves symmetric with respect to the inversion $Y \rightarrow -Y$. Therefore, all information about the bifurcation of
periodic orbits and the birth of equilibria can be inferred by investigating the mutual positions
of the boundary set of the lemon
$$
\L\equiv\{(X,Z)\in\real^2\ :\ \ X=\pm(L^2-Z^2)\} \ ,
$$
and the family of curves, parametrized by the level set $h$ of the Hamiltonian, as
$$
\{(Z,X(Z))\in\real^2\ :\ \ \H_{exp}(Z,T;L_0)=h\} \ .
$$
From \equ{H} we can express the coordinate $X$ as
$$
X(Z,T;L_0,h)={{h-H_1(Z,T;L_0)}\over {H_2(Z,T;L_0)}}\ .
$$

\vskip.2in

\begin{figure}[h]
\centering
%\vglue-5cm
\hglue0.1cm
\includegraphics[width=7truecm]{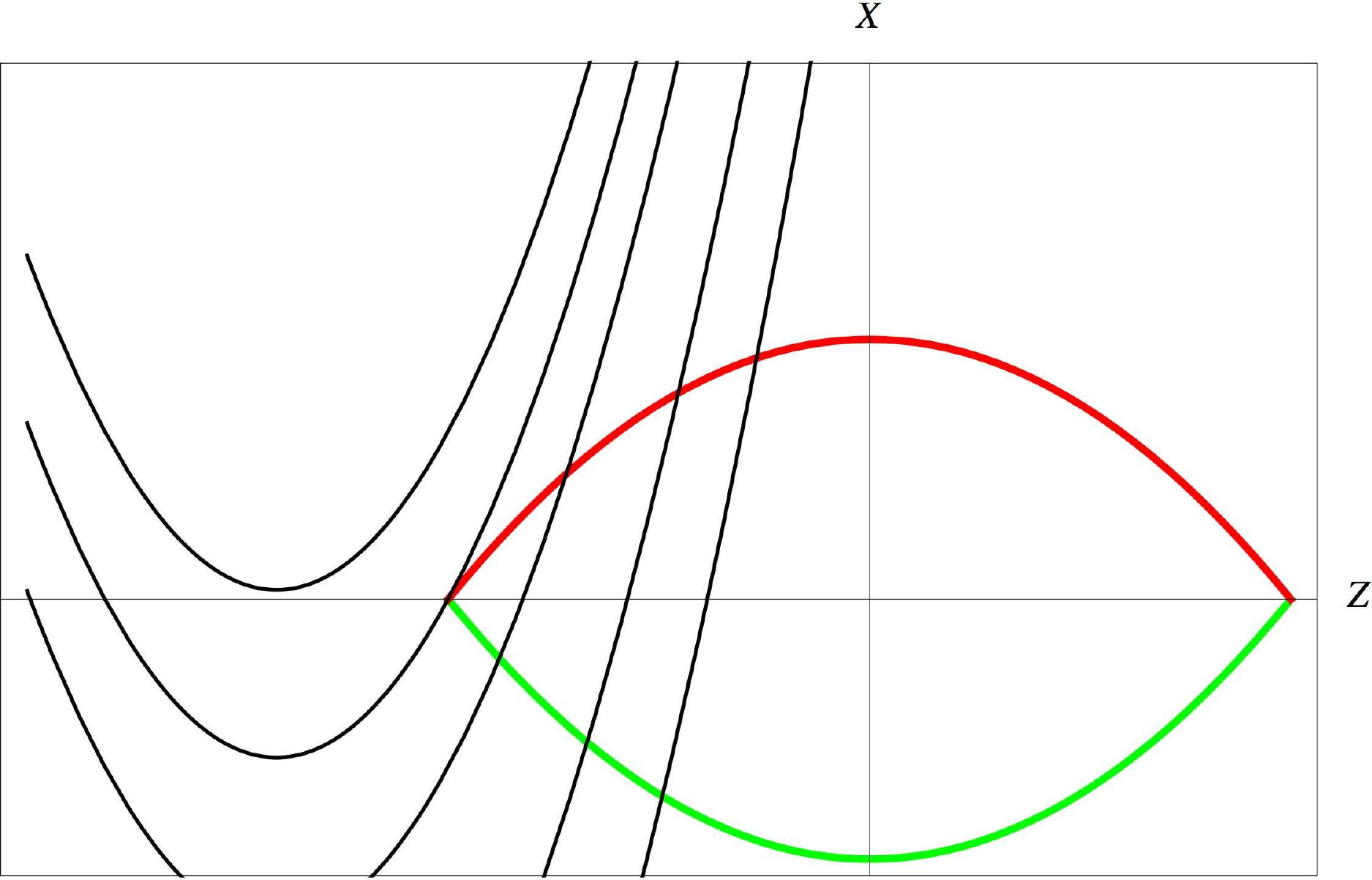}
\includegraphics[width=7truecm]{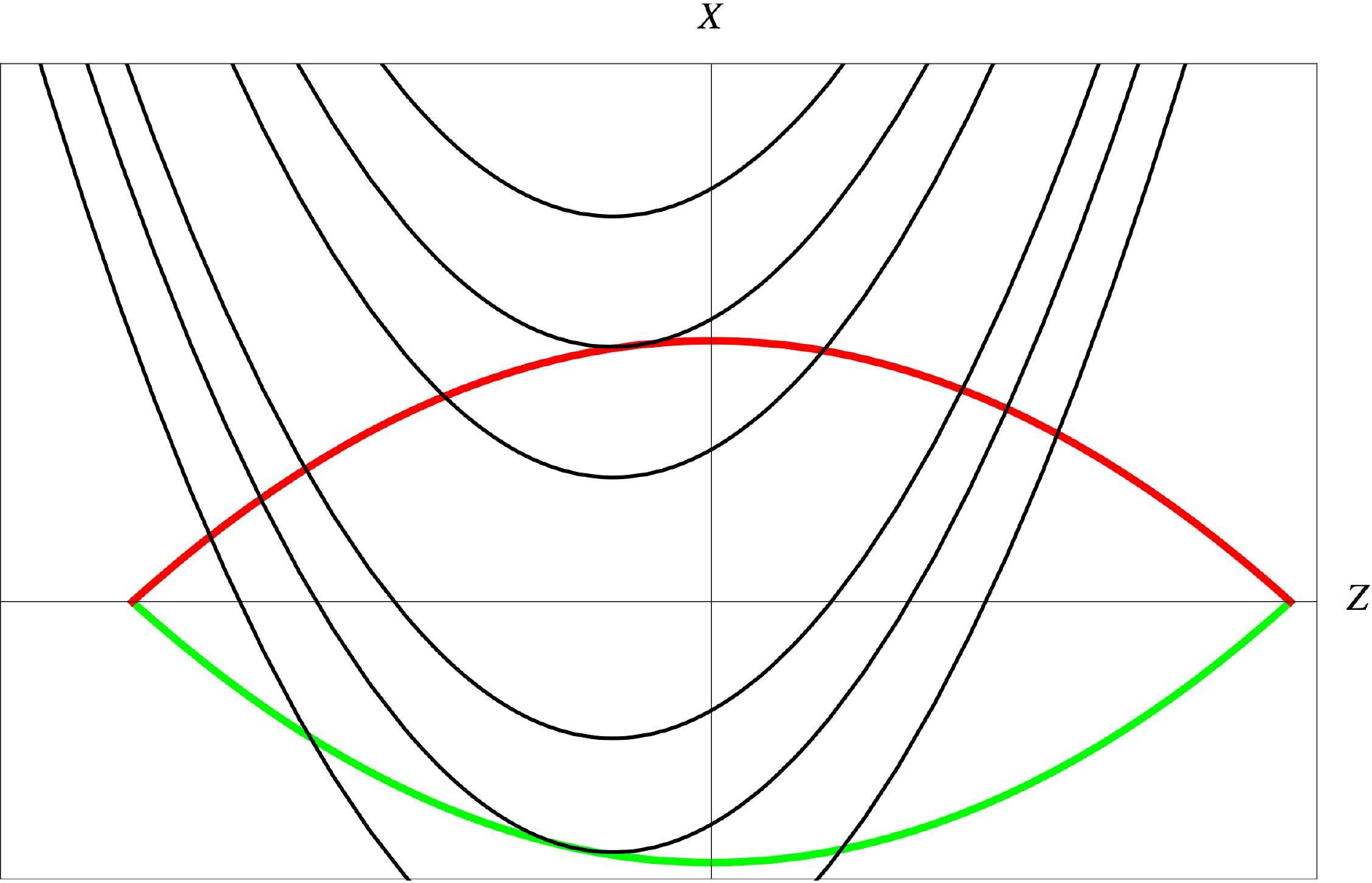}
%\vglue-15cm
\caption{The tangency between the reduced phase space and the energy surface provides
the bifurcation of the equilibria.}
\label{fig:lemon}
\end{figure}

\vskip.2in

The equilibrium points are found by imposing the tangency between the boundary of the reduced phase space $\L$ and the
function $X=X(Z,T;L_0,h)$ for a given energy level $\H_{exp}=h$ that is well approximated by a family of parabolae,
sliding parallel to the $Z$-axis by varying the energy value $h$ and parallel to the $X$-axis by varying the value of the second integral $T$. The left panel in Figure~\ref{fig:lemon} shows the case with no contacts when $T<T_A,T_B$. When both thresholds are passed (right panel), the first contact point with the upper branch of the lemon
provides the first bifurcation (stable, because it is an external contact, see \cite{PM14}), while the last contact point yields the second bifurcation (unstable, being an internal contact). The intermediate case is easily guessed as the case in which the vertex of the parabola remains to the left of the corner of the lemon.

Making explicit computations to determine the first and second thresholds
one obtains the same results as in Table~\ref{table:case1}.

\subsection{A cartographic study of the resonance $\dot{\omega}+\dot{\Omega}=0$ by using the FLIs }\label{sec:FLI}

\begin{figure}[h]
\centering
\vglue0.1cm
\hglue0.1cm
\includegraphics[width=5truecm,height=4truecm]{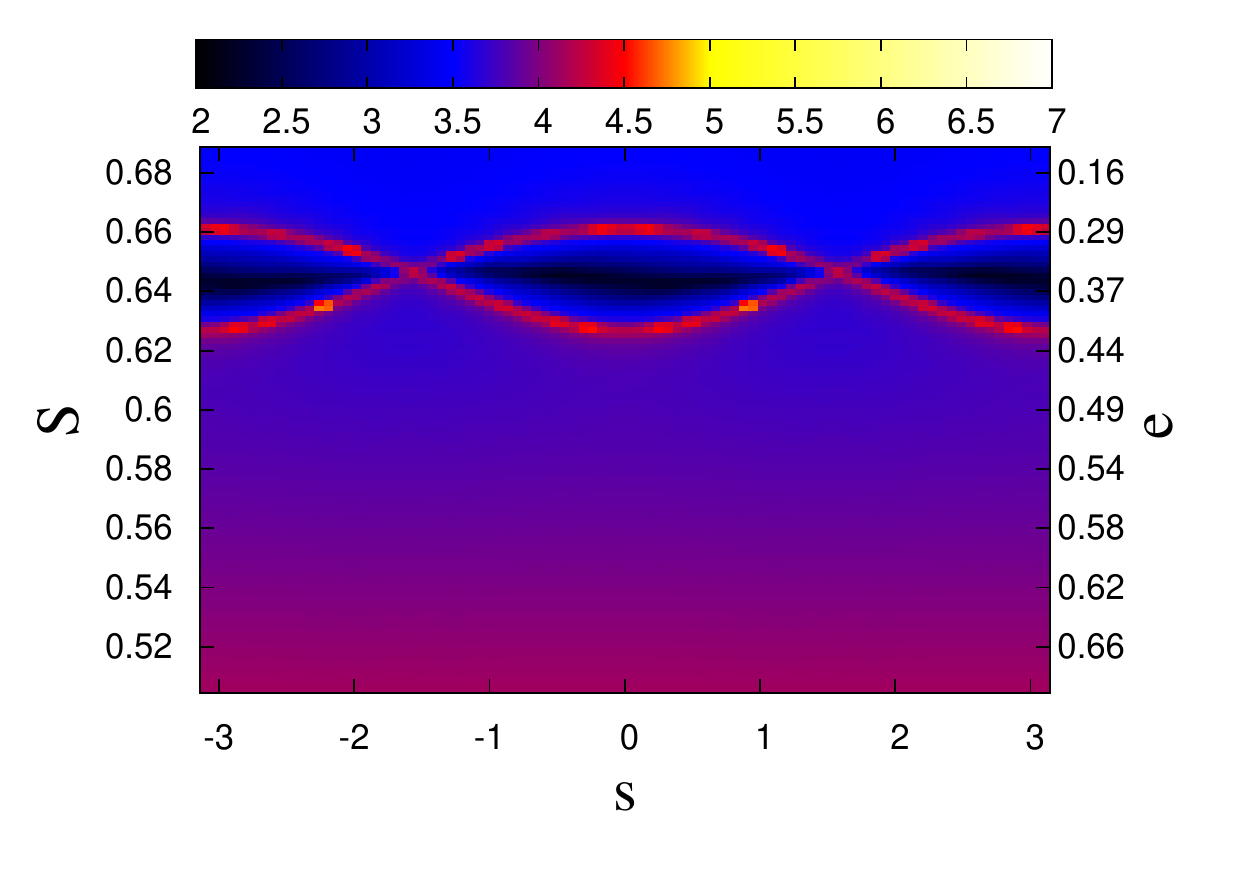}
\includegraphics[width=5truecm,height=4truecm]{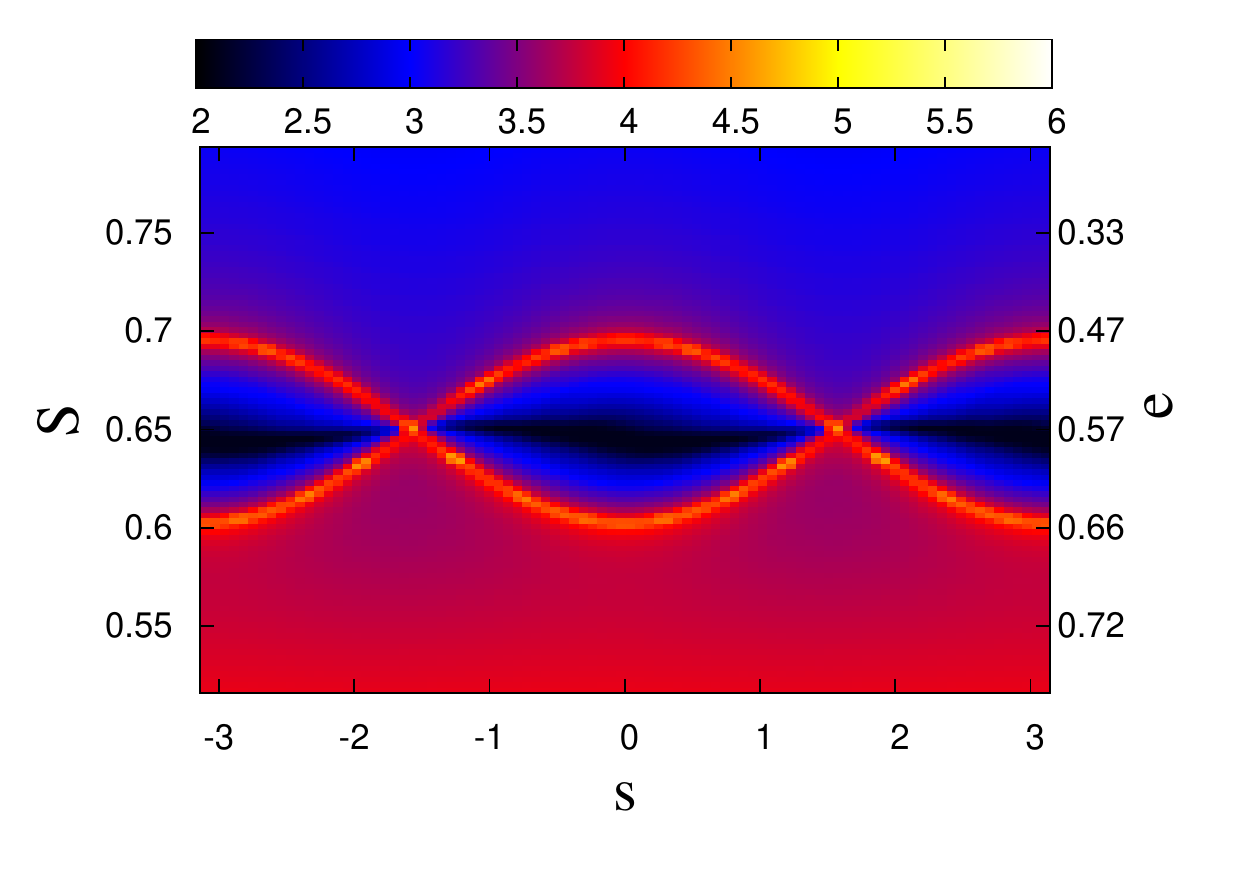}
\includegraphics[width=5truecm,height=4truecm]{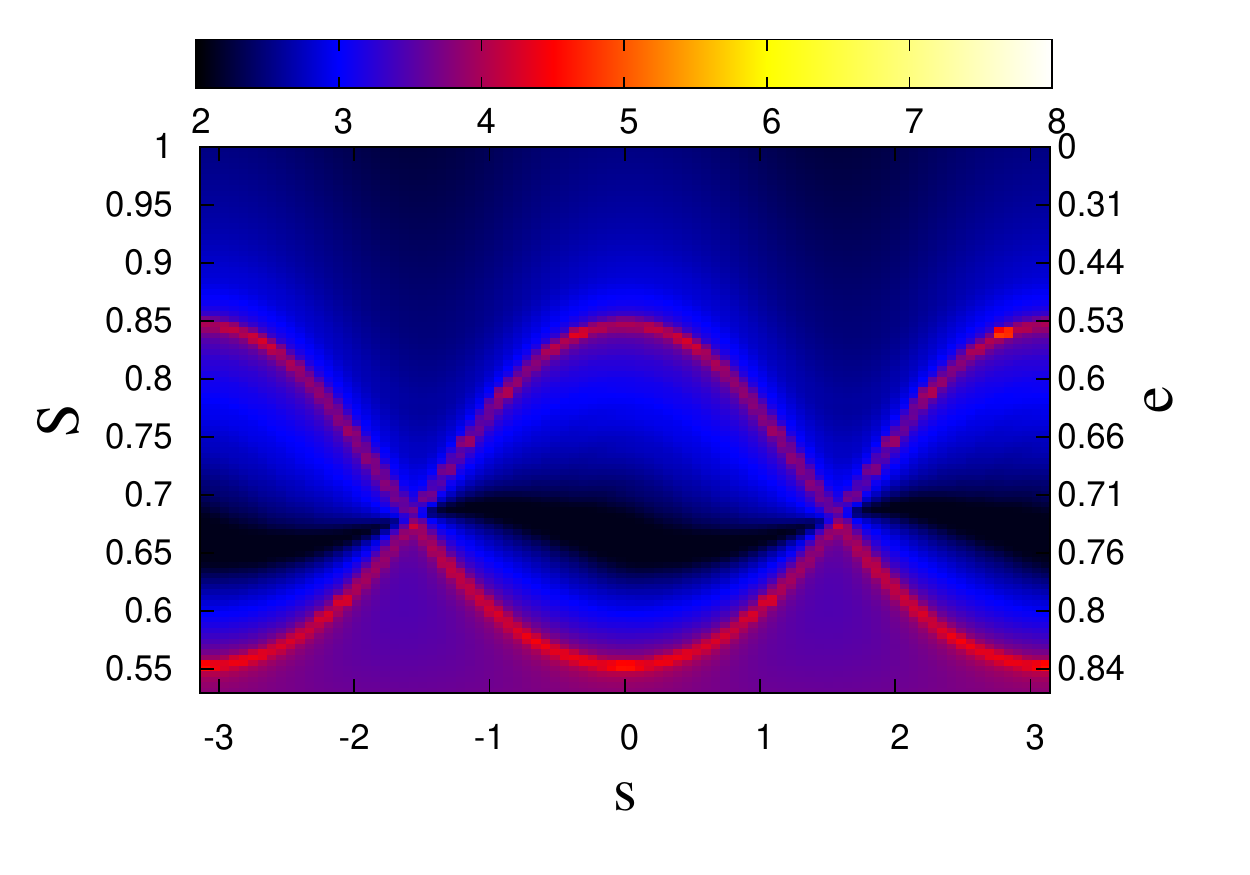}\\
\vglue-0.6cm
\includegraphics[width=5truecm,height=4truecm]{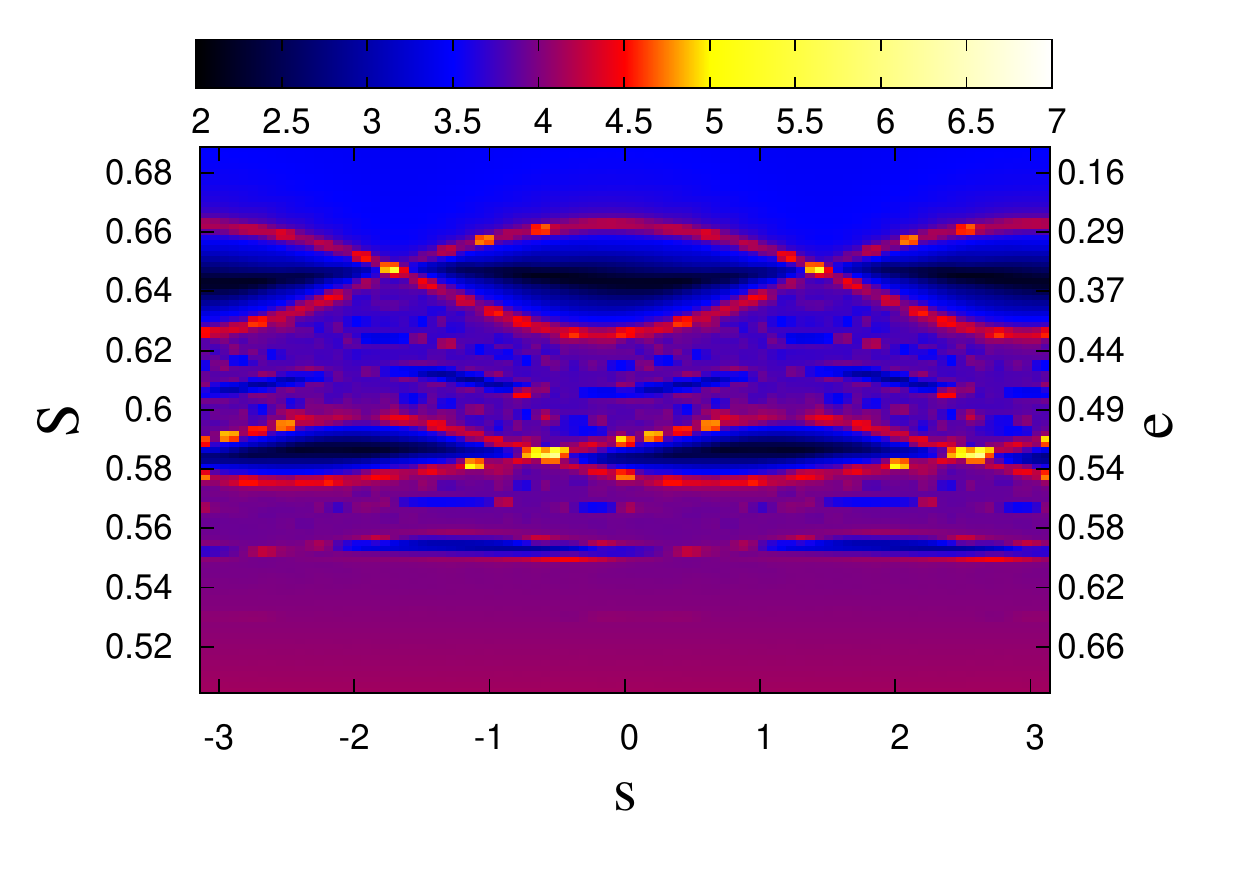}
\includegraphics[width=5truecm,height=4truecm]{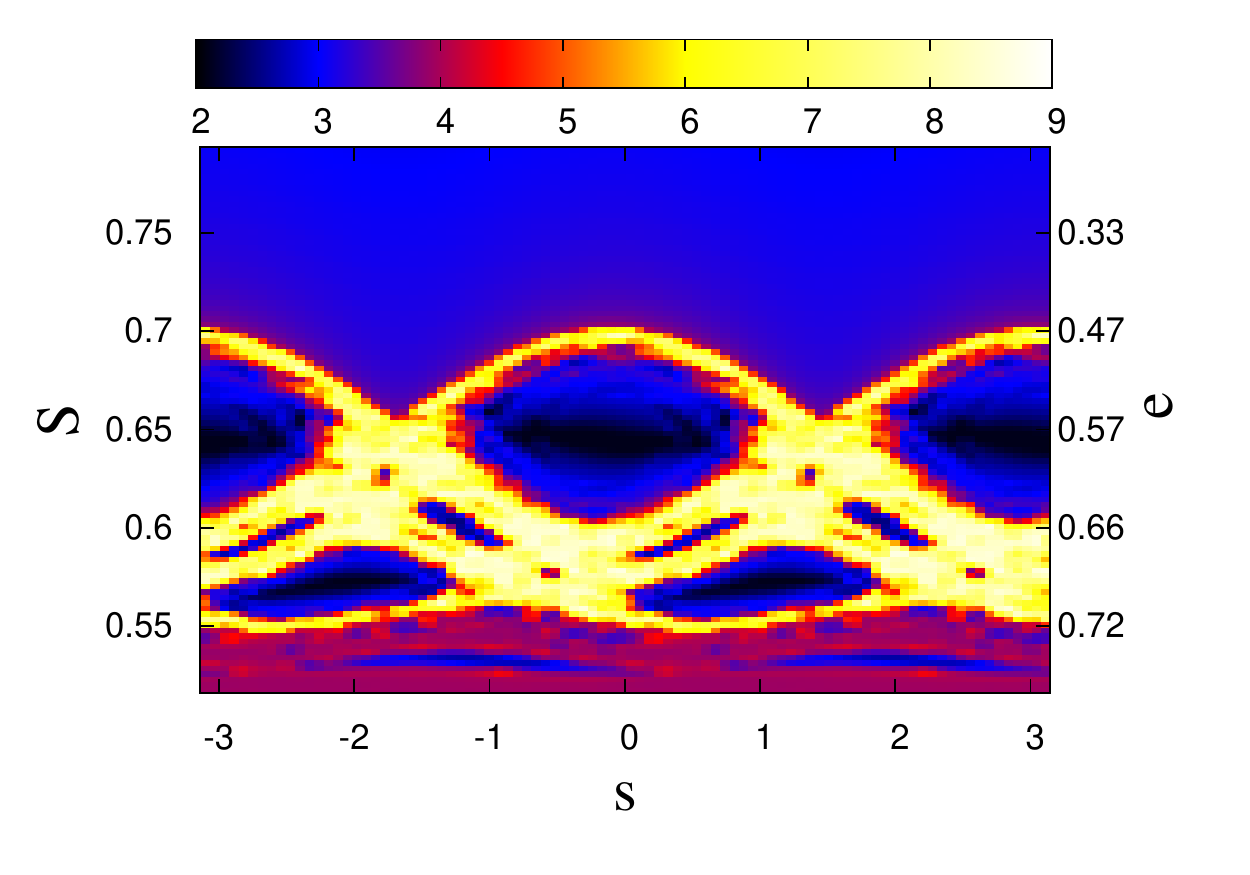}
\includegraphics[width=5truecm,height=4truecm]{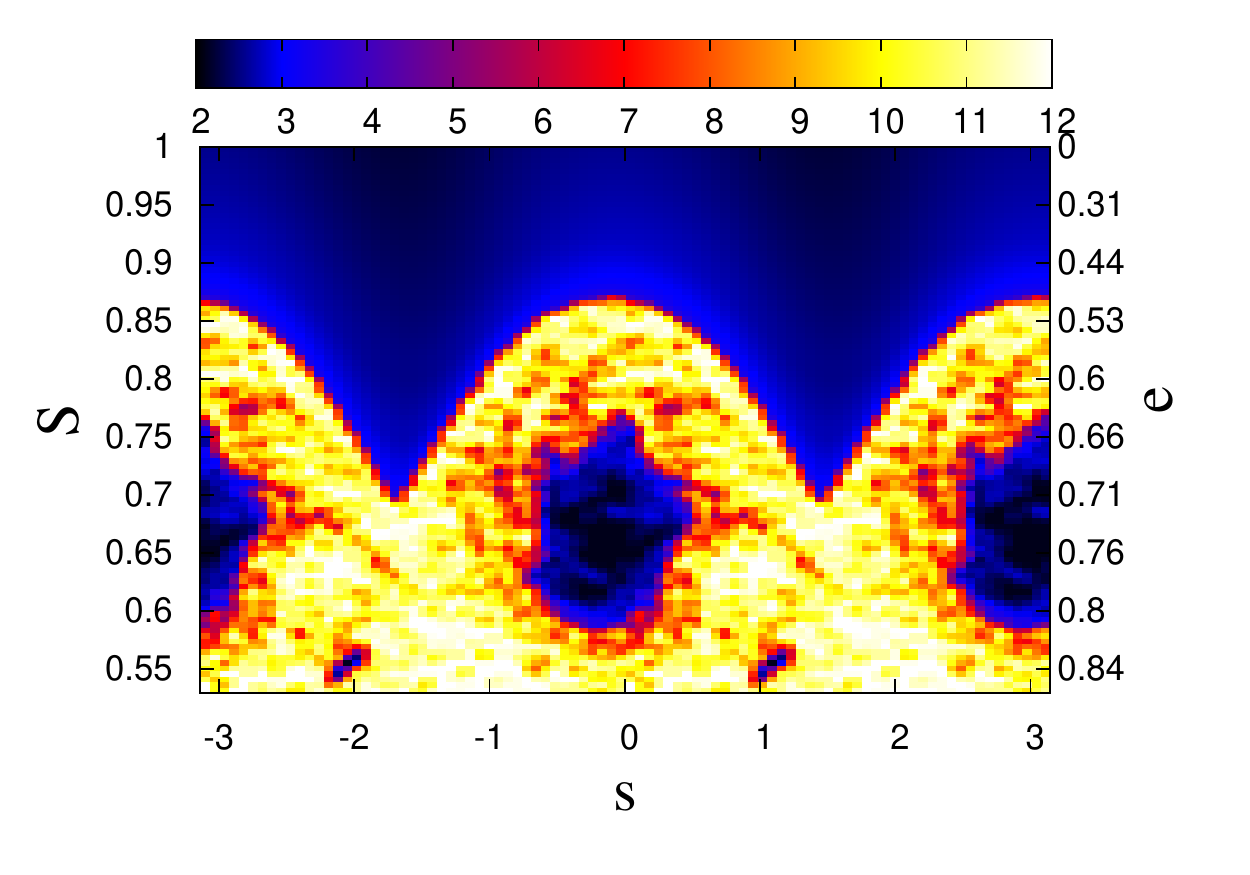}\\
%\vglue-0.4cm
\caption{FLI map for the secular resonance $\dot{\omega}+\dot{\Omega}=0$ with $T=-0.2$, and
$a=20\,000\, km$ (left panels), $a=26\,560\, km$ (center panels), $a=42\,164\, km$ (right panels).  Top panels correspond to the RM2 model, while bottom panels are computed for the RM1 model. The total time span is 465 years
(equal to $25 \cdot 18.6$ years).}
\label{fli_20_gps_geo}
\end{figure}

\begin{figure}[h]
\centering
\vglue0.1cm
\hglue0.1cm
\includegraphics[width=5truecm,height=4truecm]{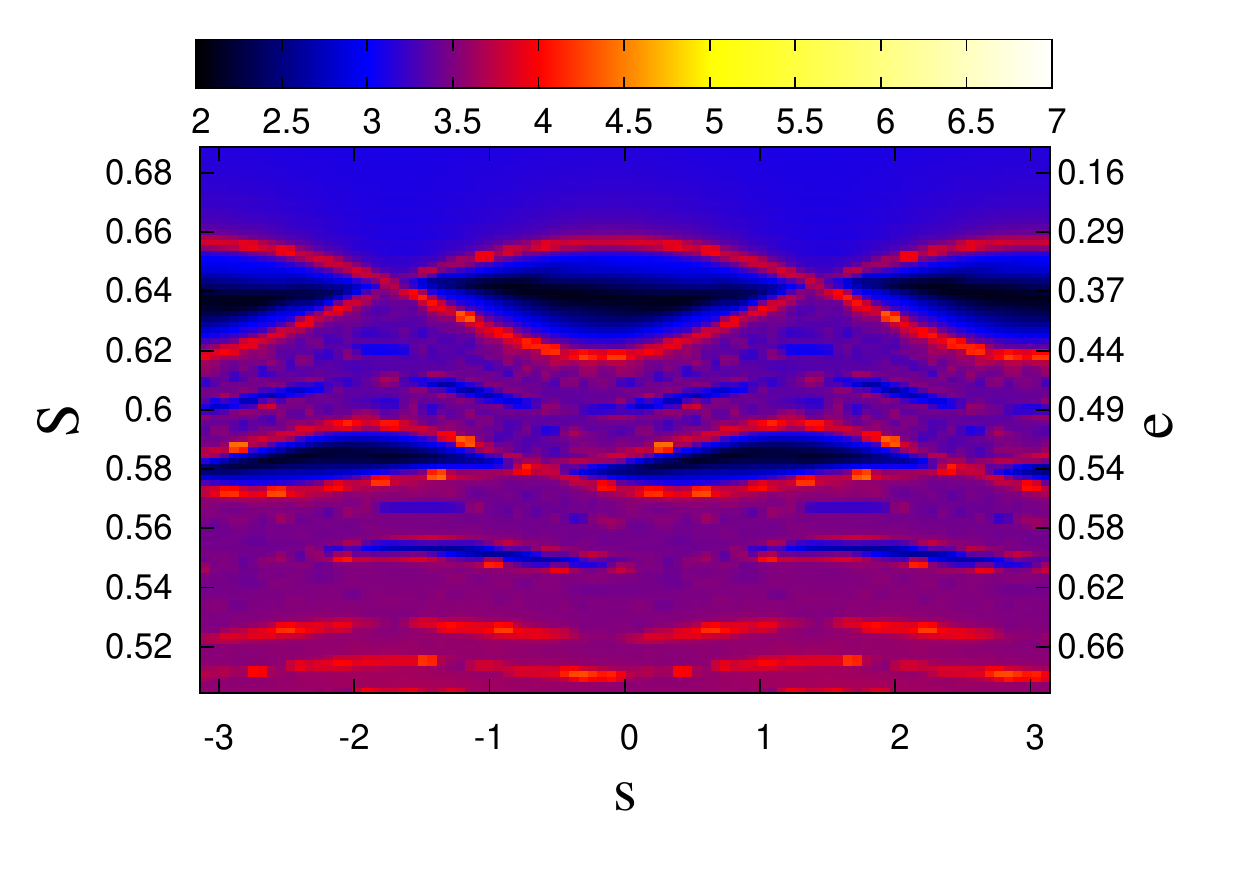}
\includegraphics[width=5truecm,height=4truecm]{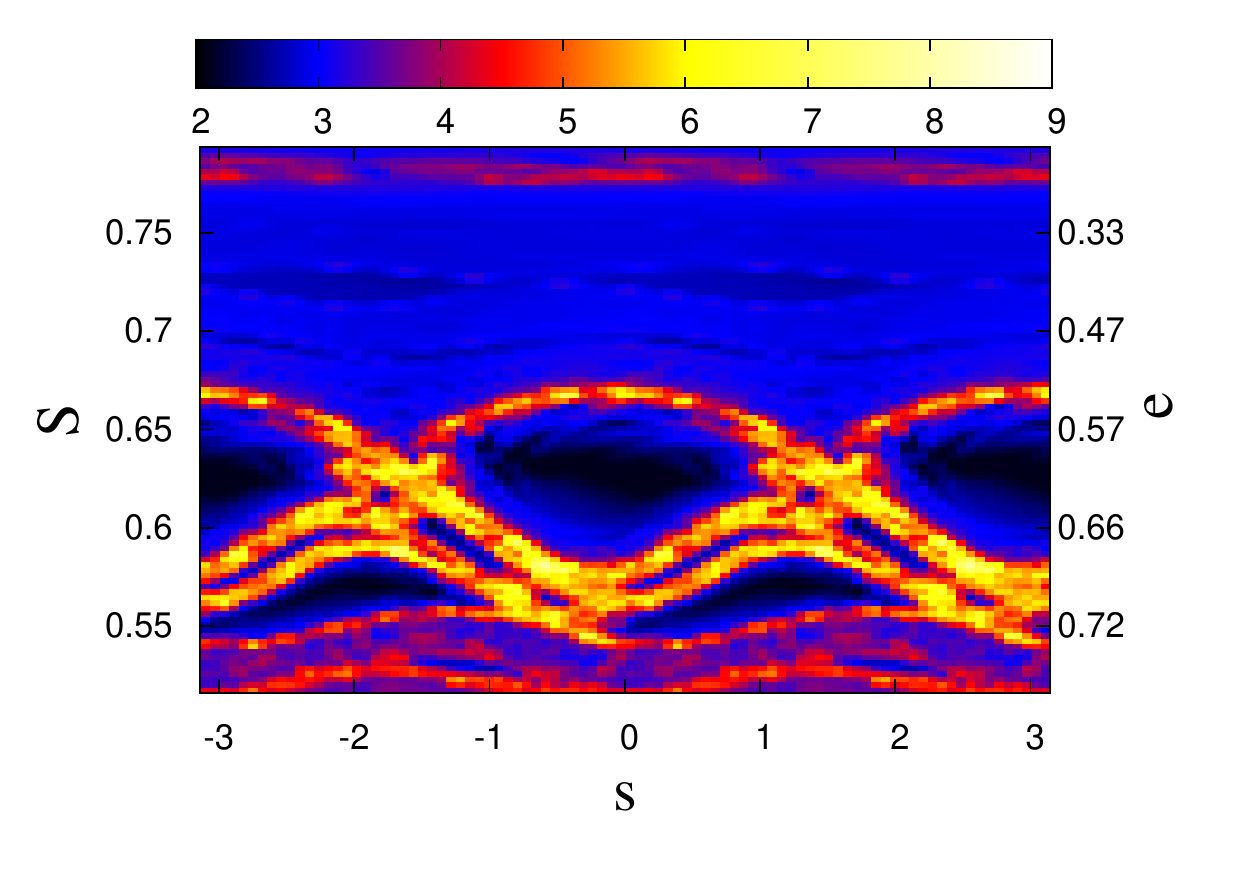}
\includegraphics[width=5truecm,height=4truecm]{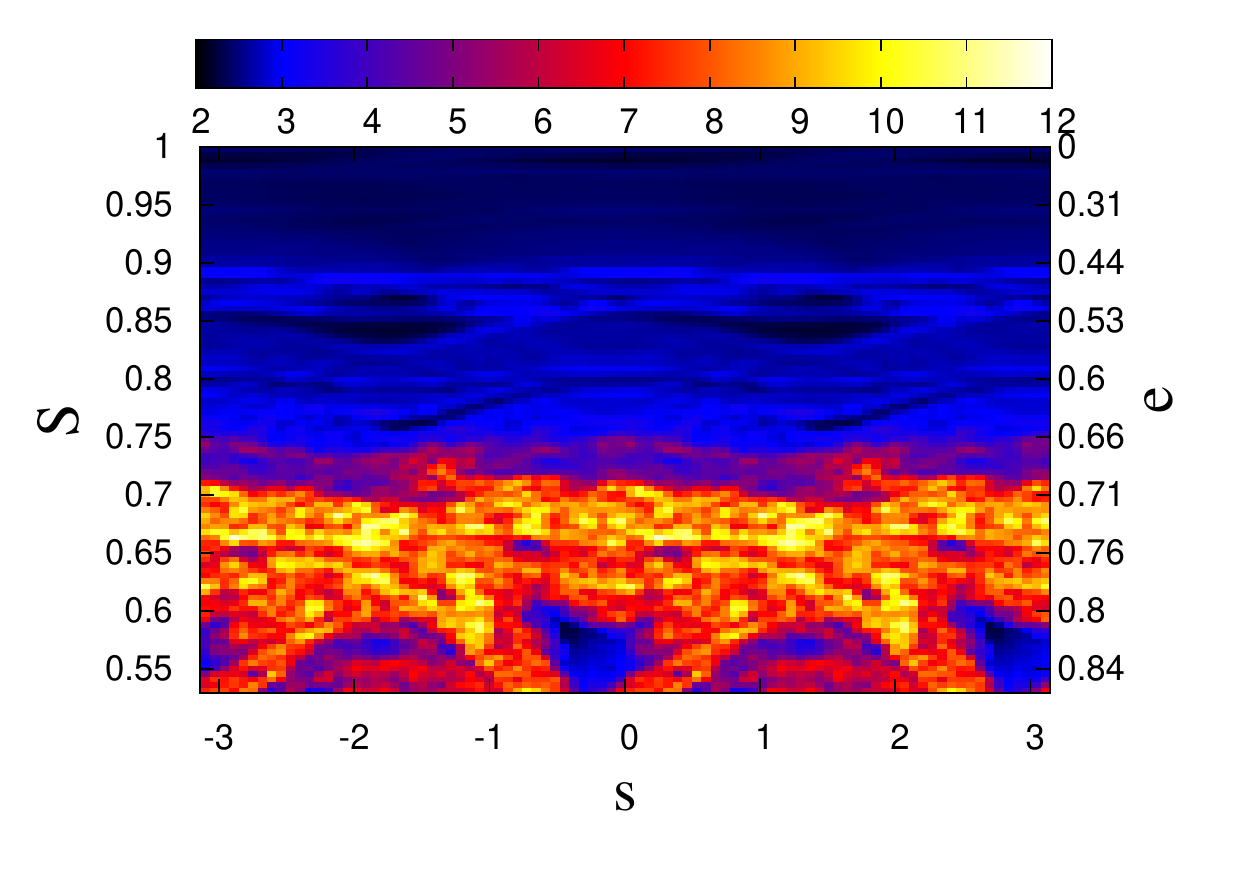}\\
\vglue-0.6cm
\includegraphics[width=5truecm,height=4truecm]{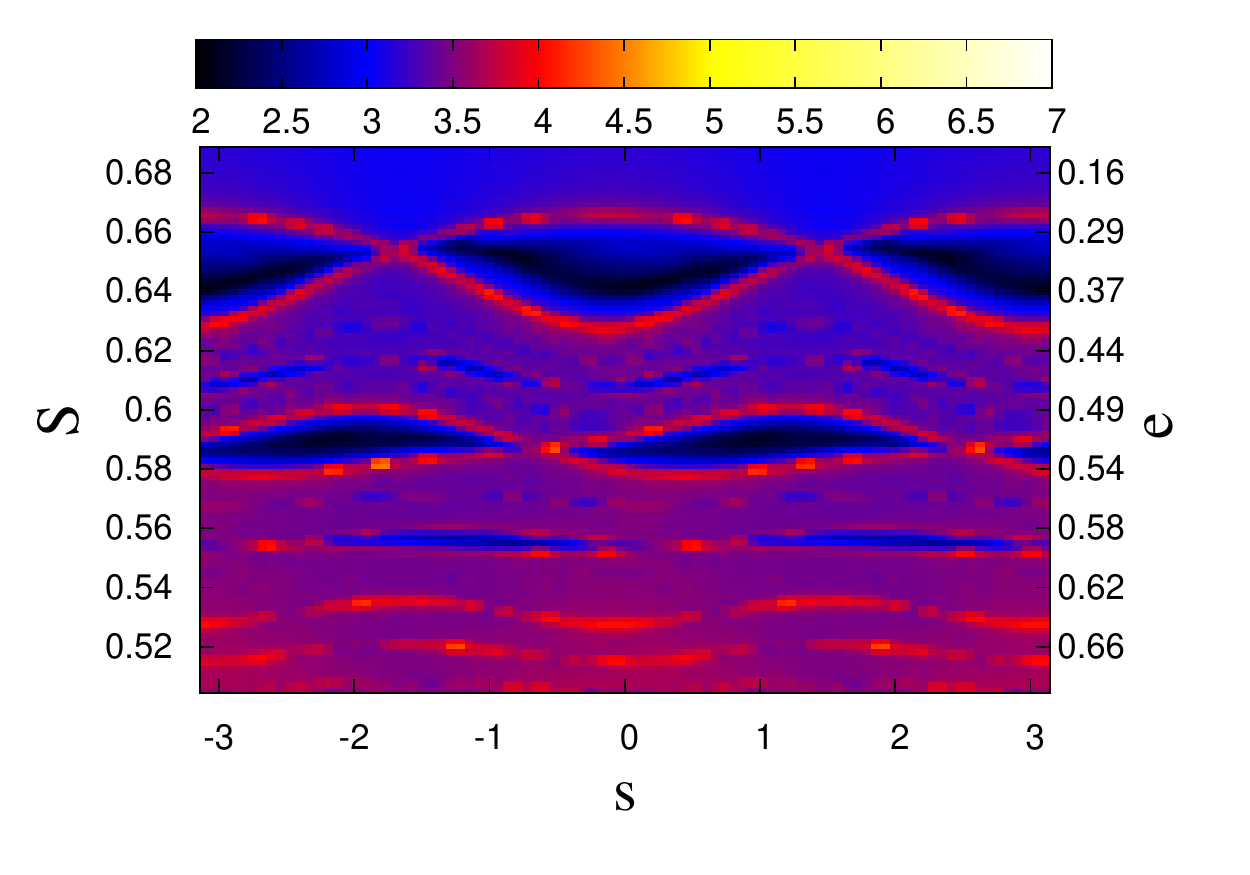}
\includegraphics[width=5truecm,height=4truecm]{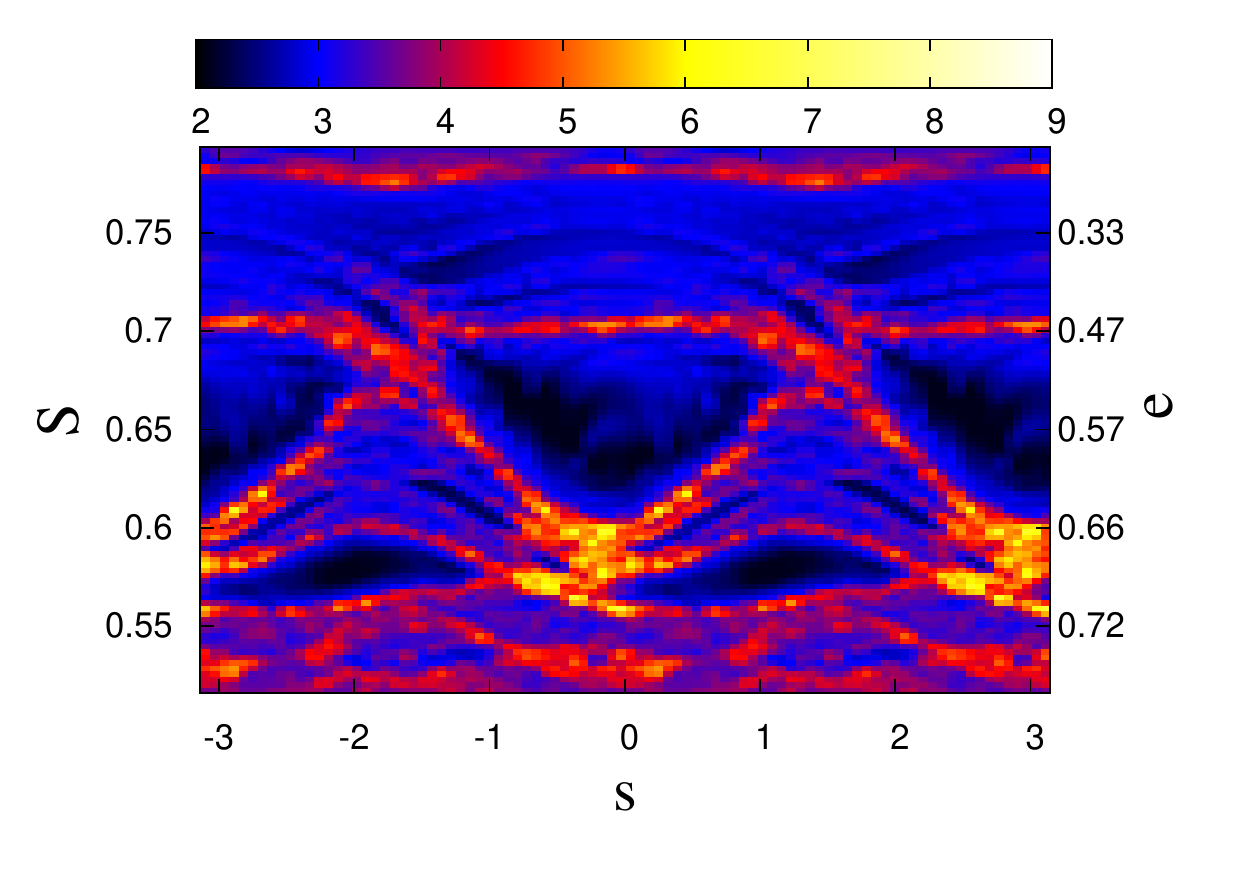}
\includegraphics[width=5truecm,height=4truecm]{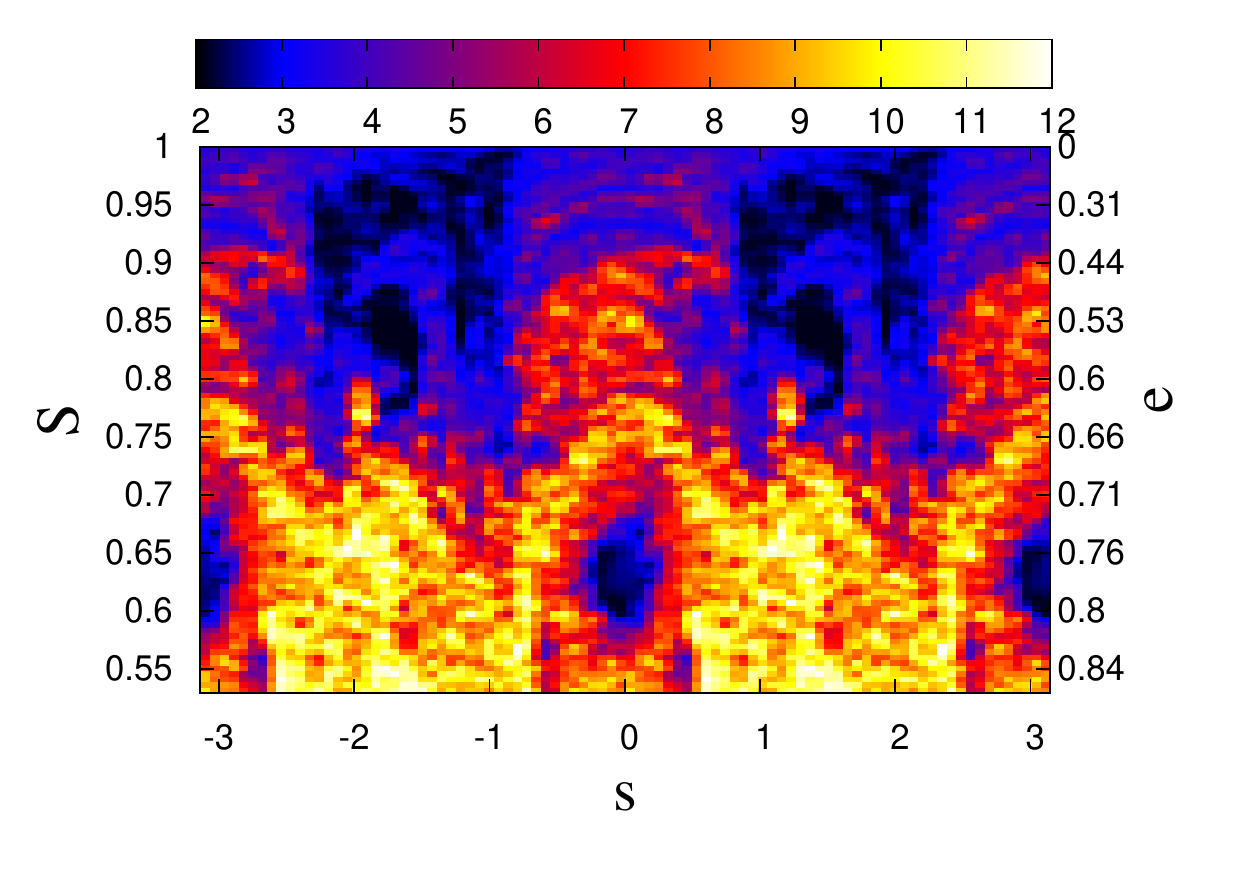}\\
%\vglue-0.6cm
\caption{FLI map for the secular resonance $\dot{\omega}+\dot{\Omega}=0$ with $T=-0.2$ and
$a=20\,000\, km$ (left panels), $a=26\,560\, km$ (center panels), $a=42\,164\, km$ (right panes). The initial $\Omega$ in the top panels is $\Omega=0^o$, while the bottom panels are obtained for $\Omega=180^o$. The panels correspond to the DAH model. The total time span is 465 years
(equal to $25 \cdot 18.6$ years).}
\label{fli_20_gps_geo_Om180}
\end{figure}

\begin{figure}[h]
\centering
\vglue0.1cm
\hglue0.1cm
\includegraphics[width=6truecm,height=4.5truecm]{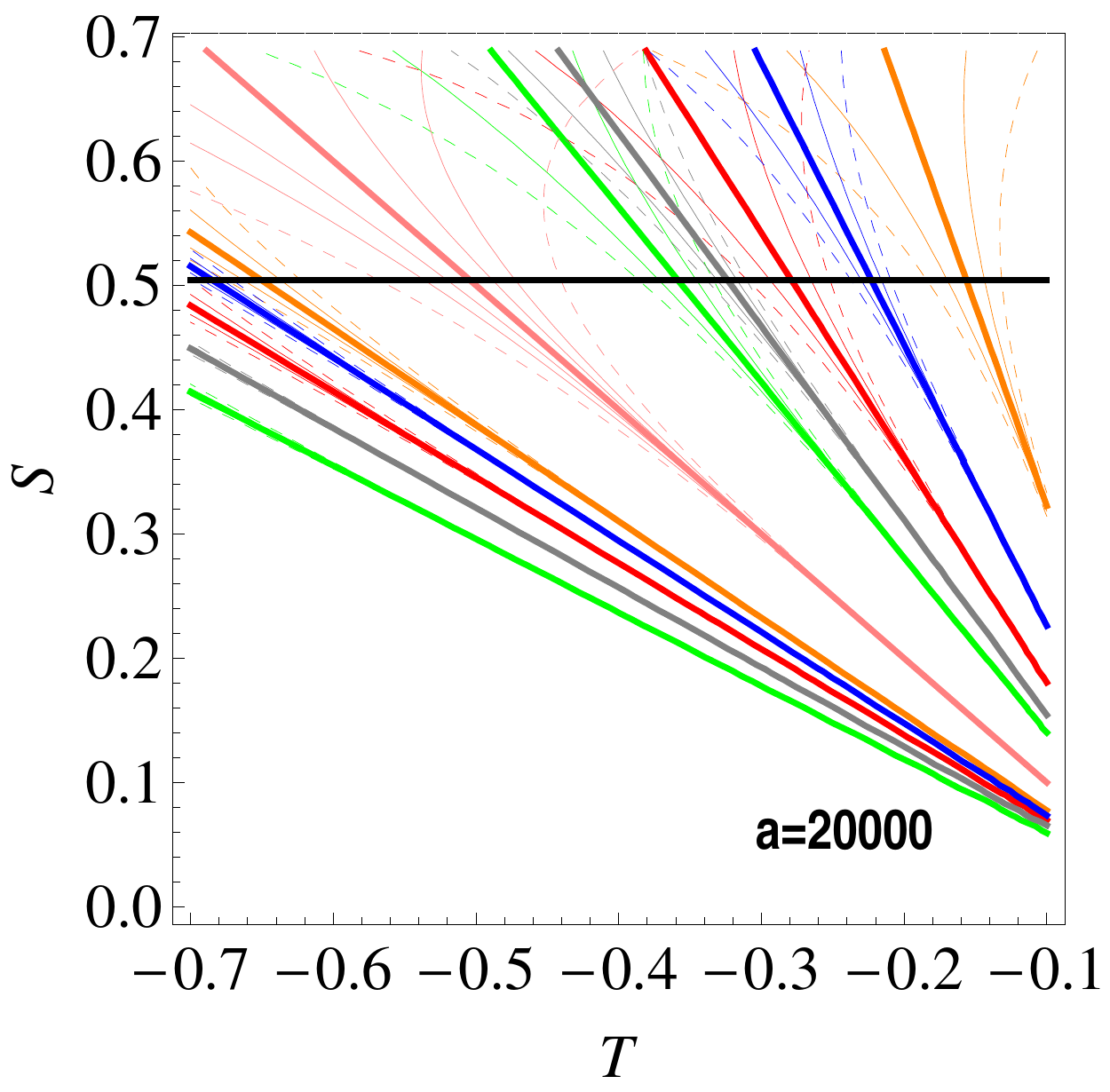}
\includegraphics[width=6truecm,height=4.5truecm]{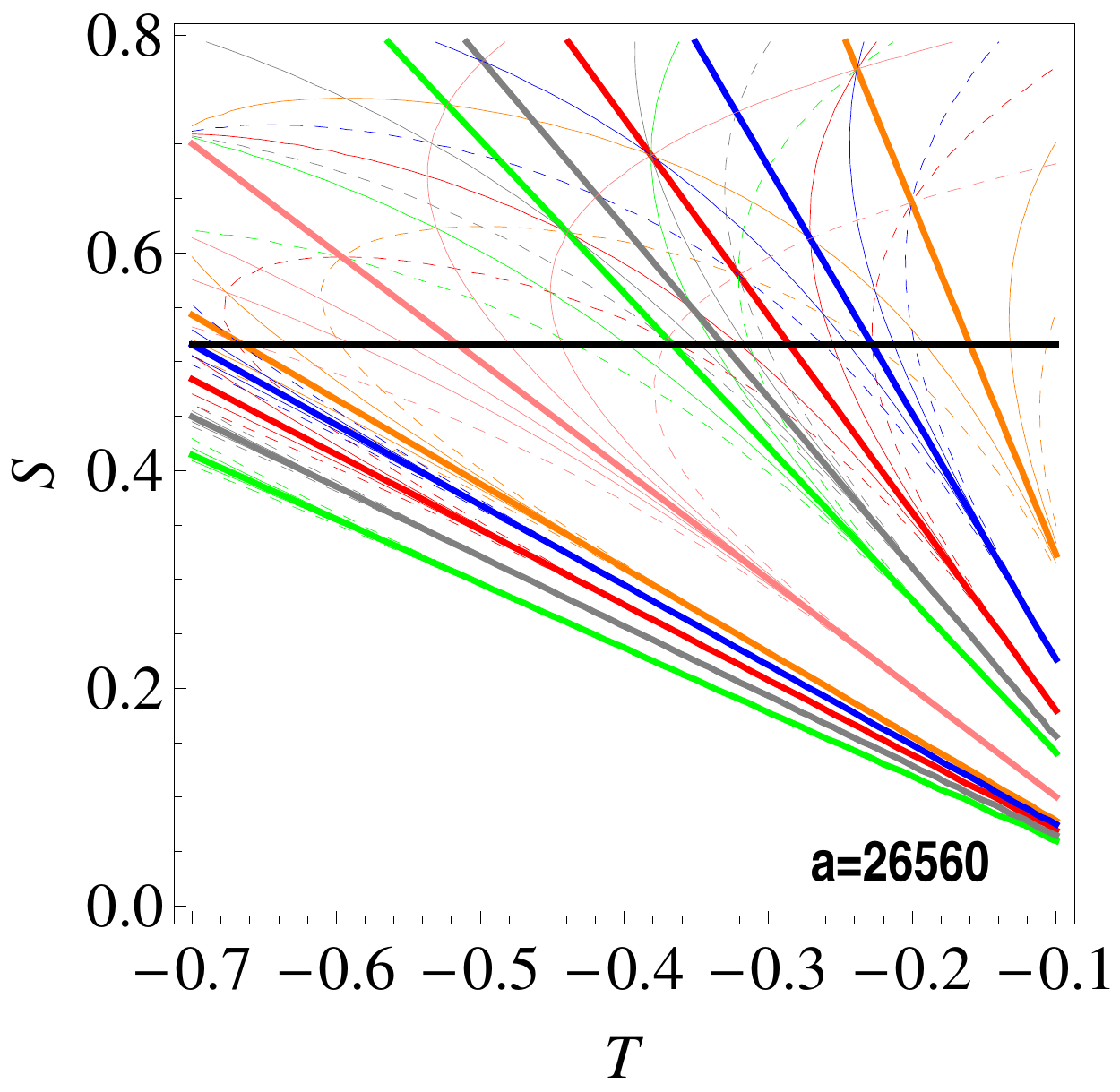}
\includegraphics[width=6.3truecm,height=4.7truecm]{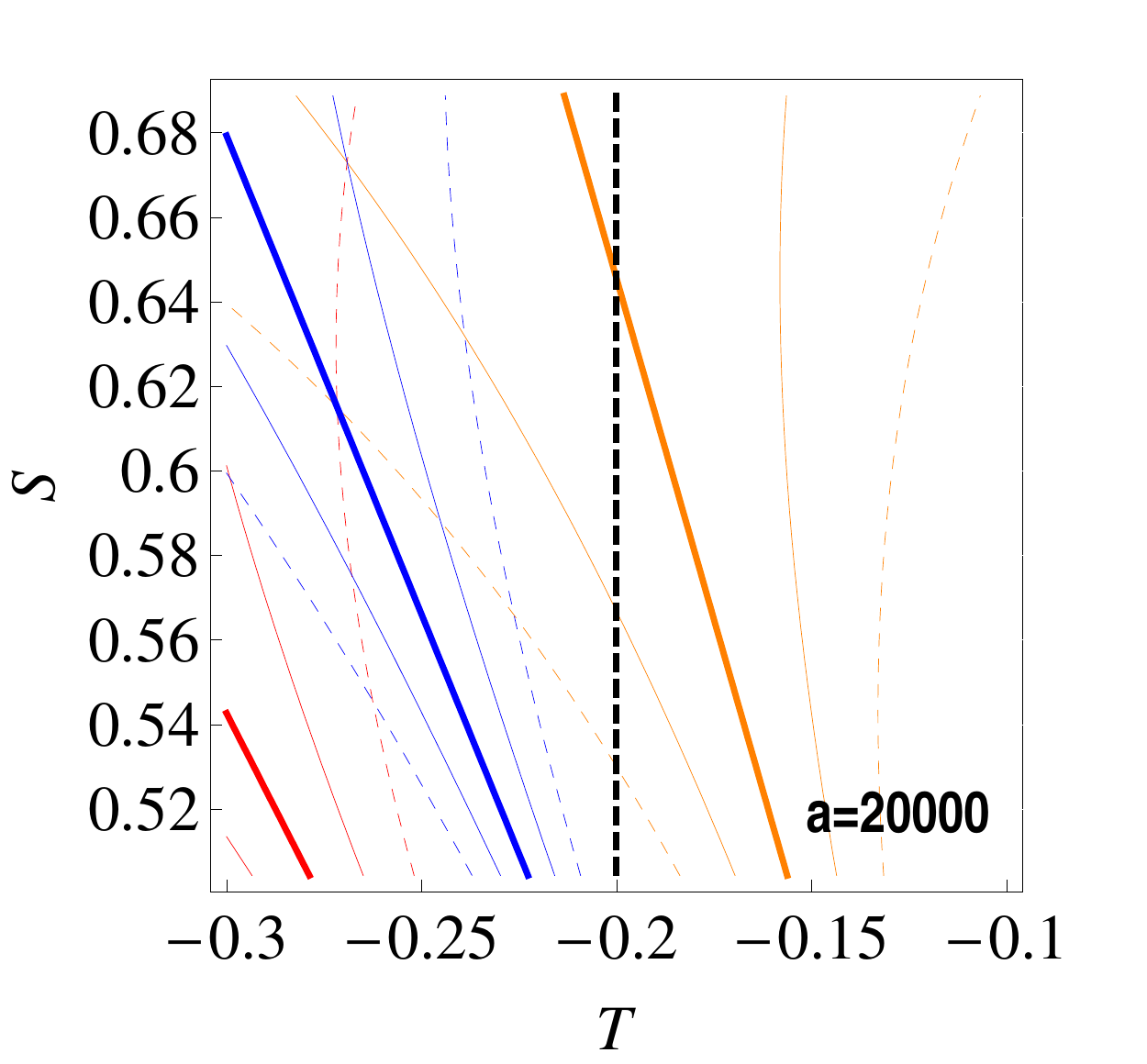}
\includegraphics[width=6.3truecm,height=4.7truecm]{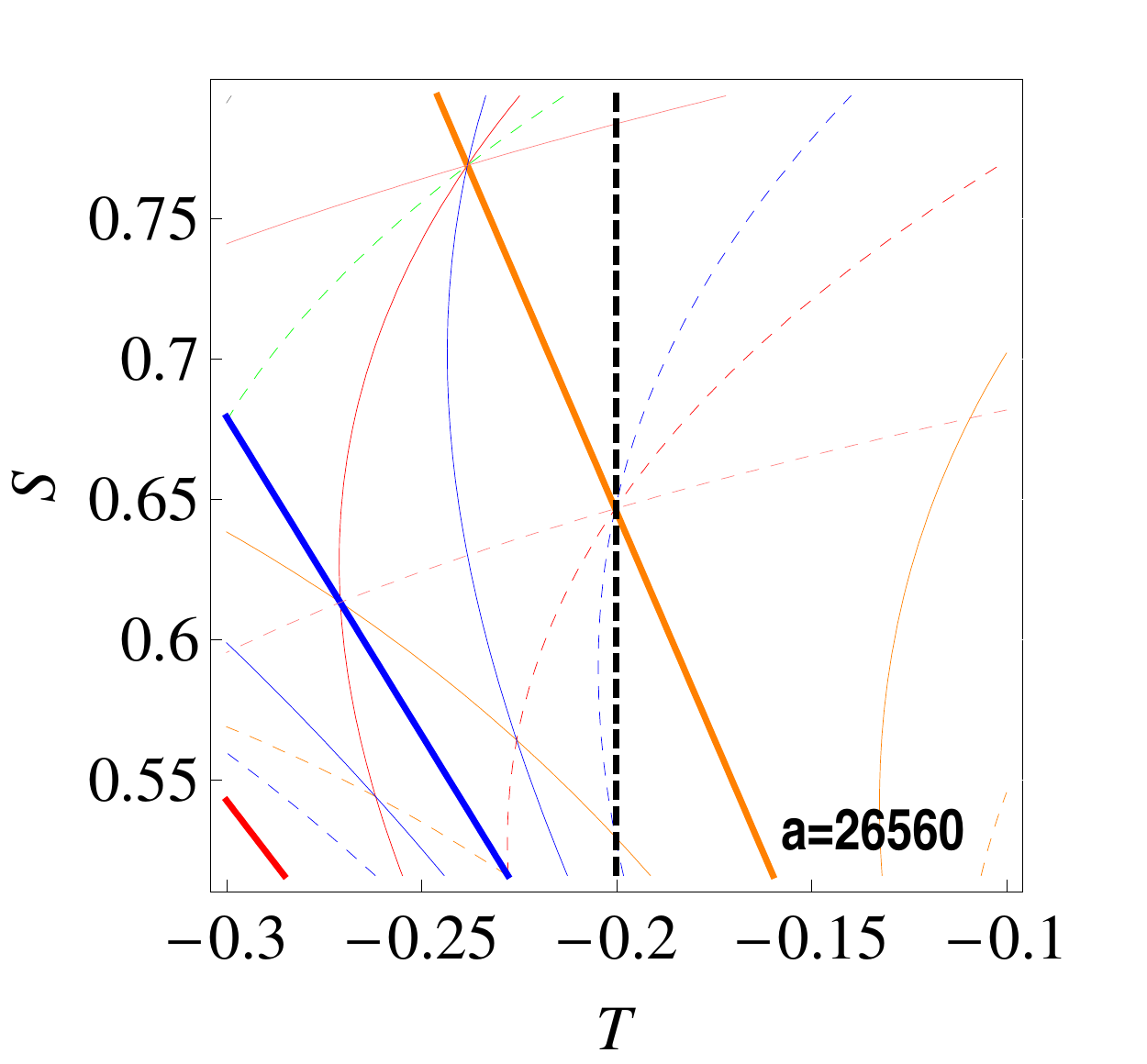}
\vglue0.5cm
\caption{The web structure of lunisolar resonances in the space of the actions for $a= 20\,000 $ km (left panels) and  $a= 26\,560 $ km (right
panels). The thick curves represent the location of the resonances depending only on inclination ($k_4$=0 in \eqref{secresmoon} and \eqref{secressun}, or $\kappa=0$ in \eqref{secresmoonbis}):
$\dot{\Omega}=0$ (pink color, $i=90^\circ$), $\dot{\omega}-\dot{\Omega}=0$ (green color, $i=73.2^\circ$, $i=133.6^\circ$), $2
\dot{\omega}-\dot{\Omega}=0$ (grey color, $i=69.0^\circ$), $i=123.9^\circ$, $\dot{\omega}=0$ (red color, $i=63.4^\circ$, $i=116.6^\circ$),
$2\dot{\omega}+\dot{\Omega}=0$ (blue color, $i=56.1^\circ$, $i=111^\circ$) and $\dot{\omega}+\dot{\Omega}=0$ (orange color, $i=46.4^\circ$,
$i=106.9^\circ$).  The thin curves give the position of the resonances $ (2-2 p) \dot{\omega}+ m \dot{\Omega}+ \kappa \dot{\Omega}_M=0$ with
$p,m \in \{0,1,2\}$, and $\kappa \in \{-2,2\}$ (dashed lines) or $\kappa\in \{-1,1\}$ (continuous lines). The horizontal black lines (top panels) are obtained for $S=S_{min}$, while the vertical black dashed lines (bottom panels) correspond to the value of $T$ used in computing the
Figures~\ref{fig:case1plots} (right panel), \ref{fli_20_gps_geo} and \ref{fli_20_gps_geo_Om180}.  Top panels are obtained for $S\in [0, S_{max}]$, whereas in the bottom plots $S$ varies from $S_{min}$
to $S_{max}$, as explained in the text.}
\label{WEB_structure}
\end{figure}

The reduced model studied in Section~\ref{sec:casestudy1} is derived under the assumptions {\bf H1} and {\bf H2},
which express
that the resonance $\dot{\omega}+\dot{\Omega}=0$ is isolated (namely, it does not interact with the other
resonances and therefore the Hamiltonian may be averaged over the fast angle $\eta$) and that $\Omega_M$ is constant.
The purpose of this Section is to discuss in detail these assumptions and to test whether the reduced model
provides reliable results on the bifurcations of periodic orbits.
To this end, we consider three models to which we refer as DAH, RM1 and RM2, where:

a) DAH is the full model based on \red{the doubly--averaged} Hamiltonian \eqref{Hres}.
\red{We underline that this model provides satisfactory results when compared with higher--fidelity models, which consider higher order terms in the lunisolar perturbations as well as in the geopotential (compare also with \cite{DRADVR15})};

b) RM1 is the reduced model characterized by the Hamiltonian \eqref{Hres} averaged also
over $\eta$ (namely, it takes into account the effects of $J_2$, $\overline \R_{Moon}^{(res)}$ and
$\overline \R_{Sun}^{(res)}$, where the contributions due to the Moon and Sun are given by
\eqref{R_moon_bigOmegaMoon} and \eqref{Rsunaverage});

c) RM2 is described by the third Hamiltonian in \eqref{H3}.

\vskip.1in

In order to compare these models, we compute the Fast Lyapunov Indicators, hereafter FLIs (\cite{FLG1997}),
which are suitable chaos indicators allowing us to find the equilibrium points and to distinguish between resonant, stable
and chaotic motions. \red{Figures~\ref{fli_20_gps_geo} and \ref{fli_20_gps_geo_Om180} show the FLI values in the $(s, S)$ plane, where
$s \in [-\pi, \pi]$, $S \in [S_{min}, S_{max}]$, for $T=-0.2$ and for $a=20\,000\, km$ (left panels),
$a=26\,560\, km$ (middle panels) and $a=42\,164\, km$ (right panels). Figure~\ref{fli_20_gps_geo} presents the results obtained for the models RM2 and RM1, while Figure~\ref{fli_20_gps_geo_Om180} shows the FLI values for the DAH model.}

\red{Moreover, in order to offer an analytical background for the numerical results depicted in Figures~~\ref{fli_20_gps_geo} and \ref{fli_20_gps_geo_Om180}, we show in Figure~\ref{WEB_structure} the structure of lunisolar resonances in the space of the actions $(T,S)$, for $a=20\,000\, km$ (left panels) and $a=26\,560\, km$ (right panels). The analytical estimate of the location of each resonance defined by the relation \eqref{secresmoonbis} is obtained by approximating $\dot{\omega}$ and $\dot{\Omega}$ with the expressions \eqref{omega12} which consider only the effect of $J_2$.
We avoid presenting similar plots for $a=42\,164\, km$, because in the GEO region the secular effects of the Moon and Sun are comparable to those of $J_2$.}

\red{Before describing in detail the results shown in Figures~\ref{fli_20_gps_geo} and \ref{fli_20_gps_geo_Om180}, let us discuss first the
meaning of Figure~\ref{WEB_structure}. The colored curves provide the location of each resonance; thick lines are associated to the inclination--dependent--only lunisolar resonances $ (2-2 p) \dot{\omega}+ m \dot{\Omega}=0$, $p,m=0,1,2$, while the thin curves give the position of the resonances $ (2-2 p) \dot{\omega}+ m \dot{\Omega}+ \kappa \dot{\Omega}_M=0$, with
$p,m \in \{0,1,2\}$, and $\kappa \in \{-1,1\}$ (continuous curves) or $\kappa\in \{-1,1\}$ (dashed curves). Upper panels show the resonant structures for $S\in [0, S_{max}]$. These plots contain also the horizontal black line $S=S_{min}$, where $S_{min}$ is computed from the condition that the distance of the perigee cannot be smaller than the radius of the Earth (see the relation \eqref{Smin}). Therefore, the interval of interest is $[S_{min}, S_{max}]$ and the bottom panels of Figure~\ref{WEB_structure} magnify the region associated to the orbits that do not collide with the Earth,  at least within a small interval time. The vertical black dashed lines correspond to the value $T=-0.2$, used in computing the FLI maps.}

\red{Let us return now to Figures~\ref{fli_20_gps_geo} and \ref{fli_20_gps_geo_Om180}. We stress that besides the values of $S$,
displayed on the vertical axis (on the left), in each plot we show the eccentricity values (on the right), computed by using the relations \eqref{Delaunayvar} and \eqref{RV}. Since $T$ is fixed, in a similar way we may compute and display the inclination values.}

From \eqref{Rmoonsunaverage} it follows that the RM2 model used to investigate the bifurcations admits
just a single resonant term, and the corresponding phase space is similar to a pendulum \red{(see Figure~\ref{fli_20_gps_geo},
top panels).}

Taking into account the variation of $\Omega_M$, namely if we consider the model RM1,
then from \eqref{R_moon_bigOmegaMoon} it is evident that the resonant part contains multiple terms, whose arguments are
of the form $2\, \sigma \pm \kappa \, \Omega_M$ with $\kappa=1,2$. In analogy to the case of the so-called minor resonances
for space debris (\cite{CGext}), we may say that the secular resonance $\dot{\omega}+\dot{\Omega}=0$ splits into a multiplet of resonances,
which leads to the phenomenon of splitting or overlapping of resonances \red{(compare also with \cite{ElyHowell, RARV15})}. These phenomena are clearly explained
in~\cite{CGext}, but in the framework of tesseral resonances.

From \eqref{R_moon_bigOmegaMoon} it turns out
that the two resonant terms with arguments $2\, \sigma $ and $2\, \sigma - \, \Omega_M$ are much larger in magnitude
than the others. As a consequence, it is reasonable to expect that the islands associated to $2\, \sigma $ and
$2\, \sigma - \, \Omega_M$ are much larger than the islands associated to the other resonant terms.
This aspect is evident from Figure~\ref{fli_20_gps_geo}, bottom left panel, which shows two larger islands having
the centers approximately at $S=0.64$ (corresponding to $2\, \sigma $) and at $S=0.58$ (associated to $2\, \sigma - \, \Omega_M$). \red{A smaller island, corresponding to $2\, \sigma - \, 2\Omega_M$ is located approximately at $S=0.55$. In fact, the bottom left panel of Figure~\ref{WEB_structure} provides an analytical argument for the existence of these islands; the vertical dashed black line $T=-0.2$ intersects three orange curves, corresponding to the following resonances: $\dot{\sigma}=0$ (the thick line), $2\dot{\sigma}-\dot{\Omega}_M=0$ (the continuous curve) and  $\dot{\sigma}-\dot{\Omega}_M=0$ (the dashed curve). As a conclusion, we may say that
the bottom left panel of Figure~\ref{fli_20_gps_geo},
obtained for $a=20\,000\, km$, shows
a splitting phenomenon, since the resonant islands are clearly separated.} For larger semimajor axis, that is
$a=26\,560\, km$, the resonant islands are not so clearly separated as shown in Figure~\ref{fli_20_gps_geo},
bottom middle panel, or they completely overlap for $a=42\,164\, km$ as in Figure~\ref{fli_20_gps_geo}, bottom right panel.

%The splitting and overlapping phenomena can be investigated as in~\cite{CGext, CGmajor} by evaluating the amplitude of
%the resonances and the location of the centers of the resonant islands. \blue{An estimate of the amplitude of resonances in the %eccentricity--inclination plane can be found in \cite{DRADVR15}.}

\red{Finally, in  Figure~\ref{fli_20_gps_geo_Om180} we give the FLI values obtained by using the non--autonomous
two degrees of freedom Hamiltonian \eqref{Hres}. The top panels show the results obtained for $\Omega=0^{\circ}$, while
the bottom panels present the results for $\Omega=180^{\circ}$. For $a=20\,000\, km$ the bottom left panel of Figure~\ref{fli_20_gps_geo} and the left plots of Figure~\ref{fli_20_gps_geo_Om180} are almost
identical, thus showing that the average over $\eta$ provides a reliable model: the resonance $\dot{\omega}+\dot{\Omega}=0$ is isolated
from the other resonances and $\eta$ (i.e., $\Omega$) is a fast angle.
For $a=26\,560\, km$ the bottom middle panel of Figure~\ref{fli_20_gps_geo} and the middle plots of Figure~\ref{fli_20_gps_geo_Om180} show some small differences as effect of both the variation of $\Omega$ and the interaction between various resonances (see the bottom right panel of Figure~\ref{WEB_structure}). Despite these differences, the patterns shown by these plots are very similar in structure,
suggesting that, with a certain degree of approximation, the RM1 model provides reliable results.
For a larger semimajor axis, that is for
$a=42\,164\, km$, $\Omega$ is slower than in the case of the MEO region; therefore, a marked difference
is visible when comparing  the bottom right panel of Figure~\ref{fli_20_gps_geo} and the right plots of Figure~\ref{fli_20_gps_geo_Om180}.}

The analysis of Figures~\ref{fli_20_gps_geo} and \ref{fli_20_gps_geo_Om180} show that the model RM2, used to study the bifurcations, yields reliable
results for the dynamics related to the bigger island,
provided that the main resonant island does not interact with possibly existing smaller islands.
As far as the
secular resonance $\dot\omega+\dot\Omega=0$ is concerned, we have a good performance of the model RM2, especially for
not too large values of the semimajor axis. Indeed, the equivalent of Figure~\ref{fig:case1plots}, right panel, but computed for
$a=20\,000$ km, overlaps almost completely with the bigger (upper) islands shown in Figures~\ref{fli_20_gps_geo} and \ref{fli_20_gps_geo_Om180},
 left panels. Another consequence of the comparison of the panels of Figures~\ref{fli_20_gps_geo} and \ref{fli_20_gps_geo_Om180} is that a non-zero rate
of variation of $\Omega_M$ provokes substantial changes in the plots given by RM2 and RM1. This remark is in
agreement with the results found in \cite{RARV15}.

In conclusion, Figures~\ref{fli_20_gps_geo} and \ref{fli_20_gps_geo_Om180} show that \red{the long--term} dynamics of space debris is very complicated
and it is influenced by various effects. However, without a study of the simplest models introduced in
Section~\ref{sec:model2}, the real dynamics would be difficult to explain.

\section{Bifurcations of other secular resonances of type \equ{secresi}}\label{sec:other}
In Section~\ref{sec:casestudy1} we have shown the details for the computation of the bifurcation thresholds for the resonance
$\dot\omega+\dot\Omega=0$. The other three resonances appearing in \equ{secresi} can be treated in a similar way and are
analyzed in this Section.

We premise that, whereas the case corresponding to $-\dot\omega+\dot\Omega=0$ is a straightforward generalization of the case $\dot\omega+\dot\Omega=0$ in the SFM2 class (see Remark~\ref{rem:SFM} for the definition of the SFM2), the $\pm 2 \dot\omega+\dot\Omega=0$ cases belong to the so-called {\it extended fundamental model for
\red{second--order} resonances}, denoted as EFM2 in Remark~\ref{rem:SFM}
(see \cite{lemaitre84,B2}).
The cases corresponding to EFM2 are characterized by the appearance of an additional critical point in the reduced problem corresponding to a new family of periodic orbits of the system.

%We can see an example of the $2 \dot\omega+\dot\Omega=0$ case in Figure~\ref{fig:case2oOplots}.

%\begin{figure}[h]
%\centering
%\includegraphics[width=10truecm]{2omegaOmegaplot.eps}
%\caption{}
%\label{fig:case2oOplots}
%\end{figure}

For the resonance $-\dot\omega+\dot\Omega=0$, one needs to make the symplectic transformation of coordinates
\beqano
\sigma&=&\omega-\Omega\ ,\qquad\ \  S=G\ ,\nonumber\\
\eta&=&\Omega\ ,\qquad\qquad\ \ T=H+G\ .
\eeqano
For the resonance $-2\dot\omega+\dot\Omega=0$, one can make the change of coordinates:
\beqano
\sigma&=&2\omega-\Omega\ ,\qquad S={G\over 2}\ ,\nonumber\\
\eta&=&\Omega\ ,\qquad\qquad\ \ T=H+{G\over 2}\ .
\eeqano
For the resonance $2\dot\omega+\dot\Omega=0$, one makes the transformation of variables:
\beqano
\sigma&=&2\omega+\Omega\ ,\qquad S={G\over 2}\ ,\nonumber\\
\eta&=&\Omega\ ,\qquad\qquad\ \ T=H-{G\over 2}\ .
\eeqano

\vskip.2in

The resonance $-\dot\omega+\dot\Omega=0$ can be treated in a way similar to that of Section~\ref{sec:casestudy1};
the same holds for the resonances $2\dot\omega+\dot\Omega=0$ and $-2\dot\omega+\dot\Omega=0$, when the inclination is equal to $111.0^{\circ}$ and
$69.0^{\circ}$, respectively. One has two exceptions:  $2\dot\omega+\dot\Omega=0$ at inclination $56.1^{\circ}$ and  $-2\dot\omega+\dot\Omega=0$ at inclination $i=123.9^{\circ}$. In fact, since $H=S\cos i$,
we find in the first case that $T=S(\cos i-{1\over 2})$, which is close to zero when $i=56.1^{\circ}$. As a consequence, since the expansion of the Hamiltonian
(compare with \equ{Hexp}) contains $T$ at the denominator, the method presented in Section~\ref{sec:birth} fails.
The same happens for the latter case.

It could be that different techniques (e.g., other changes of coordinates regularizing the problem) are successful, but
the analysis of this problem goes beyond the aims of the present work.

For the moment we limit ourselves to consider the other \sl non-singular \rm cases for which the results are provided
in Table~\ref{table:others}.

\vskip.2in

\red{
\begin{table}[h]
\begin{tabular}{|c|c|c|c|c|c|}
  \hline
  &inclination (in degrees)&&$T_A$ & $T_B$ \\
  \hline
$-\dot\omega+\dot\Omega=0$&73.2&Moon & 0.664659 & 0.664754  \\
$-\dot\omega+\dot\Omega=0$&73.2&Sun & 0.665149 & 0.665204 \\
$-\dot\omega+\dot\Omega=0$&73.2&Moon+Sun  & 0.664278 & 0.664423\\
  \hline
$-\dot\omega+\dot\Omega=0$&133.6&Moon & 0.159407 & 0.159667  \\
$-\dot\omega+\dot\Omega=0$&133.6&Sun & 0.159724 & 0.159875 \\
$-\dot\omega+\dot\Omega=0$&133.6&Moon+Sun & 0.159142 & 0.159538 \\
  \hline
$-2\dot\omega+\dot\Omega=0$&69.0&Moon & 0.442037 & 0.442172  \\
$-2\dot\omega+\dot\Omega=0$&69.0&Sun & 0.442480 & 0.442546 \\
$-2\dot\omega+\dot\Omega=0$&69.0&Moon+Sun & 0.441693 & 0.441897 \\
  \hline
$2\dot\omega+\dot\Omega=0$&111.0&Moon & -0.442172 & -0.442037 \\
$2\dot\omega+\dot\Omega=0$&111.0&Sun & -0.442546 & -0.44248 \\
$2\dot\omega+\dot\Omega=0$&111.0&Moon+Sun & -0.441897 & -0.441693\\
  \hline
 \end{tabular}
 \vskip.1in
 \caption{Bifurcation values $T_A$, $T_B$ for the GPS orbit corresponding to the cases
 in which only the Moon is considered, only the Sun is considered, both Sun and Moon are considered.
 The resonances were listed in \equ{secresi} and the corresponding inclinations appeared in \equ{inc}.}\label{table:others}
\end{table}
}

\vskip.2in

\subsection{On the accuracy of the expansion of the Hamiltonian}\label{sec:accuracy}
In Section~\ref{sec:birth} we have considered the example of the resonance $\dot\omega+\dot\Omega=0$
and we have expanded the Hamiltonian up to the second order in $S-S_0$ as in \equ{Hexp}. That case was
in fact sufficiently regular that the second order expansion was indeed enough to get an accurate description
of the full Hamiltonian. However, this might not be the same for other resonances
and it must be checked for each specific resonance.
As a case study,
we consider now the resonance $-2\dot\omega+\dot\Omega=0$ with inclination equal to $69^{\circ}$.

\begin{figure}[h]
\centering
\vglue0.1cm
\hglue0.1cm
\includegraphics[width=4.8truecm,height=3.8truecm]{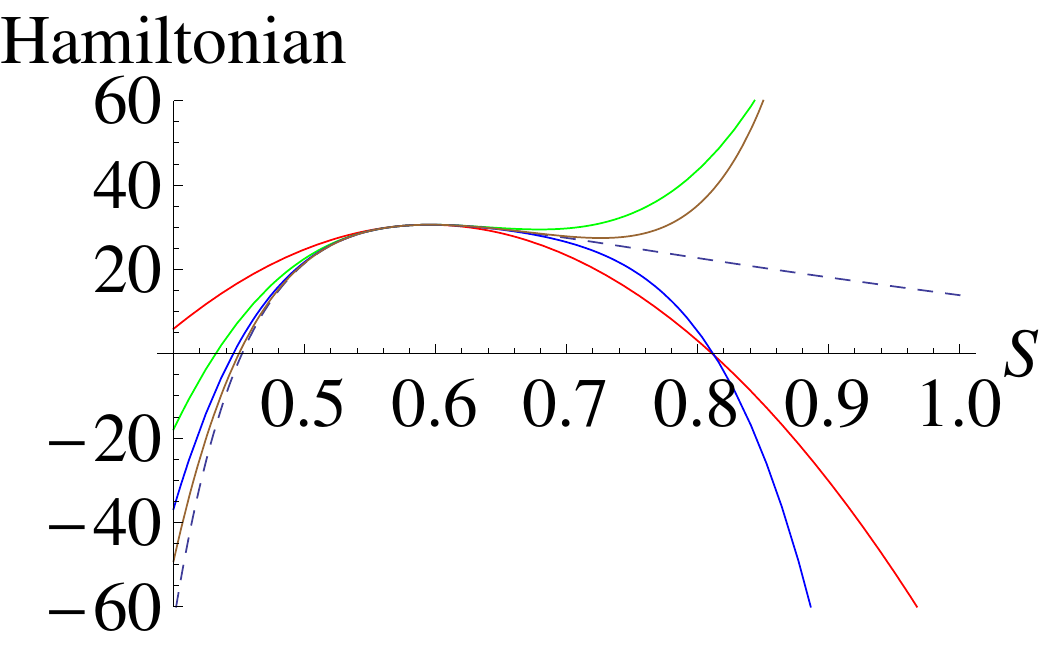}
\includegraphics[width=5truecm,height=4truecm]{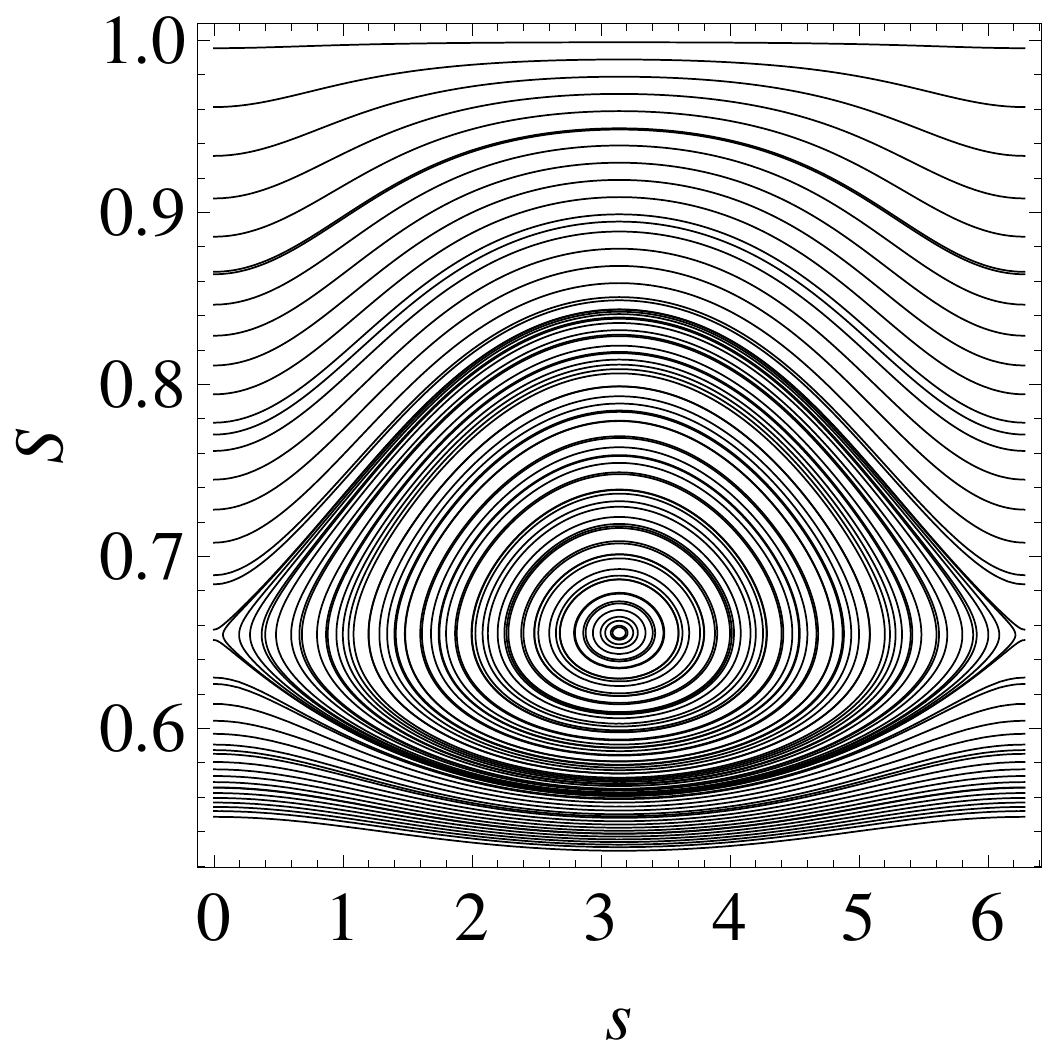}
\includegraphics[width=5truecm,height=4truecm]{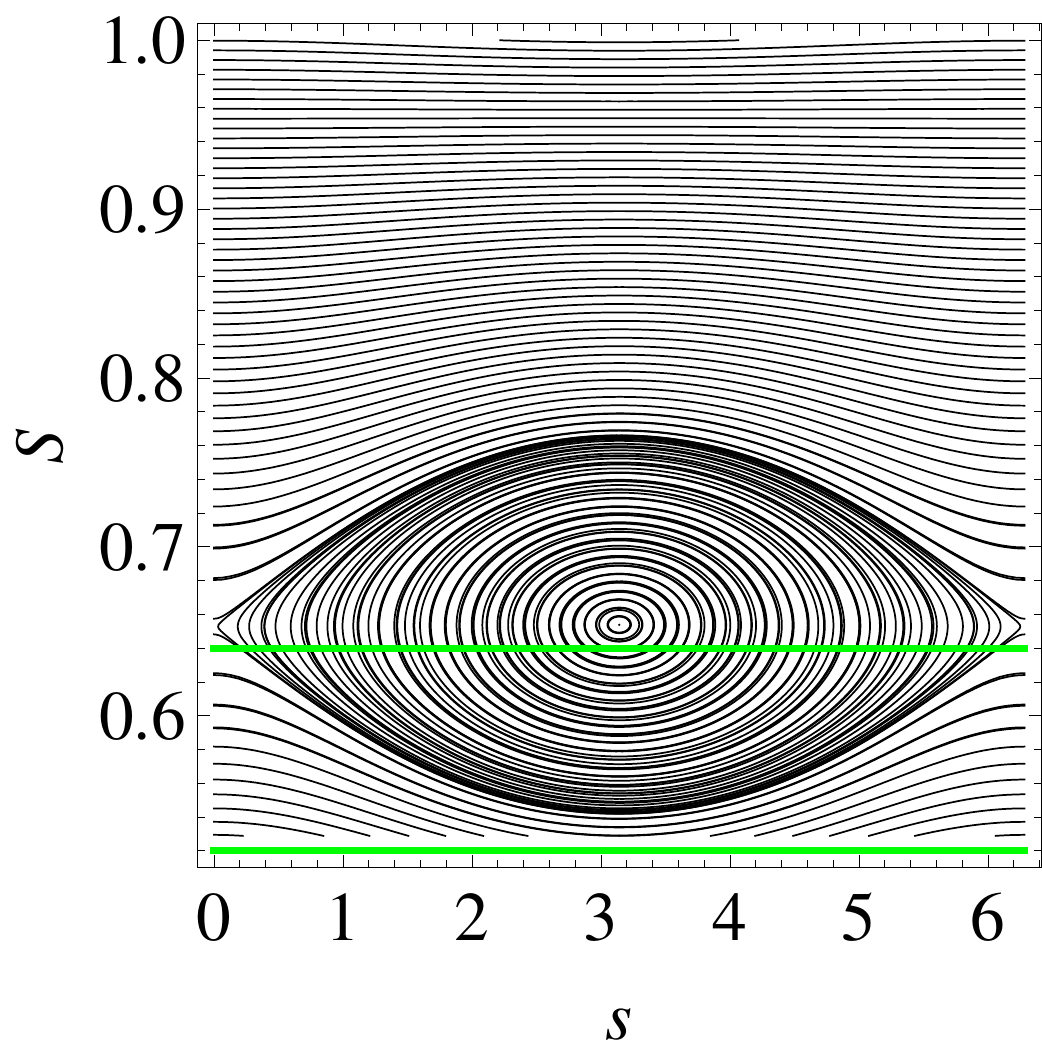}\\
\vglue0.5cm
\includegraphics[width=5truecm,height=4truecm]{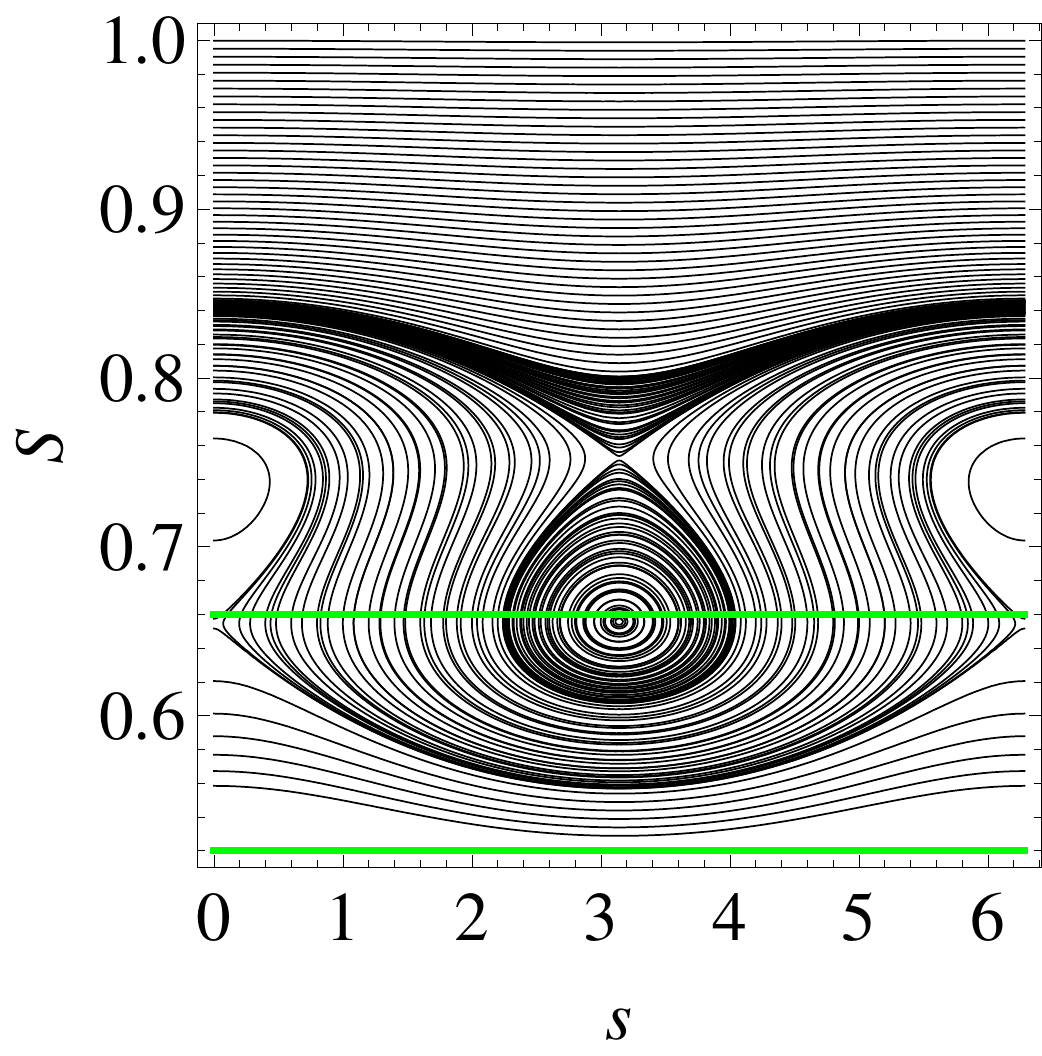}
\includegraphics[width=5truecm,height=4truecm]{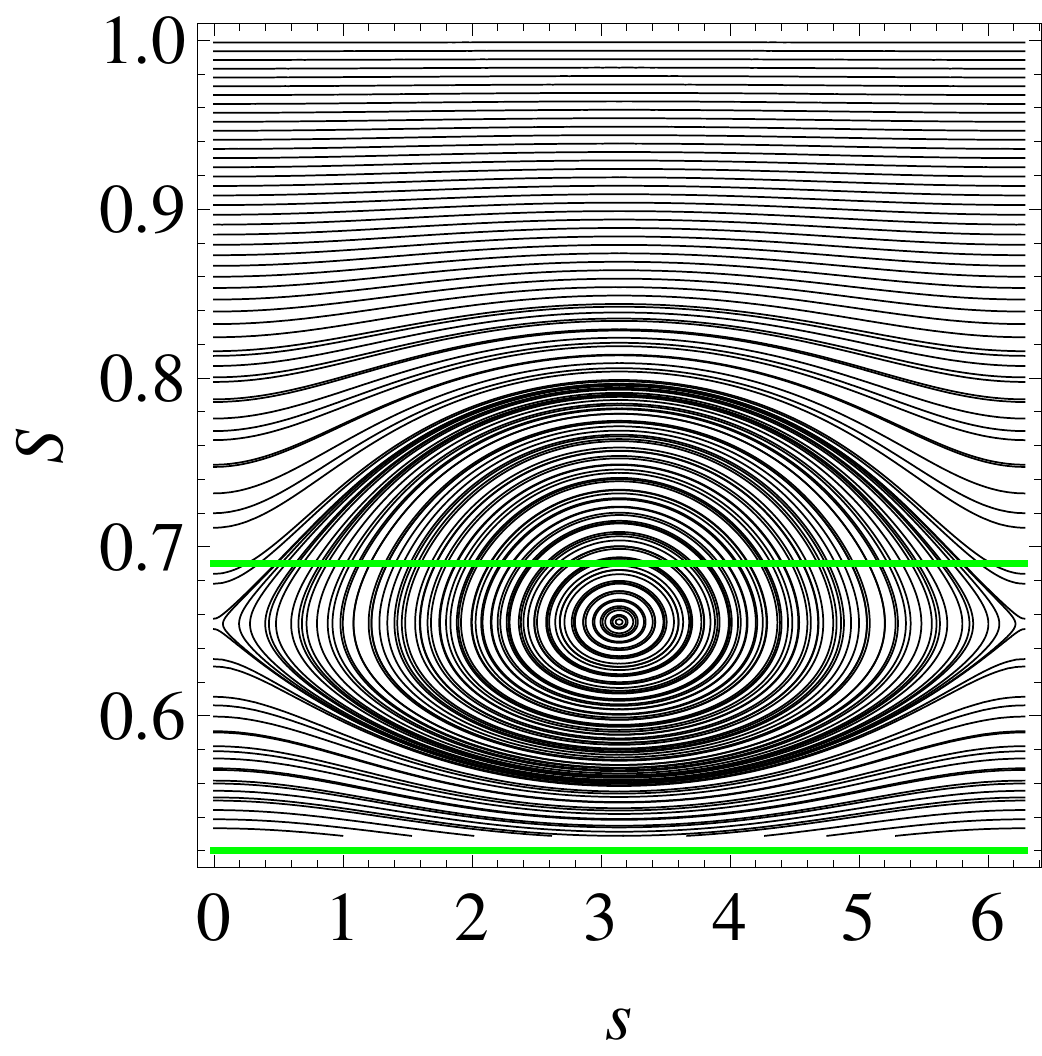}
\includegraphics[width=5truecm,height=4truecm]{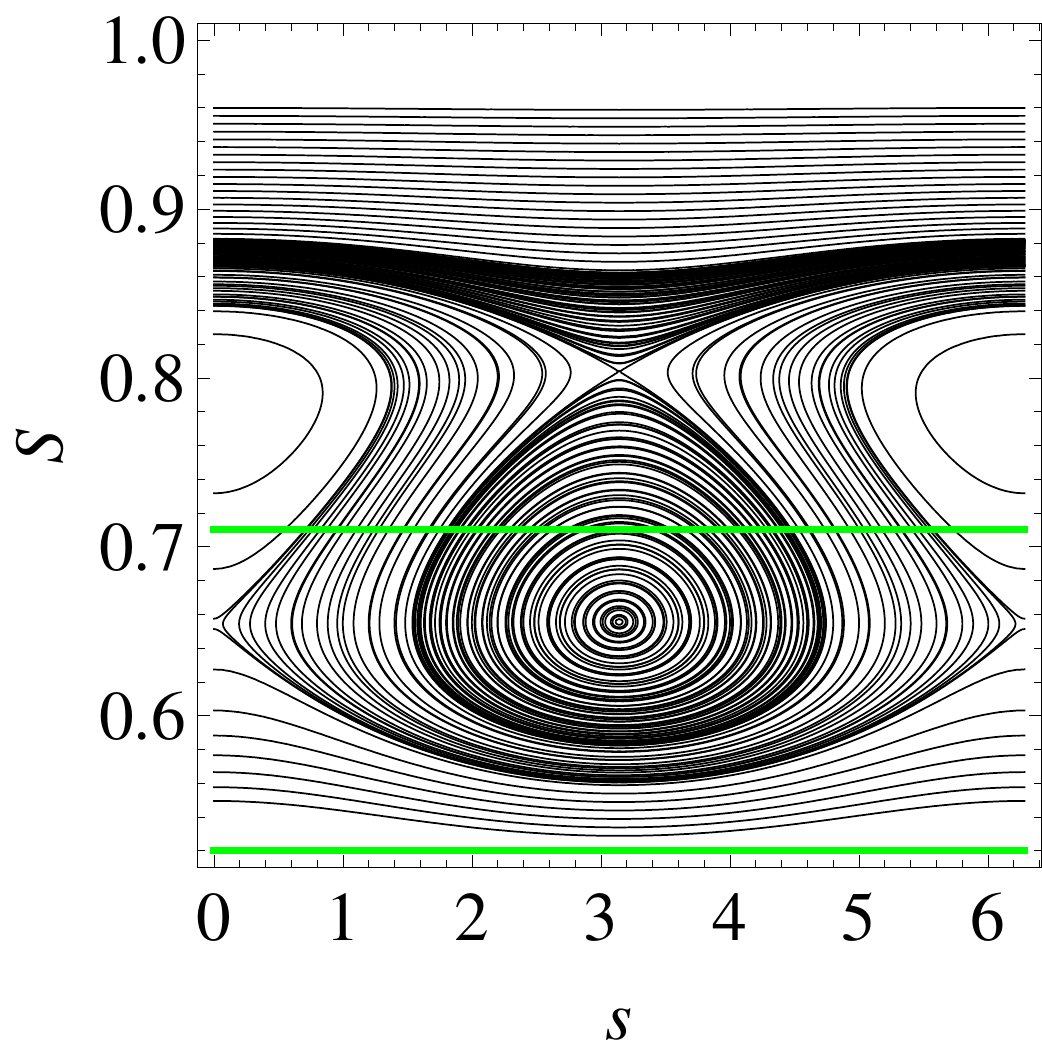}
\vglue0.5cm
\caption{Resonance $-2\dot\omega+\dot\Omega=0$, case $i=69^{\circ}$. Upper left panel: graph of the true lunisolar Hamiltonian
(dotted curve) and those of the truncated expansions ($2^{nd}$ red, $3^{rd}$ green, $4^{th}$ blue, $5^{th}$ brown)
$s=\pi$, $L_0=1$, $T=0.5$. Upper middle panel: phase space portrait in the plane $(s,S)$ for the true Hamiltonian.
Upper right panel: phase portrait for the $2^{nd}$ order expansion.
Lower panels from left to right: phase portraits for the expansions to order 3, 4, 5.
}
\label{fig:accuracy}
\end{figure}

We proceed to test the accuracy of the expansions by computing the series up to the orders 2, 3, 4, 5
around $S_0$. In Figure~\ref{fig:accuracy}, upper left panel, we draw the graph of the true Hamiltonian
(dotted curve) and those of the truncated expansions for specific values of the coordinates,
precisely $s=\pi$ \red{(where $s=\sigma+0.1412$)}, $L_0=1$ and taking $T=0.5$, a value larger
than $T_A$ in Table~\ref{table:others}. As one can see, increasing the order of the expansion
one gets that the Hamiltonian graph is reproduced within a larger interval in the variable $S$.
This interval is marked by green lines in the subsequent panels of Figure~\ref{fig:accuracy},
thus showing how the true dynamics is reconstructed with increasing accuracy as the order of the expansion gets larger.
The upper middle plot shows the phase space portrait in the plane $(s,S)$ for the full Hamiltonian,
while the subsequent plots are obtained using expansions of the Hamiltonian at orders 2, 3, 4, 5.
Notice that the upper green line gets higher as the order of the expansion increases, thus showing that
higher orders reproduce the true Hamiltonian in a larger region.

\vskip.2in

\vskip.2in

\section{Bifurcations of type $(i)$: critical inclination secular resonances}\label{sec:critical}
In this Section we investigate the secular resonances of type $(i)$, which are obtained by solving
the equation $\dot\omega=0$. This resonance occurs at the so-called \sl critical inclinations, \rm
whose values are $i=63.4^{\circ}$ and  $i=116.4^{\circ}$  (see \cite{HughesI}). To be consistent with the previous notation,
we make
the trivial (identity) transformation of variables:
\beqano
\sigma&=&\omega\ ,\qquad \ S=G\ ,\nonumber\\
\eta&=&\Omega\ ,\qquad\  T=H\ .
\eeqano
We discuss just the case $i=63.4^{\circ}$, since the other one can be treated similarly.

It is important to stress that special care must be taken when dealing with such resonance.
In fact, the equivalent of the Hamiltonian expansion in \equ{Hexp} up to second order provides only
a rough result, when one compares the phase space portraits of the Hamiltonian expanded to second
order and the full Hamiltonian.

\begin{figure}[h]
\centering
\vglue0.1cm
\hglue0.1cm
\includegraphics[width=5truecm,height=4truecm]{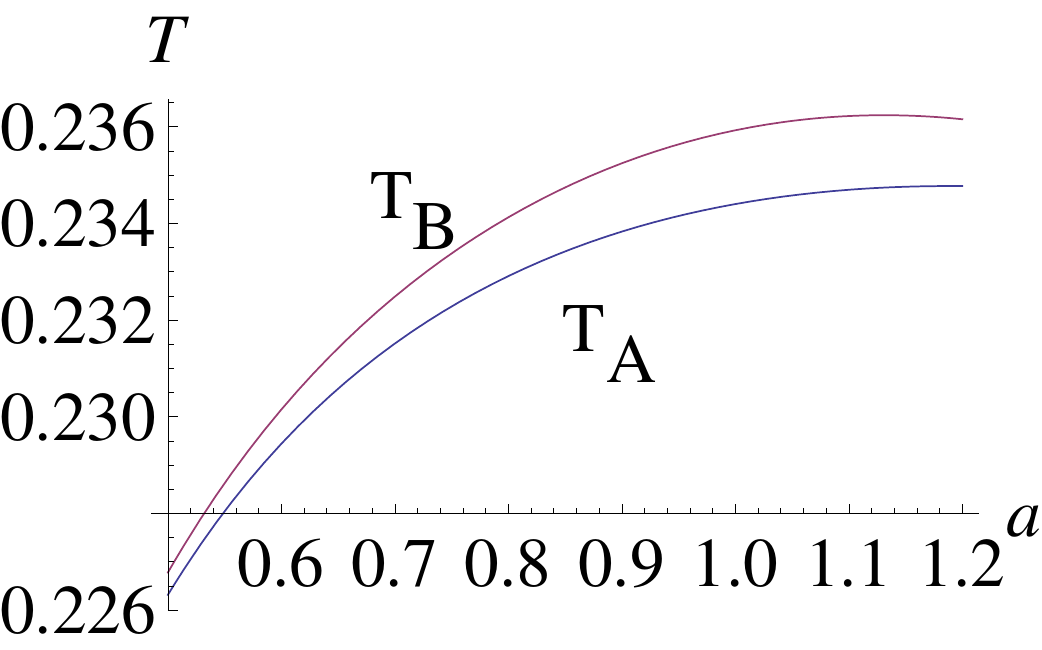}
\includegraphics[width=5truecm,height=4truecm]{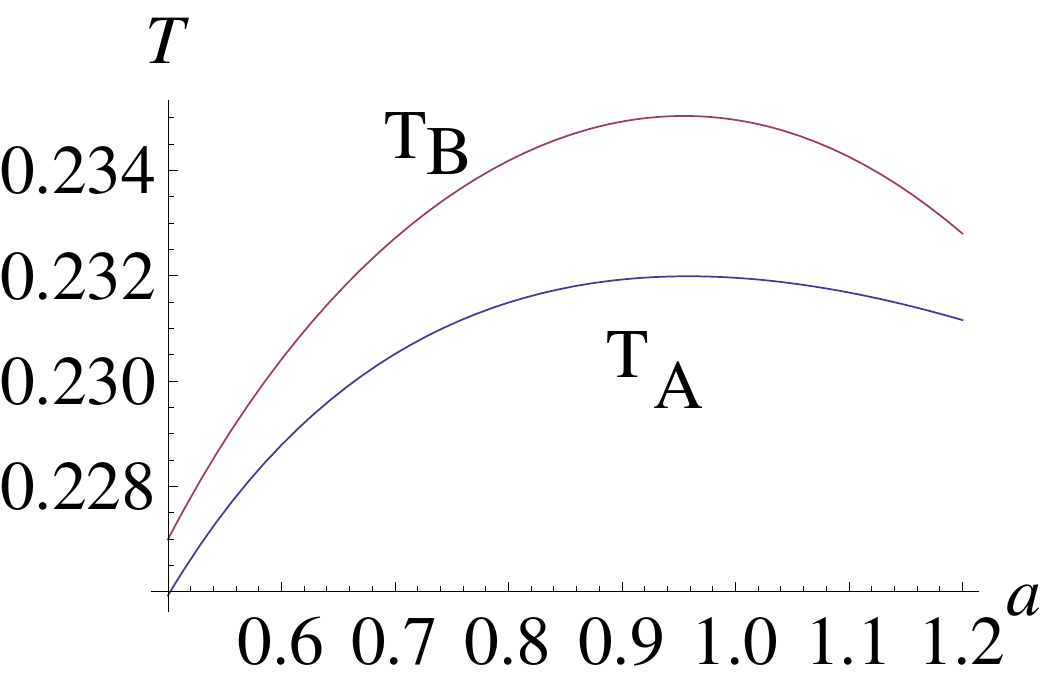}
\includegraphics[width=5truecm,height=4truecm]{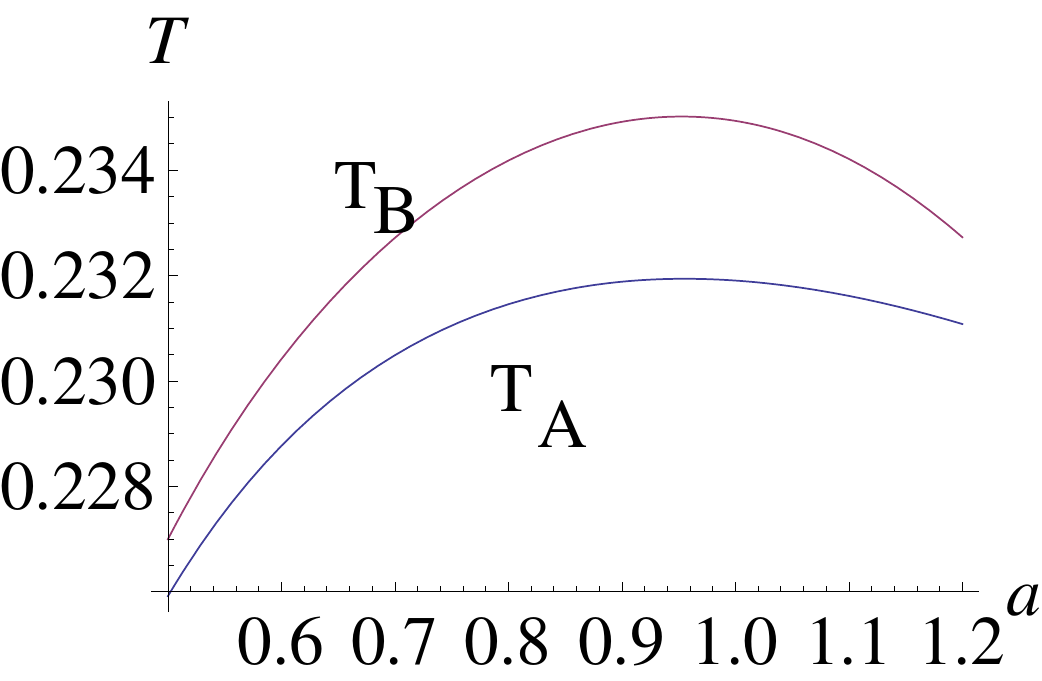}
\vglue0.5cm
\caption{Critical inclination secular resonance.
Plots of the bifurcation curves $T_A$, $T_B$ in \equ{T11} as a
function of the semimajor axis for the case $i=63.4^{\circ}$. Left: the Hamiltonian includes just the effect of the Sun;
middle: the Hamiltonian includes only the Moon with an expansion to order 2; right: the Hamiltonian includes only the
Moon with an expansion to order 4.}
\label{fig:bifcrit}
\end{figure}

To improve the results, we need to perform an expansion to higher order and therefore we proceed to expand
up to order 5 in $S_0$. The equation for $\dot S=0$
provides the equilibrium values $s=0$ and $s=\pi/2$. On the other hand, the equation $\dot s=0$ for
$s=0$ or $s=\pi/2$ provides \red{a third--order equation} in $S$. One finds that in both cases, namely $s=0$ and $s=\pi/2$,
two solutions are complex conjugated, while the third solution is real. After an expansion in $T$ up to the
order 6 of the real solutions, one can proceed as in Section~\ref{sec:birth} to compute the bifurcation values
and the corresponding phase space portraits. In particular, for $i=63.4^{\circ}$, Figure~\ref{fig:bifcrit} provides the
bifurcation curves for $T_A$ and $T_B$ (see \equ{T11}) for the second (middle panel) and fourth (right panel)
order expansions. The left panel of Figure~\ref{fig:bifcrit} shows the contribution due just to the Sun.

\vskip.2in

\vskip.2in

\begin{figure}[h]
\centering
\vglue0.1cm
\hglue0.1cm
\includegraphics[width=4.8truecm,height=3.8truecm]{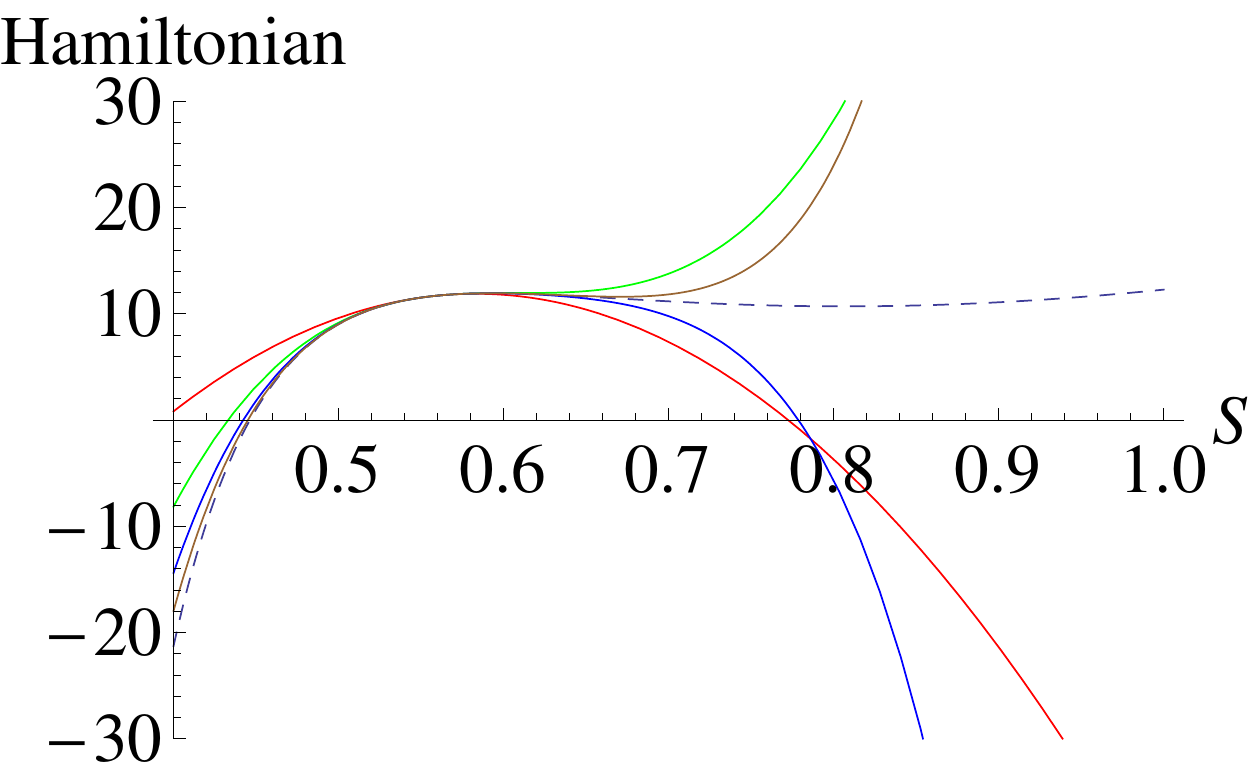}
\includegraphics[width=5truecm,height=4truecm]{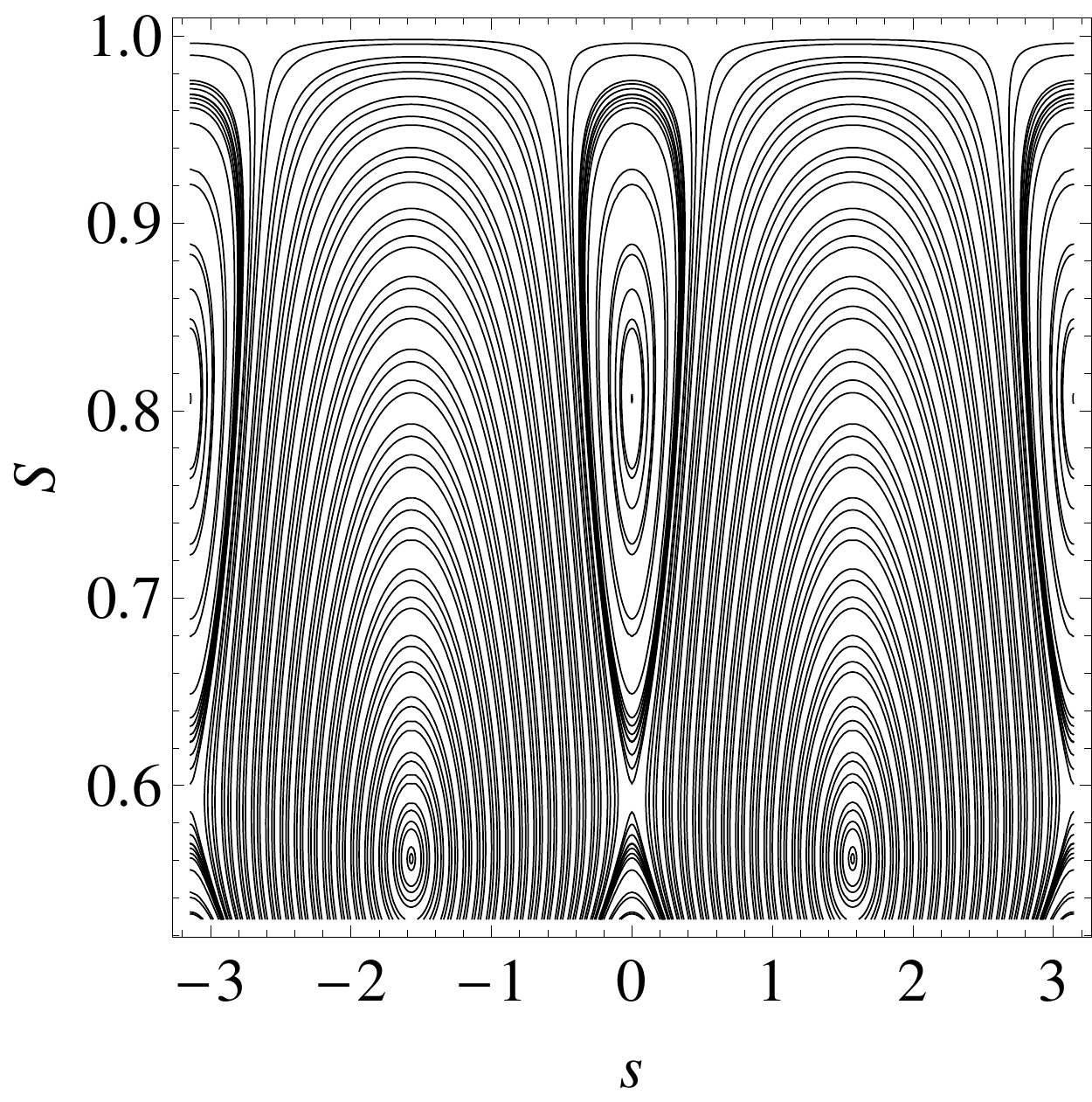}
\includegraphics[width=5truecm,height=4truecm]{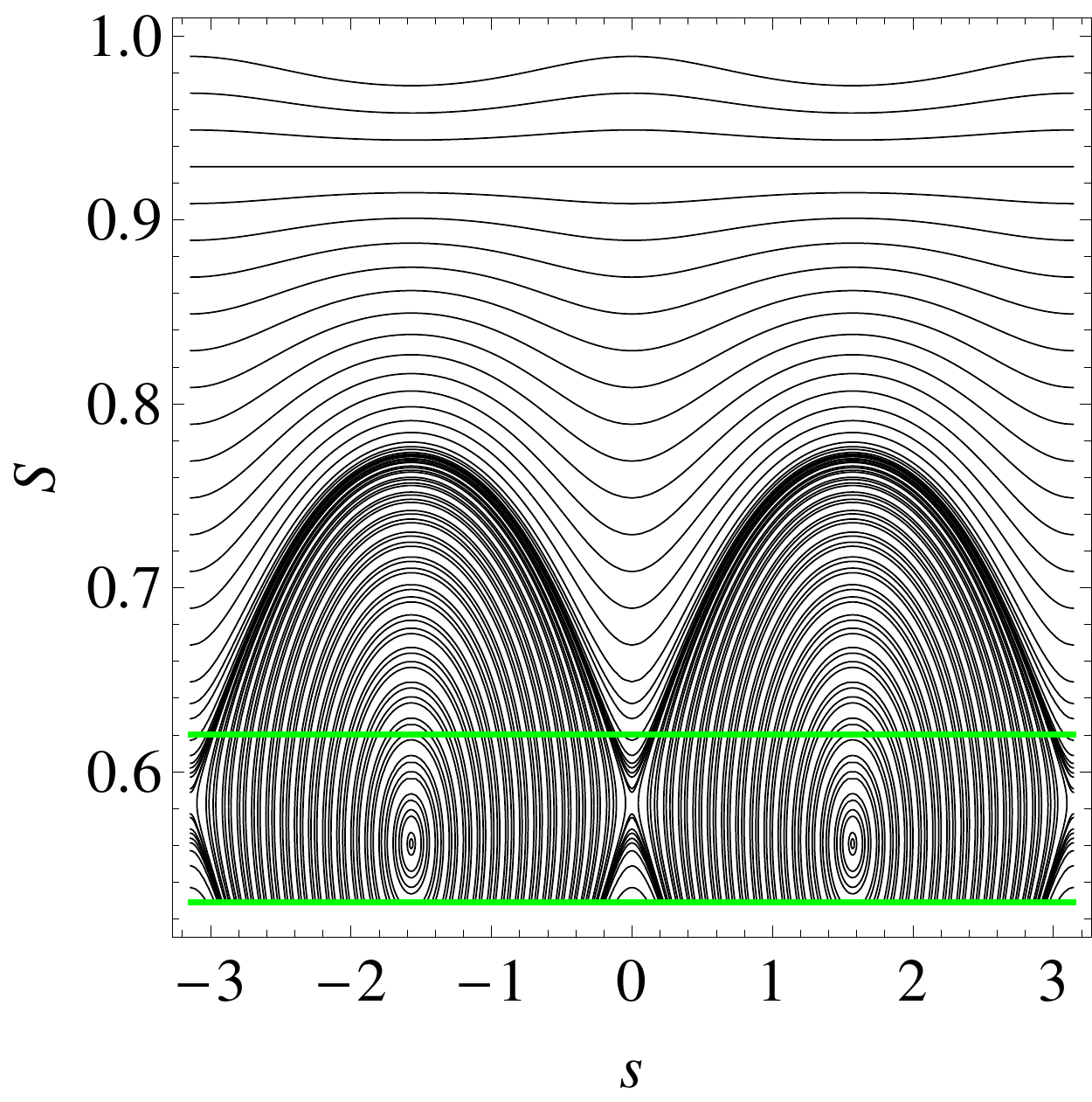}\\
\vglue0.5cm
\includegraphics[width=5truecm,height=4truecm]{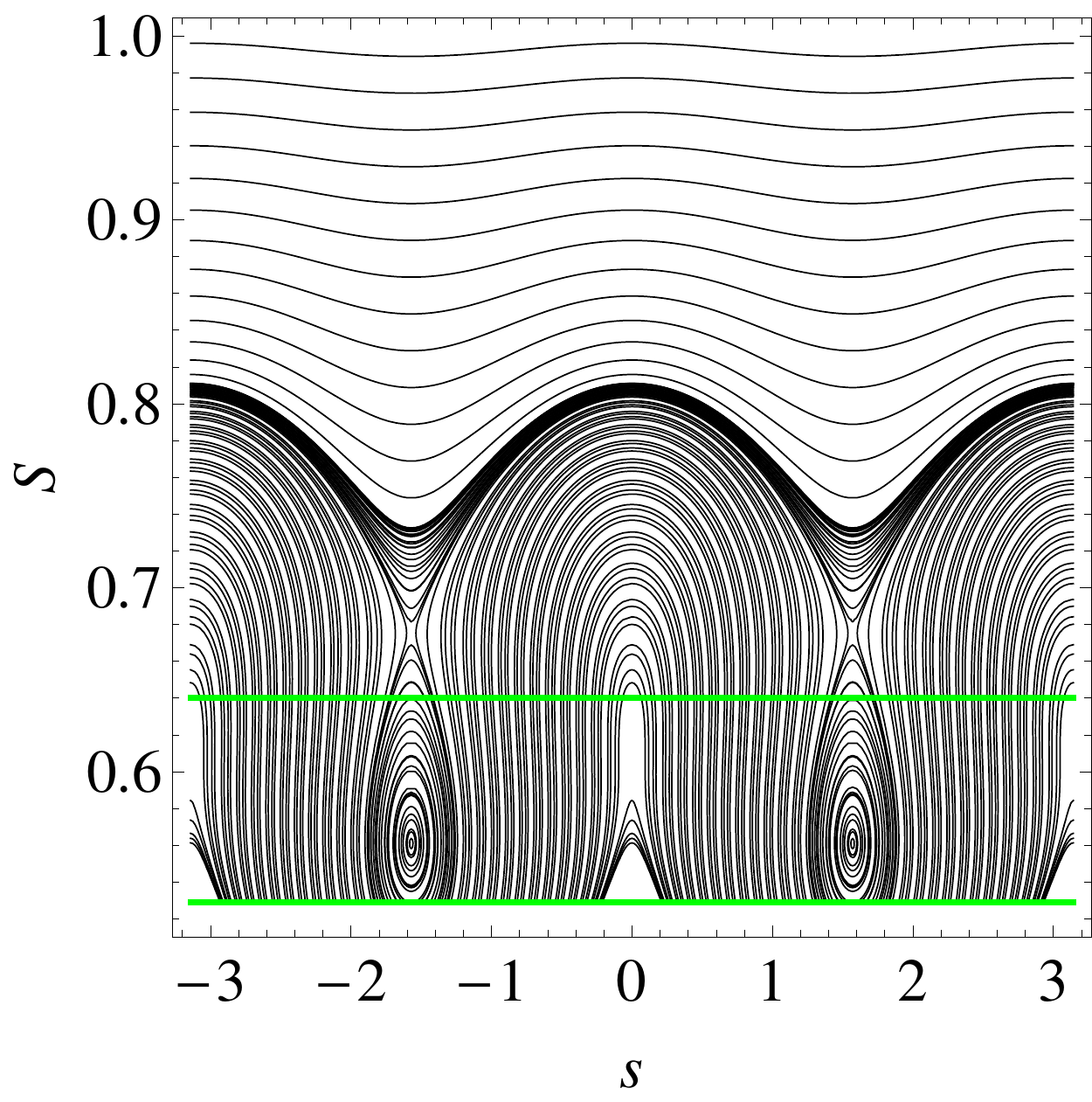}
\includegraphics[width=5truecm,height=4truecm]{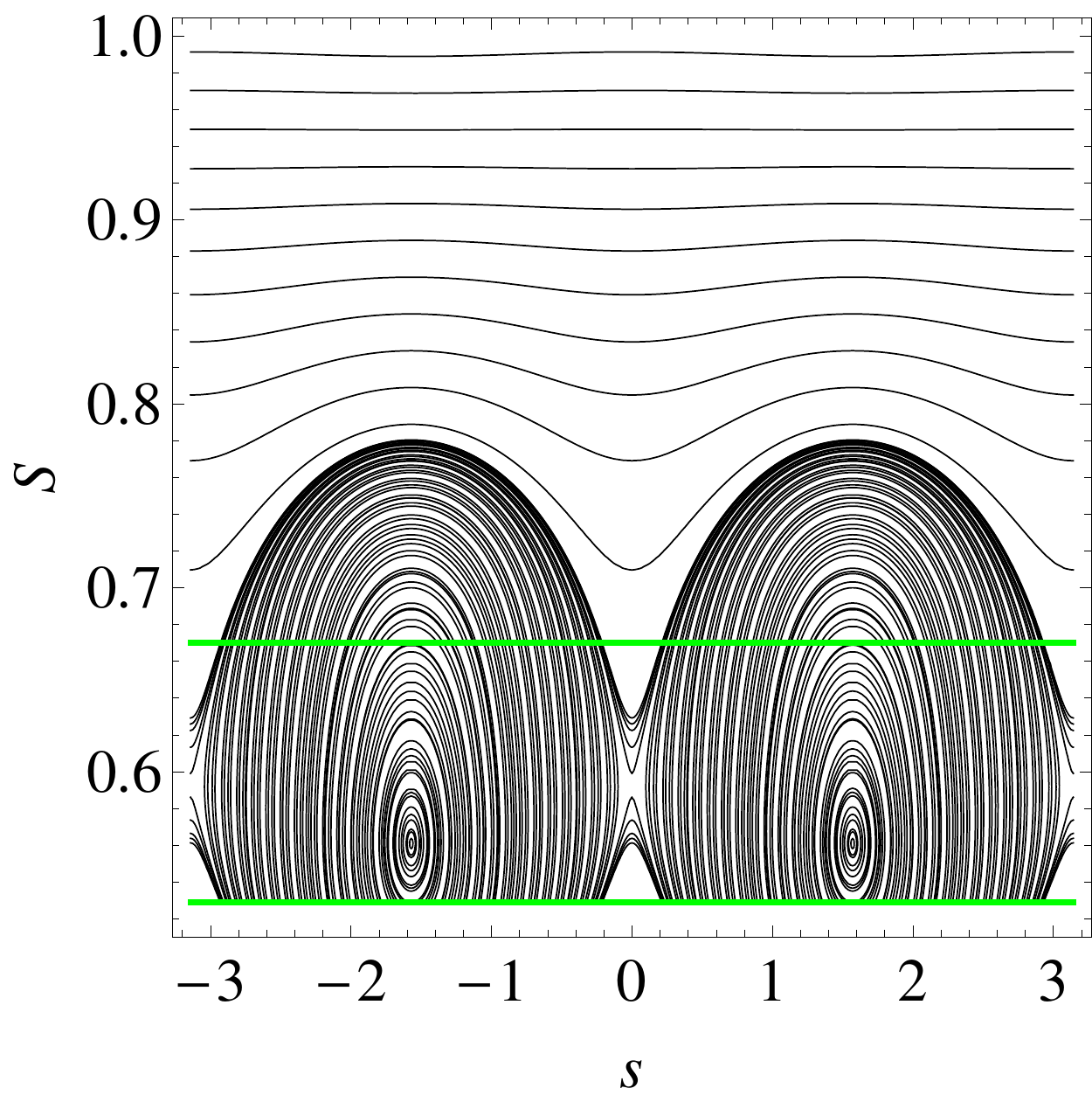}
\includegraphics[width=5truecm,height=4truecm]{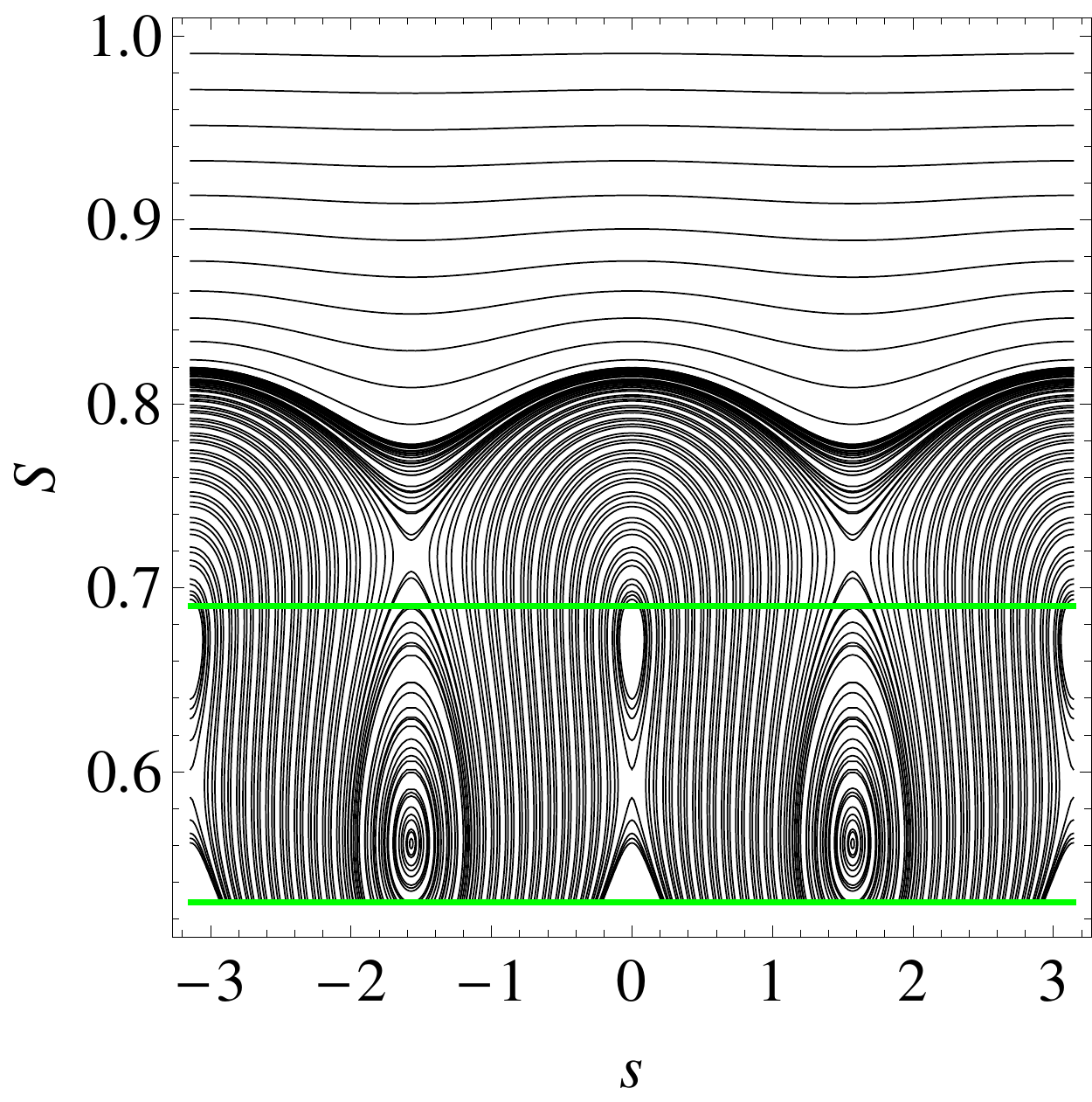}
\vglue0.5cm
\caption{Critical inclination resonance, the case $i=63.4^{\circ}$. Phase space portraits in the plane $(s,S)$ with Sun and Moon included in
the model.
Upper left panel: graph of the true Hamiltonian
(dotted curve) and those of the truncated expansions ($2^{nd}$ red, $3^{rd}$ green, $4^{th}$ blue, $5^{th}$ brown)
$s=\pi$, $L_0=1$, $T=0.25$. Upper middle panel: phase space portrait in the plane $(s,S)$ for the true Hamiltonian.
Upper right panel: phase portrait for the $2^{nd}$ order expansion.
Lower panels from left to right: phase portraits for the expansions to order 3, 4, 5.
}
\label{fig:critplots}
\end{figure}

The phase space portraits for the critical inclination case
are presented in Figure~\ref{fig:critplots}. As in Section~\ref{sec:accuracy} we test
the accuracy of the computations by comparing the truncations of the Hamiltonian with
different degrees of expansion, from 2 to 5.

As already remarked in Figure~\ref{fig:accuracy}, we notice that as the order of the expansion gets larger,
the Hamiltonian graph is reproduced with a better accuracy. The interval where the true and expanded functions
agree is reported within green lines in the phase portraits of Figure~\ref{fig:critplots}.

From an inspection of the true phase portrait given by the upper middle panel
of Figure~\ref{fig:critplots}, we observe that the critical inclination case presents
further equilibrium points (see the elliptic equilibria located at about \red{ $S=0.803$
for $s=0,\pi$)}. This phenomenon is close to what observed in the EFM2 model
discussed in Remark~\ref{rem:SFM}.

\vskip.2in

\vskip.2in

\section{Bifurcations of type $(ii)$: polar secular resonances}\label{sec:polar}

In the case of polar resonances corresponding to $\dot\Omega=0$, the transformation of variables reduces to the identity:
\beqano
\sigma&=&\Omega\ ,\qquad S=H\ ,\nonumber\\
\eta&=&\omega\ ,\qquad\  T=G\ .
\eeqano
The treatment of polar resonances is made complicated by several factors.
First of all, it is difficult to combine the effects of both the Moon and Sun, due to the fact that the corresponding expansions
contain terms which are qualitatively different. On the other hand, even if we consider the effects of only one body,
at the exact resonance one has $S=0$, so that the second order expansion of the Hamiltonian should be performed
around $S_0=0$. Hence, one has (physical) bounds on $T$, but not on $S$ which is independent of $T$.
Therefore the method can only be applied by deciding to impose a limitation to the range of variability
of the $S$ coordinate, which in turn becomes a constraint on the variation of the inclination.

\begin{figure}[h]
\centering
\vglue0.1cm
\hglue0.1cm
\includegraphics[width=5truecm,height=3truecm]{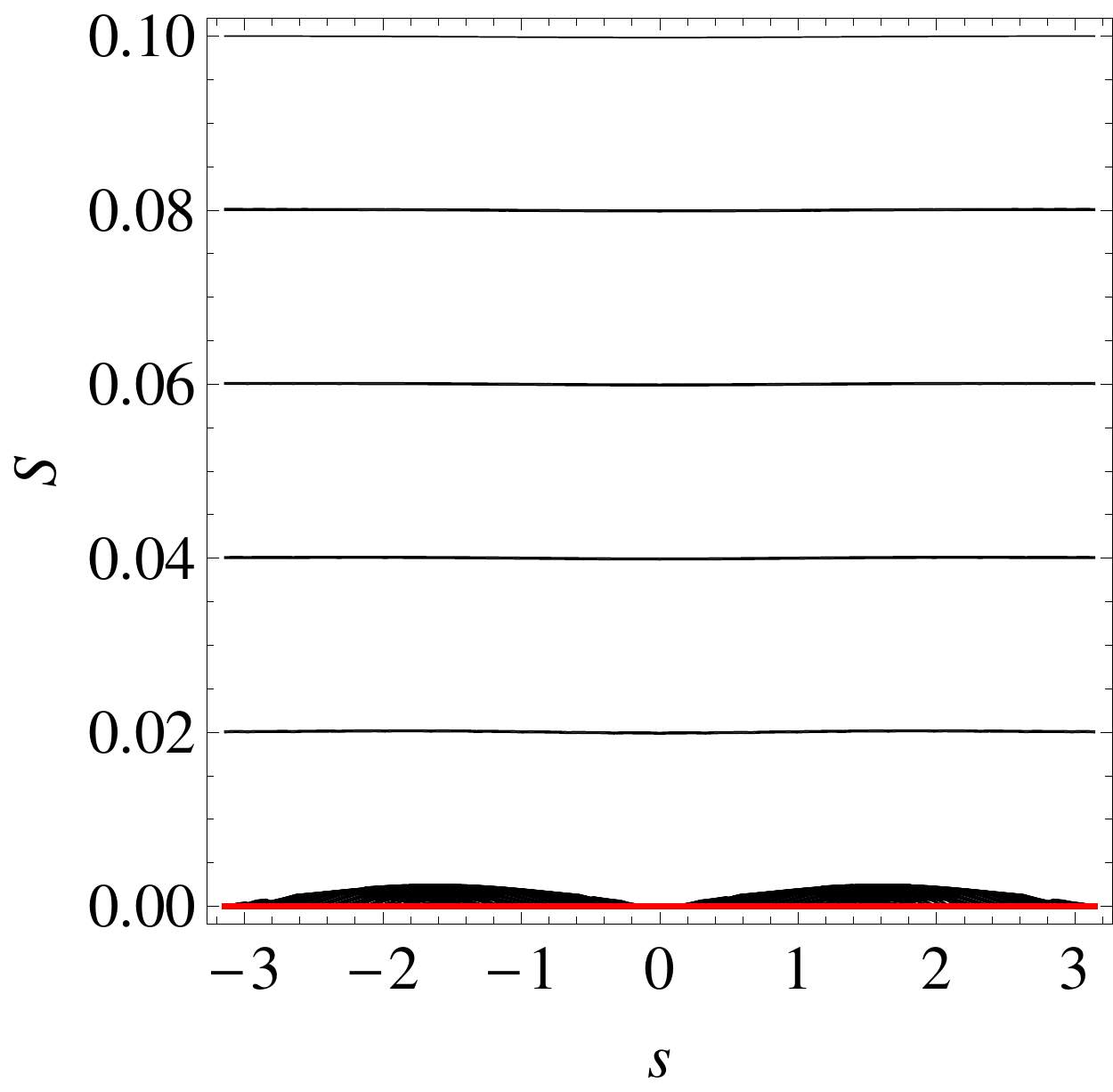}
\includegraphics[width=5truecm,height=3truecm]{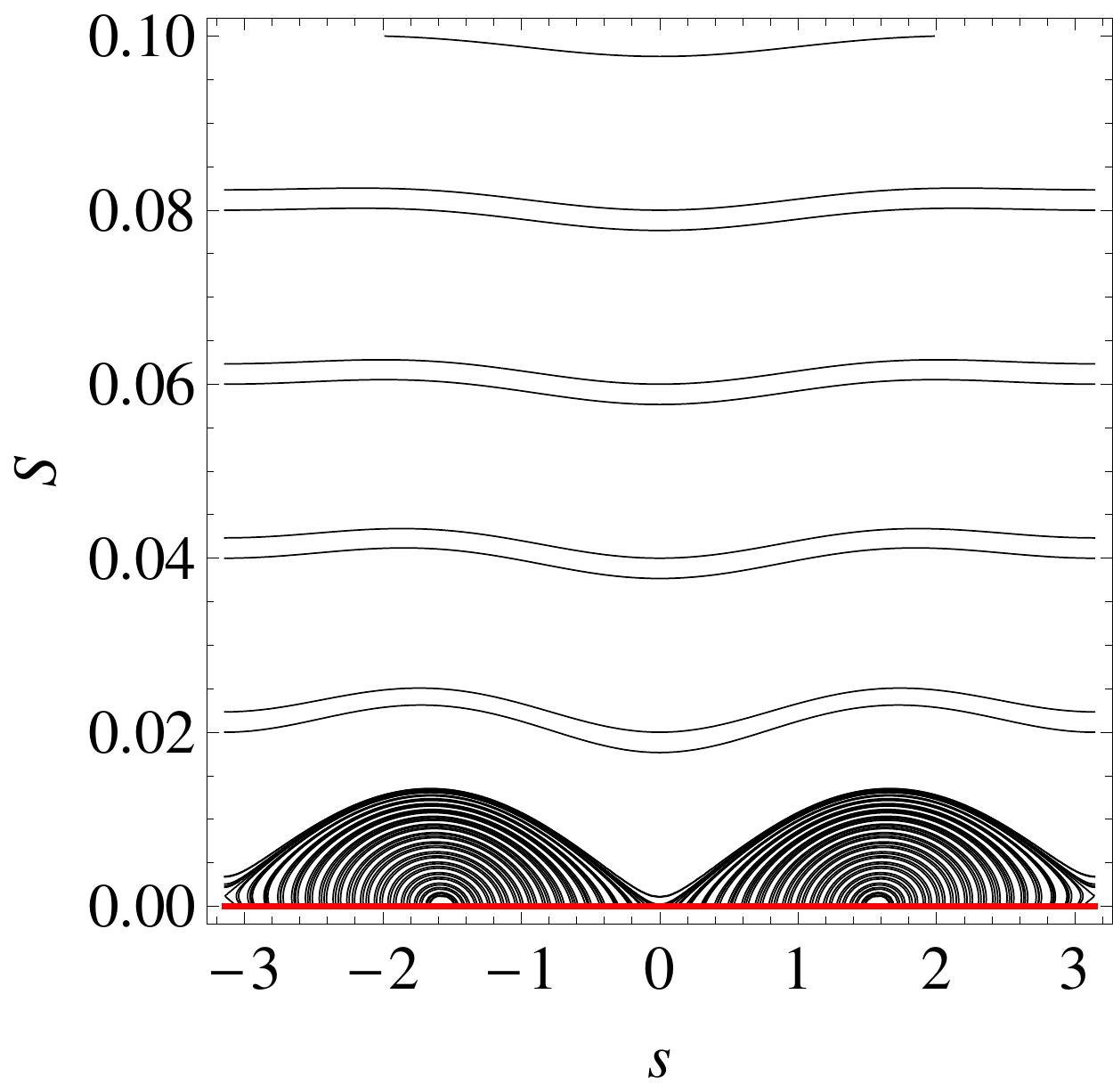}
\includegraphics[width=5truecm,height=3truecm]{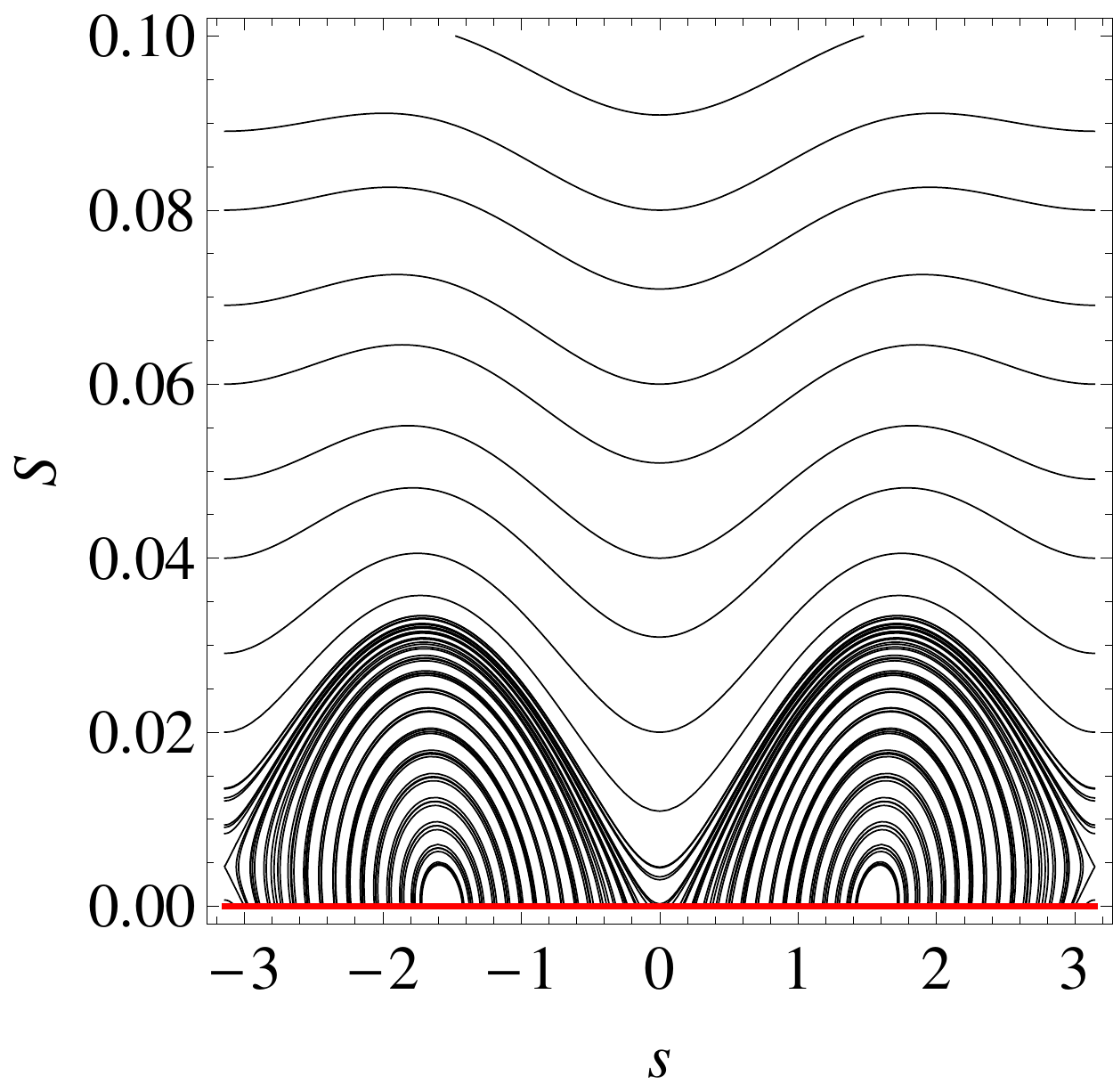}
\vglue0.5cm
\caption{Polar secular resonance:
phase space portraits corresponding to the GPS value $a=a_{GPS}$ in the plane $(s,S)$ for the Hamiltonian including the effect of the
Moon: $T=0.2$ (left), $T=0.4$ (middle), $T=0.6$ (right).}
\label{fig:polar}
\end{figure}

Let us proceed according to this strategy and assume to
consider just the effect of the Moon. Then, the corresponding Hamiltonian takes the form
$$
\H_{Moon}^{polar}(S,T,s)=h_0(T)+S^2 h_1(T)+S\ \sqrt{1-{S^2\over T^2}}\ f_0(T)\ \cos s+
(g_0(T)+S^2 g_1(T))\ \cos 2s
$$
for some functions $h_0$, $h_1$, $f_0$, $g_0$, $g_1$ and for \red{$s=\sigma-0.212$}. After expanding to second order
$\H_{Moon}^{polar}$ around $S_0=0$, one obtains
$$
\H_{Moon,exp}^{polar}(S,T,s)=h_0(T)+S^2 h_1(T)+S\ f_0(T)\ \cos s+
(g_0(T)+S^2 g_1(T))\ \cos 2s\ .
$$
Hence, the equations of motion are
\beqano
\dot S &=&S\ f_0(T)\ \sin s+2 \Big(g_0(T)+S^2 g_1(T)\Big)\ \sin 2s\nonumber\\
\dot s &=&f_0(T)\ \cos s+ 2S \Big(h_1 (T) + g_1(T)\ \cos 2s\Big)\ .
\eeqano
Therefore, in addition to the two standard solutions $(S_A,0)$ and $(S_B,\pi)$ we have also a \sl fixed \rm
(in the sense of being independent of $T$) pair of solutions, given by
$$
S=0, \quad s = \pm \pi/2.$$
These elliptic equilibria do not play any role for generating new periodic orbits. However, they give a well definite shape
to the structure of the phase space and in particular they determine the nature of the bifurcating families, since these are \sl both stable\rm.
We then proceed to compute their locations in the plane $(S,s)$, using
the technique implemented for the other resonances. To this end, we fix a value $S_{min}$ close to
zero, say $S_{min}=0.01$.
The computation of the bifurcation thresholds corresponding to $s=0$ and $s=\pi$ provides the
values $T_A=0.313$ and $T_B=0.450$. Plotting the graphs between the lower threshold $S=0$ and
an arbitrary upper limit, say $S=0.1$, we obtain the birth of equilibria for values like those shown
in Figure~\ref{fig:polar}.

\vskip.2in

\vskip.2in

\section{Some remarks on the non--averaged problem}\label{sec:nonaveraged}
In Sections~\ref{sec:casestudy1}, \ref{sec:other}, \ref{sec:critical}, \ref{sec:polar}, we have based our
discussion on the analysis of the averaged system, described by a one--dimensional
(averaged) Hamiltonian function, that
we write in compact form as
\beq{Hav}
\H_{av}=\H_{av}(S,T,\sigma)
\eeq
(compare with \equ{Hexp} where $s$ is used in place of $\sigma$). The coordinates $(S,\sigma)$ admit, after the second bifurcation, a hyperbolic point,
while $T$ is an integral of motion. For a given level $T=T_0$, the product of the hyperbolic point times the
level set $T=T_0$ is a whiskered torus\footnote{We remark that the motion on a whiskered torus is a Diophantine
rotation, while the remaining directions are as hyperbolic as allowed by the symplectic structure of the model.
We have a NHIM, when the tangent direction is dominated by the hyperbolic directions.}.

As $T$ varies we get a NHIM foliated by whiskered tori. Thus, we may wonder what happens when we add the
perturbation to \equ{Hav}. We try to answer this question at least from a qualitative point of view,
being aware that a quantitative analysis requires proper mathematical statements and
several additional computations, which might be the object of study of a future work.

\vskip.1in

We start by noticing that the non--averaged system has the form
\beq{Hnav}
\H_{nav}(S,T,\sigma,\eta)=\H_{av}(S,T,\sigma)+\varepsilon \H_p(S,T,\sigma,\eta)\ ,
\eeq
where $\H_p$ is the perturbing function, given by the part depending on the angles of
$\overline \R_{Moon}^{(res)}$, $\overline \R_{Sun}^{(res)}$ in \equ{R_moon_bigOmegaMoon}, \equ{Rsunaverage},
provided that $\Omega_M$, $\Omega_S$ are considered constants as in assumption {\bf H2}.
It turns out that the parameter $\varepsilon$ is of the order of the square of the eccentricity
and therefore it can be considered small, at least for ranges of the coordinates that correspond
to nearly circular orbits.

Given the expression of the Hamiltonian in \equ{Hnav}, we may use a perturbation approach, which allows
us to state that the NHIM persists for $\varepsilon$ small enough (see, e.g., \cite{Fenichel71, Fenichel74, HirschPS77, Pesin04}).
Under a suitable non--degeneracy condition and a Diophantine assumption on the frequency,
the two--dimensional Hamiltonian $\H_{nav}=\H_{nav}(S,T,\sigma,\eta)$ admits invariant tori
(\cite{Kolmogorov,Arnold,Moser}), provided that $\varepsilon$ satisfies some
smallness constraint.
On a three--dimensional energy level, the two--dimensional KAM tori provide a stability property
in the sense of confinement of the motion.

When the assumption on the constancy of the rates of
variation of $\Omega_M$, $\Omega_S$ is removed, we end up with a
non--autonomous Hamiltonian function of the form \equ{Hres}, that
we can write as \beq{Ht}
\H_{non-aut}(S,T,\sigma,\eta,t)=\H_{av}(S,T,\sigma)+\varepsilon
\H_p(S,T,\sigma,\eta) +\varepsilon \H_t(S,T,\sigma,\eta,\gamma t)\ ,
\eeq
where $\H_t$ is the time--dependent part arising from the
variation of the perturbers' angular elements and $\gamma$ is a
parameter depending on the rates of variation of the longitudes of
the ascending nodes of the Moon and Sun. Precisely, $\gamma$ is nearly
zero for the Sun and it is about equal to $-0.053^{\circ}/day$ for
the Moon, thus showing that the time-dependent part due to the
motion of the ascending node of the Moon is relevant on the time
scale of about 18 years. The 2--dimensional, time--dependent
Hamiltonian $\H_{non-aut}$ does not admit a confinement through
KAM tori, since the phase space has dimension 5 and the KAM tori
have dimension 3. A mechanism giving rise to Arnold's diffusion
(\cite{ArnoldDiff}) can be triggered by the existence of
transition chains generated by the intersection of the unstable manifold of a
torus and the stable manifold of another. The onset of Arnold's
diffusion is subject to the fulfillment of some non--degeneracy
assumptions as well as some conditions on the so--called
Melnikov's integrals (see, e.g., \cite{Holmes}), ensuring the
occurrence of transversal intersections. As an alternative to
Melnikov's method for the study of Arnold's diffusion, one may
compute the so--called \sl scattering map \rm (\cite{DLS1,DLS2}),
which describes the behavior of the excursions between the tori.

\section{Conclusions}\label{sec:conclusions}
The study of lunisolar resonances has raised a renewed interest, thanks to the increasing
awareness of their effects on the space debris orbiting at different altitudes around
our planet. Several works have underlined the importance of the influence of the Moon and Sun
on objects located in MEO and GEO (see, e.g., \cite{rossi2008}). Therefore, a careful
investigation of lunisolar secular resonances is mandatory, also in view of practical
applications.

A first step toward such investigation is represented by the analysis of those resonances
which depend just on the inclination (\cite{HughesI}). These resonances occur in physically
relevant regions, where several space debris are observed to orbit around the Earth.

\red{Understanding the structure of the phase space, even for the simplest mathematical models as those
considered in this work, can offer a simple explanation for the long-term evolution of space debris.
For example, the long-term growth in eccentricity, observed for disposal orbits of various satellites, such as GPS, GLONASS, and GALILEO (see \cite{chaogick}),  may be viewed as a natural effect of the lunisolar resonances. As shown by the
phase--portraits in  Figures~\ref{fig:case1plots}, \ref{fig:accuracy} and \ref{fig:critplots},
and the FLI maps of Figures~\ref{fli_20_gps_geo} and \ref{fli_20_gps_geo_Om180}, inside the libration region,
the action $S$ varies periodically. Since the eccentricity $e$ is related to $S$, it follows naturally
that the eccentricity varies in time. Moreover, if the libration region take a large portion
of the phase space, as in the top middle panel of Figure~\ref{fig:critplots}, then an orbit having a very small initial eccentricity (or $S \simeq 1 $) could become a collision orbit.}
\\

Motivated by the intrinsic physical interest, we have analyzed the bifurcations associated with the
lunisolar secular resonances; the model describing such resonances obviously includes also the effect of the
geopotential. Our study supports that the role of the Moon is definitely relevant and
cannot be neglected within a careful study of space debris dynamics (compare, e.g.,
with Figure~\ref{fig:case1}). We have provided two methods to analyze the occurrence
of bifurcations (see Sections~\ref{sec:birth} and \ref{sec:alternative}).
Both methods are fast and simple to implement; they provide the explicit
values of the orbital elements at which two different bifurcations take place.
Typically, the first bifurcation is associated to the birth of an elliptic
equilibrium point for physically relevant values of the altitude (namely,
above the Earth's radius). The second bifurcation shows the occurrence of
the elliptic equilibrium point together with a pair of hyperbolic equilibria.\\

Care must be taken in analyzing the different classes of lunisolar resonances.
In fact, it should be stressed that when the inclination is such that one resonant variable
becomes degenerate, then the method fails. Moreover, in some cases a higher accuracy
is necessary to get reliable results. This means that the expansions of the Hamiltonian describing the model
must be computed to higher orders as in the case, e.g., of the critical inclination secular resonance.
In all these cases we have pitchfork bifurcations of two different families, the first being stable and the second unstable.
A peculiar case is represented by polar orbits, where both bifurcating families are stable.

As a conclusive remark, we underline that a mathematical study of the
existence of equilibria is also of practical interest. In fact, the elliptic points and their neighboring orbits
represent phase space regions where one could stably place an object, while hyperbolic equilibria
provide unstable regions, where the objects can be placed to experience large excursions
in phase space through the stable and unstable manifolds associated with the hyperbolic
dynamics.

\vskip.2in

{\bf Acknowledgements.}  We are grateful to Christoph Lhotka, Alessandro Rossi, Tere Seara, and
Alfonso Sorrentino for very useful discussions and suggestions.

\end{document}